\documentclass[twocolumn]{aastex631}

\usepackage{graphicx} 

\newcommand{\Msun}{\ensuremath{{\rm\;M_\odot}}}

\begin{document}

\title{JADES Imaging of GN-z11: Revealing the Morphology and Environment\\
of a Luminous Galaxy 430 Myr After the Big Bang}

\author[0000-0002-8224-4505]{Sandro Tacchella}
\altaffiliation{These authors contributed equally to this work.}
\affiliation{Kavli Institute for Cosmology, University of Cambridge, Madingley Road, Cambridge, CB3 0HA, UK}
\affiliation{Cavendish Laboratory, University of Cambridge, 19 JJ Thomson Avenue, Cambridge, CB3 0HE, UK}

\author[0000-0002-2929-3121]{Daniel J.\ Eisenstein}
\altaffiliation{These authors contributed equally to this work.}
\affiliation{Center for Astrophysics $|$ Harvard \& Smithsonian, 60 Garden St., Cambridge MA 02138 USA}

\author[0000-0003-4565-8239]{Kevin Hainline}
\affiliation{Steward Observatory, University of Arizona, 933 N. Cherry Ave., Tucson, AZ 85721, USA}

\author[0000-0002-9280-7594]{Benjamin D.\ Johnson}
\affiliation{Center for Astrophysics $|$ Harvard \& Smithsonian, 60 Garden St., Cambridge MA 02138 USA}

\author[0000-0003-0215-1104]{William M. Baker}
\affiliation{Kavli Institute for Cosmology, University of Cambridge, Madingley Road, Cambridge, CB3 0HA, UK}
\affiliation{Cavendish Laboratory, University of Cambridge, 19 JJ Thomson Avenue, Cambridge, CB3 0HE, UK}

\author[0000-0003-4337-6211]{Jakob M. Helton}
\affiliation{Steward Observatory, University of Arizona, 933 N. Cherry Ave., Tucson, AZ 85721, USA}

\author[0000-0002-4271-0364]{Brant Robertson}
\affiliation{Department of Astronomy and Astrophysics, University of California, Santa Cruz, 1156 High Street, Santa Cruz, CA 95064, USA}

\author[0000-0002-1714-1905]{Katherine A. Suess}
\affiliation{Department of Astronomy and Astrophysics, University of California, Santa Cruz, 1156 High Street, Santa Cruz, CA 95064, USA}
\affiliation{Kavli Institute for Particle Astrophysics and Cosmology and Department of Physics, Stanford University, Stanford, CA 94305, USA}

\author[0000-0002-2178-5471]{Zuyi Chen}
\affiliation{Steward Observatory, University of Arizona, 933 N. Cherry Ave., Tucson, AZ 85721, USA}

\author[0000-0002-7524-374X]{Erica Nelson}
\affiliation{Department for Astrophysical and Planetary Science, University of Colorado, Boulder, CO 80309, USA}

\author[0000-0001-8630-2031]{D\'{a}vid Pusk\'{a}s}
\affiliation{Kavli Institute for Cosmology, University of Cambridge, Madingley Road, Cambridge, CB3 0HA, UK}
\affiliation{Cavendish Laboratory, University of Cambridge, 19 JJ Thomson Avenue, Cambridge, CB3 0HE, UK}

\author[0000-0002-4622-6617]{Fengwu Sun}
\affiliation{Steward Observatory, University of Arizona, 933 N. Cherry Ave., Tucson, AZ 85721, USA}

\author[0000-0002-8909-8782]{Stacey Alberts}
\affiliation{Steward Observatory, University of Arizona, 933 N. Cherry Ave., Tucson, AZ 85721, USA}

\author[0000-0003-1344-9475]{Eiichi Egami}
\affiliation{Steward Observatory, University of Arizona, 933 N. Cherry Ave., Tucson, AZ 85721, USA}

\author[0000-0002-8543-761X]{Ryan Hausen}
\affiliation{Department of Physics and Astronomy, The Johns Hopkins University, 3400 N. Charles St., Baltimore, MD 21218, USA}

\author{George Rieke}
\affiliation{Steward Observatory, University of Arizona, 933 N. Cherry Ave., Tucson, AZ 85721, USA}

\author[0000-0002-7893-6170]{Marcia Rieke}
\affiliation{Steward Observatory, University of Arizona, 933 N. Cherry Ave., Tucson, AZ 85721, USA}

\author[0000-0003-4702-7561]{Irene Shivaei}
\affiliation{Steward Observatory, University of Arizona, 933 N. Cherry Ave., Tucson, AZ 85721, USA}

\author[0000-0003-2919-7495]{Christina C. Williams}
\affiliation{NSF’s National Optical-Infrared Astronomy Research Laboratory, 950 North Cherry Avenue, Tucson, AZ 85719, USA}
\affiliation{Steward Observatory, University of Arizona, 933 N. Cherry Ave., Tucson, AZ 85721, USA}

\author[0000-0001-9262-9997]{Christopher N. A. Willmer}
\affiliation{Steward Observatory, University of Arizona, 933 N. Cherry Ave., Tucson, AZ 85721, USA}

\author[0000-0002-8651-9879]{Andrew Bunker}
\affiliation{Department of Physics, University of Oxford, Denys Wilkinson Building, Keble Road, Oxford OX1 3RH, UK}

\author[0000-0002-0450-7306]{Alex J. Cameron}
\affiliation{Department of Physics, University of Oxford, Denys Wilkinson Building, Keble Road, Oxford OX1 3RH, UK}

\author[0000-0002-6719-380X]{Stefano Carniani}
\affiliation{Scuola Normale Superiore, Piazza dei Cavalieri 7, I-56126 Pisa, Italy}

\author[0000-0003-3458-2275]{Stephane Charlot}
\affiliation{Sorbonne Universit\'e, CNRS, UMR 7095, Institut d'Astrophysique de Paris, 98 bis bd Arago, 75014 Paris, France}

\author[0000-0002-2678-2560]{Mirko Curti}
\affiliation{Kavli Institute for Cosmology, University of Cambridge, Madingley Road, Cambridge, CB3 0HA, UK}
\affiliation{Cavendish Laboratory, University of Cambridge, 19 JJ Thomson Avenue, Cambridge, CB3 0HE, UK}
\affiliation{European Southern Observatory, Karl-Schwarzschild-Strasse 2, D-85748 Garching bei M{\"u}enchen, Germany}

\author[0000-0002-9551-0534]{Emma Curtis-Lake}
\affiliation{Centre for Astrophysics Research, Department of Physics, Astronomy and Mathematics, University of Hertfordshire, Hatfield AL10 9AB, UK}

\author[0000-0002-3642-2446]{Tobias J. Looser}
\affiliation{Kavli Institute for Cosmology, University of Cambridge, Madingley Road, Cambridge, CB3 0HA, UK}
\affiliation{Cavendish Laboratory, University of Cambridge, 19 JJ Thomson Avenue, Cambridge, CB3 0HE, UK}

\author[0000-0002-4985-3819]{Roberto Maiolino}
\affiliation{Kavli Institute for Cosmology, University of Cambridge, Madingley Road, Cambridge, CB3 0HA, UK}
\affiliation{Cavendish Laboratory, University of Cambridge, 19 JJ Thomson Avenue, Cambridge, CB3 0HE, UK}
\affiliation{Department of Physics and Astronomy, University College London, Gower Street, London WC1E 6BT, UK}

\author[0000-0003-0695-4414]{Michael V. Maseda}
\affiliation{Department of Astronomy, University of Wisconsin-Madison, 475 N. Charter St., Madison, WI 53706 USA}

\author[0000-0002-7028-5588]{Tim Rawle}
\affiliation{European Space Agency, Space Telescope Science Institute, Baltimore, Maryland, US}

\author[0000-0003-4996-9069]{Hans-Walter Rix}
\affiliation{Max-Planck-Institut f{\"u}r Astronomie, K{\"o}nigstuhl 17, D-69117, Heidelberg, Germany}

\author[0000-0001-8034-7802]{Renske Smit}
\affiliation{Astrophysics Research Institute, Liverpool John Moores University, 146 Brownlow Hill, Liverpool L3 5RF, UK}

\author[0000-0003-4891-0794]{Hannah {\"U}bler}
\affiliation{Kavli Institute for Cosmology, University of Cambridge, Madingley Road, Cambridge, CB3 0HA, UK}
\affiliation{Cavendish Laboratory, University of Cambridge, 19 JJ Thomson Avenue, Cambridge, CB3 0HE, UK}

\author[0000-0002-4201-7367]{Chris Willott}
\affiliation{NRC Herzberg, 5071 West Saanich Rd, Victoria, BC V9E 2E7, Canada}

\author[0000-0002-7595-121X]{Joris Witstok}
\affiliation{Kavli Institute for Cosmology, University of Cambridge, Madingley Road, Cambridge, CB3 0HA, UK}
\affiliation{Cavendish Laboratory, University of Cambridge, 19 JJ Thomson Avenue, Cambridge, CB3 0HE, UK}

\author[0000-0002-4735-8224]{Stefi Baum}
\affiliation{Department of Physics and Astronomy, University of Manitoba, Winnipeg, MB R3T 2N2, Canada}

\author[0000-0003-0883-2226]{Rachana Bhatawdekar}
\affiliation{European Space Agency, ESAC/ESAC, Camino Bajo del Castillo s/n, 28692 Villanueva de la Cañada, Madrid, Spain}
\affiliation{European Space Agency, ESA/ESTEC, Keplerlaan 1, 2201 AZ Noordwijk, NL}

\author[0000-0003-4109-304X]{Kristan Boyett}
\affiliation{School of Physics, University of Melbourne, Parkville 3010, VIC, Australia}
\affiliation{ARC Centre of Excellence for All Sky Astrophysics in 3 Dimensions (ASTRO 3D), Australia}

\author[0000-0002-9708-9958]{A. Lola Danhaive}
\affiliation{Kavli Institute for Cosmology, University of Cambridge, Madingley Road, Cambridge, CB3 0HA, UK}
\affiliation{Cavendish Laboratory, University of Cambridge, 19 JJ Thomson Avenue, Cambridge, CB3 0HE, UK}

\author[0000-0002-2380-9801]{Anna de Graaff}
\affiliation{Max-Planck-Institut f{\"u}r Astronomie, K{\"o}nigstuhl 17, D-69117, Heidelberg, Germany}

\author[0000-0003-4564-2771]{Ryan Endsley}
\affiliation{Department of Astronomy, University of Texas, Austin, TX 78712, USA}

\author[0000-0001-7673-2257]{Zhiyuan Ji}
\affiliation{Steward Observatory, University of Arizona, 933 N. Cherry Ave., Tucson, AZ 85721, USA}

\author[0000-0002-6221-1829]{Jianwei Lyu}
\affiliation{Steward Observatory, University of Arizona, 933 N. Cherry Ave., Tucson, AZ 85721, USA}

\author[0000-0001-9276-7062]{Lester Sandles}
\affiliation{Kavli Institute for Cosmology, University of Cambridge, Madingley Road, Cambridge, CB3 0HA, UK}
\affiliation{Cavendish Laboratory, University of Cambridge, 19 JJ Thomson Avenue, Cambridge, CB3 0HE, UK}

\author[0000-0001-5333-9970]{Aayush Saxena}
\affiliation{Department of Physics, University of Oxford, Denys Wilkinson Building, Keble Road, Oxford OX1 3RH, UK}
\affiliation{Department of Physics and Astronomy, University College London, Gower Street, London WC1E 6BT, UK}

\author{Jan Scholtz}
\affiliation{Kavli Institute for Cosmology, University of Cambridge, Madingley Road, Cambridge, CB3 0HA, UK}
\affiliation{Cavendish Laboratory, University of Cambridge, 19 JJ Thomson Avenue, Cambridge, CB3 0HE, UK}

\author[0000-0001-8426-1141]{Michael W. Topping}
\affiliation{Steward Observatory, University of Arizona, 933 N. Cherry Ave., Tucson, AZ 85721, USA}

\author[0000-0003-1432-7744]{Lily Whitler}
\affiliation{Steward Observatory, University of Arizona, 933 N. Cherry Ave., Tucson, AZ 85721, USA}

\correspondingauthor{Sandro Tacchella}
\email{st578@cam.ac.uk}



\begin{abstract}
We present JWST NIRCam 9-band near-infrared imaging of the luminous $z=10.6$ galaxy GN-z11 from the JWST Advanced Deep Extragalactic Survey (JADES) of the GOODS-N field.  We find a spectral energy distribution (SED) entirely consistent with the expected form of a high-redshift galaxy: a clear blue continuum from 1.5 to 4 microns with a complete dropout in F115W.  The core of GN-z11 is extremely compact in JWST imaging.  We analyze the image with a two-component model, using a point source and a S\'{e}rsic profile that fits to a half-light radius of 200 pc and an index $n=0.9$. We find a low-surface brightness haze about $0.4''$ to the northeast of the galaxy, which is most likely a foreground object but might be a more extended component of GN-z11. At a spectroscopic redshift of 10.60 \citep{bunker23}, the comparison of the NIRCam F410M and F444W images spans the Balmer jump. From population synthesis modeling, here assuming no light from an active galactic nucleus, we reproduce the SED of GN-z11, finding a stellar mass of $\sim$$10^{9}~M_{\odot}$, a star-formation rate of $\sim$$20~M_{\odot}~\mathrm{yr}^{-1}$ and a young stellar age of $\sim$$20$ Myr. As massive galaxies at high redshift are likely to be highly clustered, we search for faint neighbors of GN-z11, finding 9 galaxies out to $\sim$5 comoving Mpc transverse with photometric redshifts consistent with $z=10.6$, and a 10$^{\rm th}$ more tentative dropout only $3''$ away. This is consistent with GN-z11 being hosted by a massive dark-matter halo ($\approx8\times10^{10}~M_{\odot}$), though lower halo masses cannot be ruled out.
\end{abstract}


\section{Introduction}
\label{sec:intro}

Measuring the abundance and physical properties of the highest-redshift galaxies is crucial to understand and constrain the earliest stages of galaxy formation and evolution, including the formation of the first stars and black holes \citep{stark16, dayal18, robertson22}. The first galaxies are a sensitive probe for a range of baryonic processes (gas cooling and energetic feedback from stars and black holes), structure formation and the nature of dark matter \citep{dayal15, khimey21, gandolfi22}. 

Before the advent of JWST, these early galaxies at redshift $z>10$ have been selected using a combination of Hubble Space Telescope (HST) and Spitzer measurements  \citep[e.g.,][]{ellis13, oesch13b, mcleod16, bouwens19, bouwens21, finkelstein15, finkelstein22_hst}. One of the most distant galaxy found in this way is the spectroscopically confirmed GN-z11 \citep{oesch16, jiang21}, pushing the limits to $z \sim 11$. JWST rapidly increased the number of $z > 10$ discoveries, finding luminous $z=10-12$ galaxy candidates \citep{adams23, atek23, donnan23, finkelstein22, harikane23_uvlf, naidu22_highz} and four spectroscopically confirmed $10.3 \leq z \leq 13.2$ galaxies \citep[][]{curtis-lake23, robertson23}. 

Using data from the deep GOODS NICMOS Survey \citep{conselice11}, GN-z11 was initially identified by \citet{bouwens10_nicmos} under the designation GNS-JD2, located at 12:36:25.44, +62:14:31.3\footnote{Position based on the analysis presented in this work.}. There was no evidence that GN-z11 was detected at wavelengths other than 1.6 $\mu$m ($H$ band), but because it is close to another source, it was unclear if it was detected redward of 2 $\mu$m from the IRAC data. Therefore, \citet{bouwens10_nicmos} concluded GN-z11 could be a $z\sim9$ galaxy, but considered it unlikely since it could also be a transient source or spurious given its brightness (apparent magnitude of $\approx26$ in HST $H$ band). GN-z11 was then again identified as a $z\sim9-10$ candidate by \citet{oesch14} with designation GN-z10-1. Its redshift was later determined by HST grism spectroscopy to be $z_{\rm grism} = 11.09^{+0.08}_{-0.12}$ \citep{oesch16}. 

GN-z11 is unusually bright with $M_{\rm UV}=-21.6$ AB mag ($\mathrm{F200W}=144.4\pm2.7$ nJy, see Table~\ref{tab:nc}). For comparison, the other recent $z=10-13$ spectroscopically confirmed galaxies have $M_{\rm UV}$ in the range of $-18.4$ to $-19.3$ AB mag ($\mathrm{F200W}=6-14$ nJy; \citealt{robertson23}). GN-z11's brightness makes ground-based observations feasible. Based on the probable detections of three UV emission lines ([CIII]$\lambda\lambda1907,1909$ doublet and OIII]$\lambda1666$), \citet{jiang21} found it to be at $z = 10.957 \pm 0.001$. By modeling the spectral energy distribution (SED) using the photometry in 5 bands (HST F140W, F160W, ground-based K band and Spitzer channels 1 and 2), \citet{jiang21} constrained the stellar mass, dust attenuation and stellar age. They found that GN-z11 hosts a young stellar population with an age of $70 \pm 40$ Myr, stellar mass of $(1.3 \pm 0.6) \times 10^9~M_{\odot}$, a UV continuum slope of $\beta=-2.4\pm0.2$ and essentially no dust attenuation. Therefore, this surprisingly bright object has a relatively large stellar mass at this epoch ($\sim 420$ Myr after the Big Bang), which suggests a rapid build-up of stellar mass, but consistent with recent JWST-based measurements of other spectroscopically-confirmed galaxies at $z>10$ \citep{curtis-lake23, robertson23}.

Here we present new JWST observations of GN-z11 in the GOODS-N field. These observations have been conducted as part of the JWST Advanced Deep Extragalactic Survey (JADES). In this paper we focus on the 9-band JWST/NIRCam imaging, providing constraints on the SED, morphology and large-scale environment of GN-z11. The JWST/NIRSpec observations are presented in a companion paper \citep{bunker23}. The NIRSpec spectrum definitively measures a redshift, $z_{\rm spec}=10.60$, and provides a rich set of diagnostics of the physical properties of the galaxy.  In the listing of JADES confirmed redshifts, GN-z11 is given the designation JADES-GN-z10-0.

We present the details of the observations, data reduction and photometry measurement in Section~\ref{sec:data}. We present a detailed analysis of the morphology of GN-z11 in Section ~\ref{sec:morphology}, while we perform the SED analysis and present key results on the stellar populations in Section~\ref{sec:sed}. Section~\ref{sec:environment} discusses the large-scale environment of GN-z11. The conclusions are presented in Section~\ref{sec:conclusions}. Throughout this work, we use the AB magnitude system and assume the Planck18 flat $\Lambda$CDM cosmology \citep{planck-collaboration20} with $\Omega_m=0.315$ and $H_0=67.4$ km/s/Mpc.  It is useful to note that in this cosmology $1''$ corresponds to a transverse distance of 4.08 proper kpc and 47 comoving kpc at $z=10.6$.

\section{Data and Photometry}
\label{sec:data}

\subsection{Observations}
\label{subsec:observations}

\newcommand{\nod}{\nodata}

The NIRCam observations presented here come from the JWST Advanced Deep Extragalactic Survey (JADES), which is conducting deep JWST imaging and spectroscopy of the GOODS-S and GOODS-N fields.  The 9-band imaging of GN-z11 results from the combination of two NIRCam pointings, observations 2 and 3, from program 1181 (PI: Eisenstein) taken on UT2023-02-03.  Each pointing is a 6-point dither, conducted with a two-point subdither with the MIRI F1800W pattern\footnote{to support the parallel imaging obtained simultaneously with MIRI} in a 3-part IntramoduleX dither.  The two pointings intentionally overlap on GN-z11, but in opposite portions of NIRCam module B (detectors B3 and B2).  Hence, in most filters, GN-z11 was placed on 12 distinct and well-separated pixels.  Both pointings include 8 filters: F090W, F115W, F150W, F200W, F277W, F356W, F410M, and F444W.  One pointing additionally includes F335M, paired with an extra 6 exposures of F115W.  The exposure times are listed in Table \ref{tab:nc}.

\subsection{Data Reduction}
\label{subsec:reduction}

\begin{figure*}
    \centering
    \includegraphics[width=\textwidth]{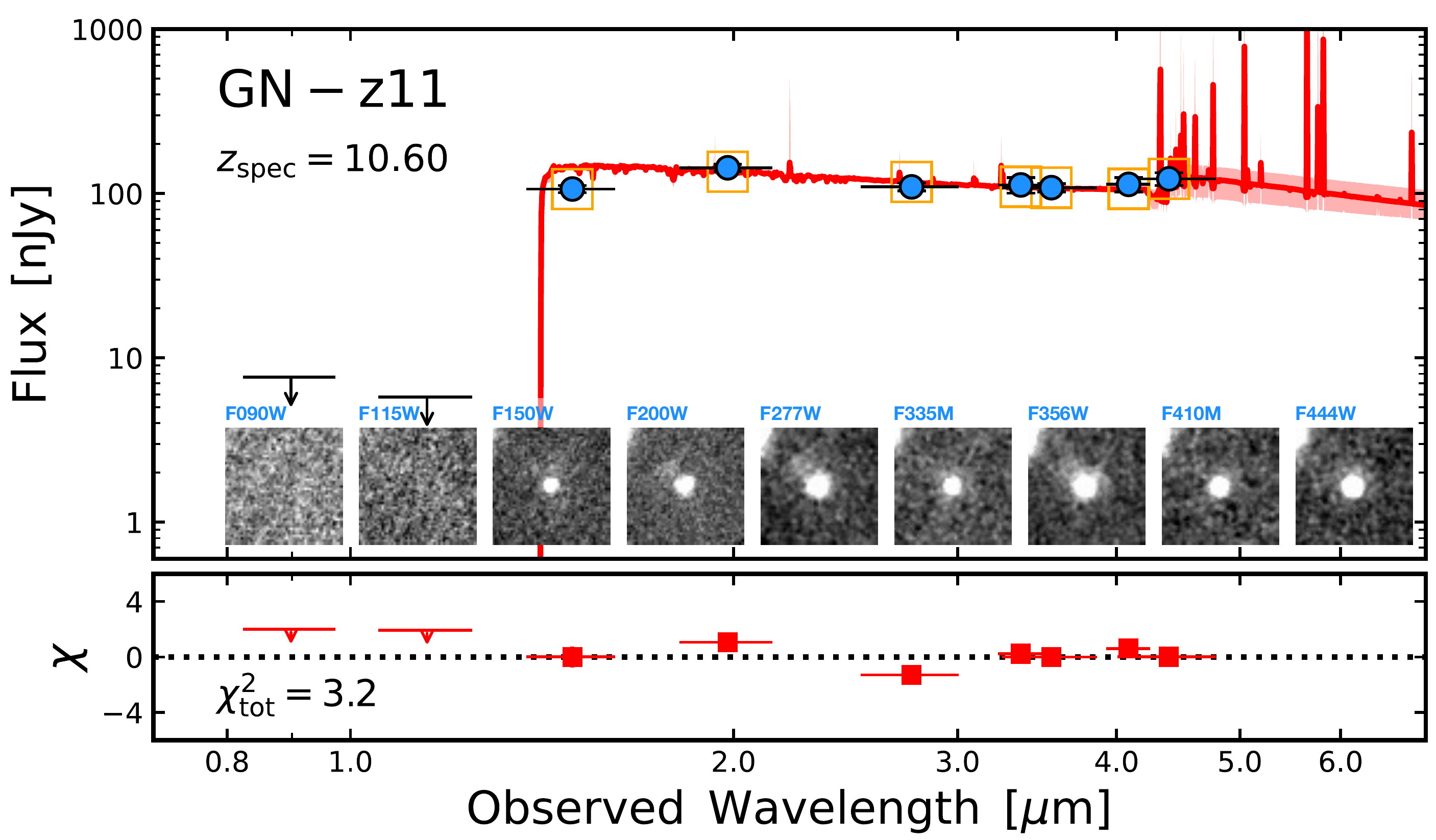}
    \caption{The spectral energy distribution and thumbnail images of GN-z11 from 9 JWST NIRCam bands. Each thumbnail square is 1.5 arcsec on a side. The blue points show the 7 detected NIRCam bands (``Galaxy'' in Table~\ref{tab:modelphotom}), while the 2$\sigma$ upper limits for F090W and F115W are indicated as downward pointing arrows. The orange squares mark the photometry from the best-fit SED model. The red solid line and the shaded region shows the median the 16th-84th percentile of the SED posterior from the Prospector modeling (Section~\ref{sec:sed}). }
    \label{fig:sed}
\end{figure*}

The details of the data reduction of the NIRCam data will be presented as part of the JADES program in Tacchella et al. (in prep.). We give here a brief overview of the main steps. We use the JWST Calibration Pipeline v1.9.2 with the CRDS pipeline mapping (pmap) context 1039. We process the raw images (\texttt{uncal} frames) with the JWST Stage 1, which performs detector-level corrections and produces count-rate images (\texttt{rate} frames). We run this step with the default parameters, including masking and correction of ``snowballs'' in the images caused by charge deposition following cosmic ray hits. 

Stage 2 of the JWST pipeline performs the flat-fielding and applies the flux calibration (conversion from counts/s to MJy/sr; \citealt{boyer22}). For the long-wavelength (LW) filters, we find that the current pipeline flats (ground flat corrected for in-flight performance) introduce artifacts in the background that become visible in the final mosaics. We therefore construct sky-flats for the Stage 2 step  by stacking, separately for each LW filter and module, $80-200$ \texttt{uncal} frames from PID 1180, 1210 1286, and the public program JEMS \citep[JWST Extragalactic Medium-band Survey; PID 1963;][]{williams23}. Since we do not have enough exposures to construct a robust sky-flat for the F335M and F410M filters, we have effectively interpolated those flat fields via {a linear combination of flat field ``components'' determined from a non-negative factorization of} the wide-band sky-flats (i.e. F227W, F356W and F444W). The rest of Stage 2 is run with the default values. 

Following Stage 2, we perform several custom corrections in order to account for several features in the NIRCam images \citep{rigby22}, including the 1/f noise \citep{schlawin20}, scattered-light effects (``wisps'') and the large-scale background. Since all of those effects are additive, we fit and subtract them. We assume a parametric model for the 1/f noise, a scaled wisp template (only for the SW channel detectors A3, A4, B3 and B4) and a constant, homogeneous large-scale background. We have constructed wisp templates by stacking all images from our JADES (PID 1180, 1210, 1286) program and several other programs that are publicly available (PIDs 1063, 1345, 1837, 2738). 

The final mosaics are constructed using Stage 3 of the JWST Pipeline. Before combining the individual exposures into a mosaic, we perform an astrometric alignment using a custom version of JWST TweakReg. We calculate both the relative and absolute astrometric correction for images grouped by visit and band by matching sources to a reference catalogue constructed from HST F814W and F160W mosaics in the GOODS-N field with astrometry tied to Gaia-EDR3 (\citealt{gaia_2021}; G. Brammer priv. comm.). We then run Stage 3 of the JWST pipeline, combining all exposures of a given filter and a given visit. We choose a pixel scale of 0.03 arcsec/pixel and drizzle parameter of pixfrac=1 for the SW and LW images\footnote{The pixel scale of the original SW and LW images are 0.031 and 0.064 arcsec, respectively.}. A careful analysis of the astrometric quality reveals an overall good alignment with relative offsets between bands of less than 0.1 short-wavelength pixel ($<3$ mas).

The resulting thumbnail images together with the SED are shown in Figure \ref{fig:sed}. These thumbnails show a compact source at $\alpha=189.106042^\circ$, $\delta=+62.242042^\circ$ that is an obvious F115W dropout, with a faint haze to the north-east. The haze is seen in several bands, including in both the SW and LW detectors, establishing it as a true on-sky signal. 

\subsection{Photometry}
\label{subsec:photometry}

\begin{table}[t]
    \begin{center}
        \caption{\label{tab:nc}Data and aperture photometry of GN-z11.}
    \noindent\begin{tabular}{cccc}\hline
        Filter &  Exposure & 0.7$''$ Flux & 0.2$''$ Flux  \\
        & (ks) & (nJy) & (nJy) \\
        \hline
        F090W &  6.18   &  $-2.9\pm4.1$  & $-1.0 \pm 1.7$ \\
        F115W &  9.92   &   $1.2\pm3.1$  &  $1.2\pm1.5$\\
        F150W &  6.18   & $115.9\pm3.3$ & $99.2\pm1.6$\\
        F200W &  6.18   &  $144.4\pm2.7$  &$135.2\pm1.5$\\
        \hline
        F277W &  6.18     & $121.7\pm4.2$ &$112.0\pm1.0$\\
        F335M &  3.09$^a$ & $132.9\pm6.3$ &$107.2\pm1.9$\\\
        F356W &  6.18     &  $123.5\pm3.9$&$106.7\pm1.0$\\
        F410M &  6.83     &  $114.9\pm4.8$&$109.9\pm1.4$\\
        F444W &  6.18     &  $133.8\pm4.5$&$121.0\pm1.3$\\
        \hline
        F435W & & $7.3\pm10.7$ & \\
        F606W & & $1.4\pm1.5$ & \\
        F775W & & $-1.3\pm6.5$ & \\
        F814W & & $-2.8\pm7.4$ & \\
        F850LP & & $5.5\pm16$ & \\
        F105W & & $-5.1\pm5.5$ & \\
        F125W & & $2.2\pm4.2$ & \\
        F140W & & $49.5\pm4.9$ & \\
        F160W & & $112.2\pm4.9$ & \\
        \hline
    \end{tabular}
    \end{center}\smallskip
    Notes: Photometry is presented here in 0.7$''$ and 0.2$''$ diameter apertures, with point-source aperture corrections. The best-fit ForcePho centroid is used as the center of the aperture photometry. The last 9 bands of HST ACS and WFC3 photometry are 0.7$''$ diameter aperture photometry from the HST Hubble Legacy Field v2.5 images of GOODS-N. 
    $^a$F335M was observed in only one of the two JADES pointings and hence has 6 dither locations instead of 12. 
\end{table}

Using an inverse-variance-weighted stack of the F277W, F335M, F356W, F410M, and F444W images, a signal-to-noise-ratio image is constructed to provide a detection image. Contiguous regions of greater than five pixels with signal-to-noise ratio (SNR) $\ge3$ were selected as potential sources. For every source location, forced photometry was performed in $0.1''$ and $0.35''$~radius circular apertures on the JADES/NIRCam and the 30mas pixel scale HST Legacy Fields mosaics \citep{illingworth13, whitaker19} for the ACS F435W, F606W, F775W, F814W, F850LP and WFC3/IR F105W, F125W, F140W, and F160W filters. We used \texttt{photutils} to perform the force photometry measurements \citep{bradley22}. An annular aperture of width $\Delta r=0.1''$ and inner radius $r=0.4''$ about each source is used to measure and remove the local background. subtracted background fluxes $1-9$ nJy, i.e. roughly $3-5\%$ of the source flux (maximum of 8\% in F150W). No PSF matching was performed on the HST and JWST bands for the forced photometry, but we perform a point-source aperture correction. The $0.1''$ ($0.35''$) aperture corrections for F090W, F115W, F150W, F200W, F277W, F335M, F356W, F410M, and F444W amount to  1.37 (1.11), 1.31 (1.10), 1.33 (1.11), 1.38 (1.10), 1.63 (1.15), 1.74 (1.14), 1.79 (1.15), 1.88 (1.13), and 2.07 (1.17).

We note that in the JWST SW images, GN-z11 lies directly on a diffraction spike from an F115W$\sim18.4$ AB star at $[\alpha,\delta] \simeq [189.10568, 62.2458]$ that necessitates the local background correction.  We report the aperture photometry in Table \ref{tab:nc}, which includes both sky and source photon contributions. The uncertainties are computed by adding in quadrature the contribution from the sky measured by placing random apertures on the images and the Poisson uncertainty from the source counts. For the long wavelength bands, these uncertainties are larger than those measured at the source locations in the \textit{JWST} pipeline ERR mosaics, which also include both sky and source photon contributions, because the random aperture sky uncertainties include the effects of correlated noise from resampling the mosaics.  Our quoted uncertainties do not include any contribution from photometric zero-point uncertainties or from large-scale gradients in the instrument flat fields. We include an error floor of 5\% in our SED and photometric redshift fits to hedge against these.

The SED of GN-z11 is displayed in Figure \ref{fig:sed} and shows the classic shape of a high-redshift Ly$\alpha$ dropout.  The continuum is strong and blue, roughly zero color in the AB system, before plummeting shortward of the break.
\citet{bunker23} provides a robust determination of $z=10.60$ through the detection of many well-detected narrow lines.  The photometric measurements are wholly consistent with this redshift.  Ly$\alpha$ is shifted to 1.41~$\mu$m, lying in the F150W filter.  We observe a rest-UV flux density of $144.4\pm2.7$ nJy in F200W.  The F115W filter lies entirely shortward of the Ly$\alpha$ wavelength and is indeed observed as a complete dropout, with a flux of $1.2 \pm 3.1$  nJy.  This is a very strong suppression of a factor of at least $\sim20$ (95\% confidence), nearly 3 magnitudes.

The last nine bands of Table~\ref{tab:nc} are the ACS and WFC3 0.7$''$ diameter aperture photometry from the HST Hubble Legacy Field v2.5 images of GOODS-N. \citet{oesch14} reported $-7\pm9$ nJy, $11\pm8$ nJy, $64\pm13$ nJy and $152\pm10$ nJy for HST WFC3 F105W, F125W, F140W, and F160W, respectively. Our measurements are in agreement with those of \citet{oesch14}, excepting F160W for which we measure a flux lower by about 25\% ($3\sigma$ lower). The Spitzer IRAC photometry of $139\pm20$ and $122\pm21$ nJy at 3.6 and 4.5 $\mu$m \citep{ashby13, oesch14} are also in good agreement.

Longward of Ly$\alpha$, the ratio between the fluxes measured in F200W and F356W indicates a rest-UV slope of $\beta=-2.41_{-0.07}^{+0.06}$, where the flux density scales as $f_\nu\propto\lambda^{\beta+2}$. This is consistent with the value determined from the prism spectrum of $\beta=-2.36\pm0.10$ \citep{bunker23}. At the longest NIRCam wavelengths, the F410M filter corresponds to 3360 to 3700~\AA\ in the rest-frame, just blueward of the Balmer jump, while the F444W filter extends to 4300 \AA.  Hence, the comparison of these two filters can be sensitive to the presence of older stars.

\section{Morphological Results}
\label{sec:morphology}

\begin{figure*}[t]
\noindent
\includegraphics[width=0.2\textwidth]{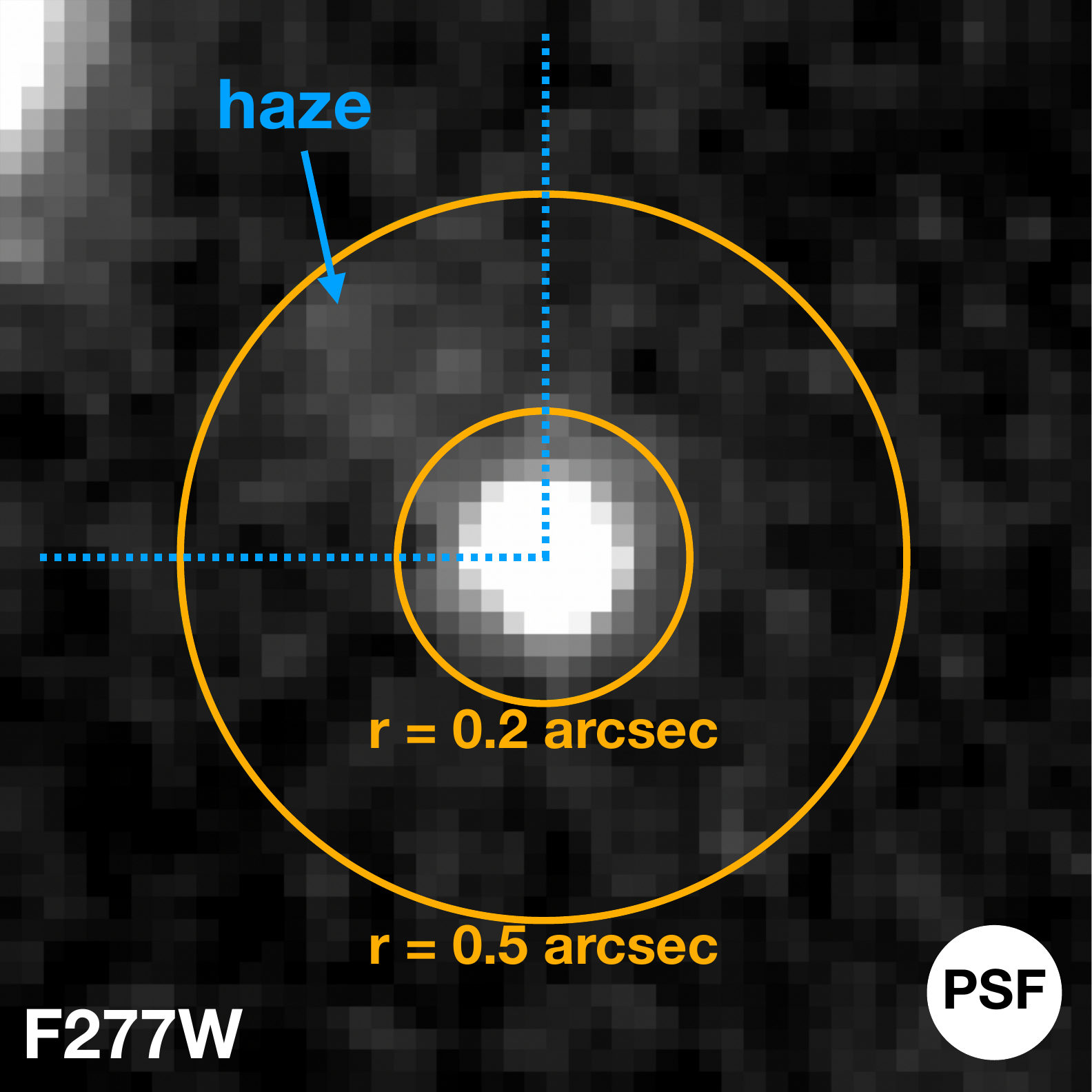} 
\includegraphics[width=0.26\textwidth]{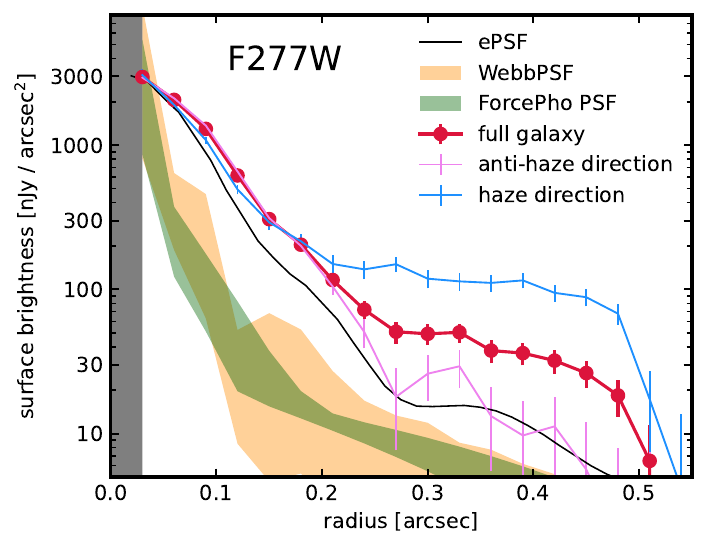} 
\includegraphics[width=0.26\textwidth]{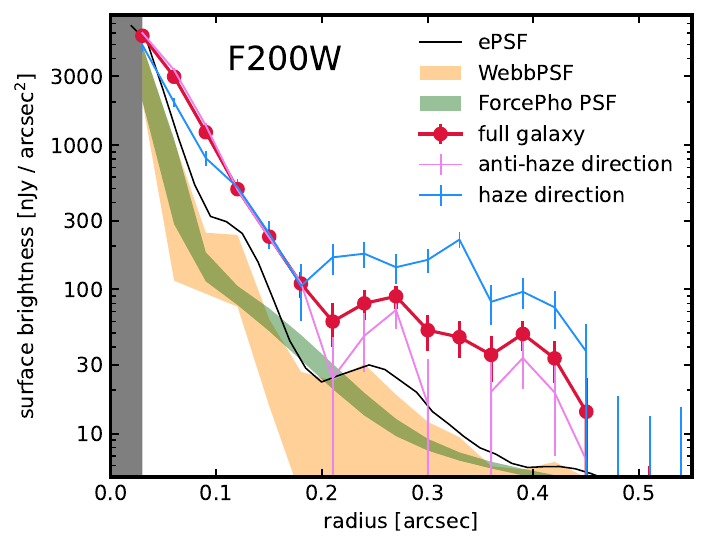} 
\includegraphics[width=0.26\textwidth]{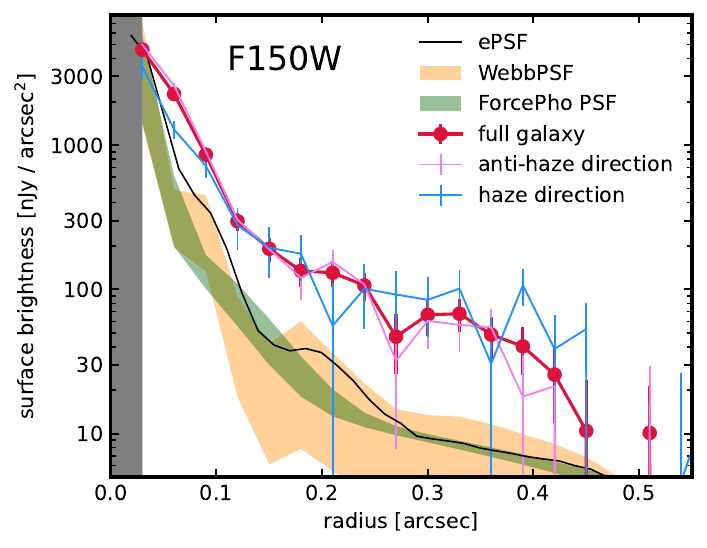} 
\caption{Encircled light profiles for GN-z11 in F277W, F200W, and F150W. The thumbnail on the left shows the image in F277W. The three panels to the right plot the surface brightness profiles in red. The thin black line shows the decline of the empirical PSF (ePSF) profiles, while the orange and green shaded regions (indicating the 16--84th percentile) show the PSF from WebbPSF and ForcePho, all normalized to the flux in the central pixel (marked as gray region). We find that GN-z11 is only marginally extended in the inner $\approx0.2$ arcsec. We model this central region with two components: a ``Point Source'' (PS) and an ``Extended'' component (see Figures~\ref{fig:FP_residual} and \ref{fig:flux_ratio_size}). To investigate the haze at a distance of 0.4 arcsec, we isolate the quadrant containing the haze and compare its profile to that of the other three. The haze is considerably less prominent in F150W. }
\label{fig:sbprofile}
\end{figure*}

\begin{figure*}[t]
\noindent
\includegraphics[width=1.0\textwidth]{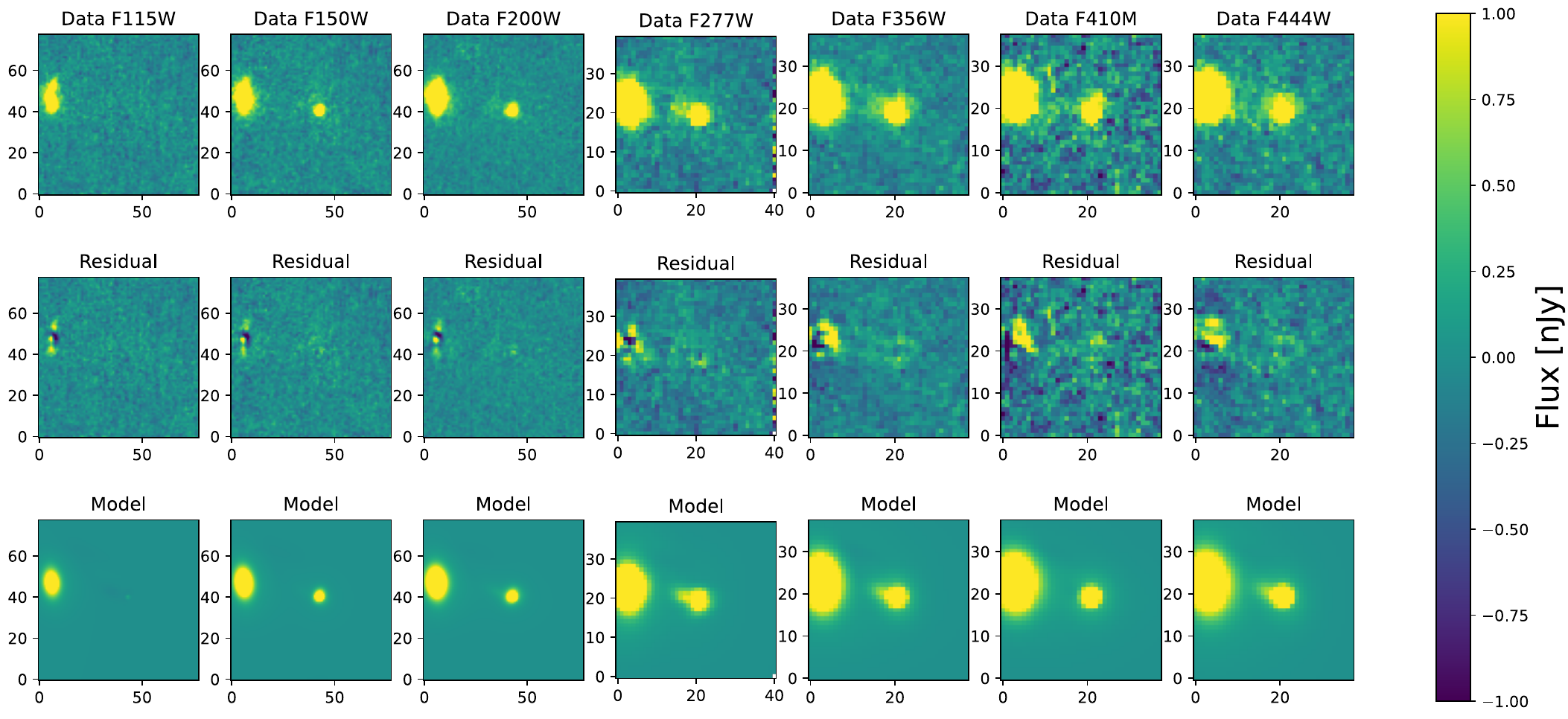} 
\caption{The scene fitting from ForcePho.  We model the immediate region of GN-z11 with 3 components (in addition to the bright source on the left): a point source, an extended S\'{e}rsic component near that source, and a second S\'{e}rsic centered on the haze. The two elliptical S\'{e}rsic components are free in index $n$, half-light radius, position angle, axis ratio, and centroid.  All three components vary freely in amplitude in each band.  A fourth S\'ersic component is used to model the brighter galaxy on the left of the thumbnails.  ForcePho fits to the individual exposures, producing a Markov Chain sampling of the likelihood of the scene. To display these residuals, we subtracted the model from the data exposures and then combined the exposures into a mosaic for each band. The residuals are very weak, indicating that the three component model has explained most of the scene.  We note that these images are displayed in pixel coordinates, unlike other images in this paper, which use the north-up mosaic. All maps are showing flux in nJy (see color bar) in each pixel.
}
\label{fig:FP_residual}
\end{figure*}

\begin{figure}[t]
\noindent
\includegraphics[width=1.0\linewidth]{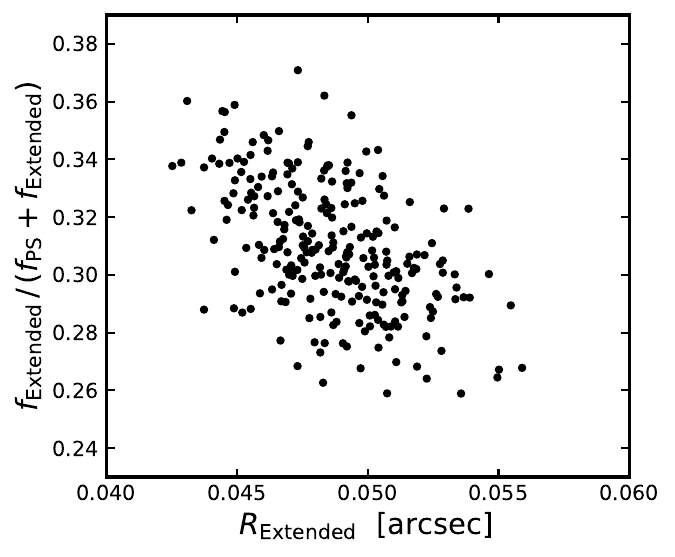} 
\caption{Posterior distribution of the fraction of $F200W$ flux in the Extended component as a function of its half-light size (i.e., major axis radius). 
We obtain this directly from the ForcePho modeling. We find an interesting covariance: the larger the Extended component, the less flux it contributes. Importantly, there is no tail towards zero size or zero flux, emphasizing that the Extended component is clearly detected and indeed extended. We find the same qualitative behaviour for all other bands.}
\label{fig:flux_ratio_size}
\end{figure}

\begin{table}[t]
    \begin{center}
        \caption{\label{tab:modelphotom}Model photometry of GN-z11}
    \noindent\begin{tabular}{ccccc}\hline
        Filter & Galaxy & PS & Extended & Haze  \\
         & (nJy) & (nJy) & (nJy) & (nJy) \\
        \hline
        F090W &  $-3.5\pm1.2$ & $1.2\pm1.0$ & $-4.6\pm5.1$ & $-5.0\pm5.5$ \\
        F115W &  $-2.6\pm1.1$ & $1.5\pm0.8$ & $-4.1\pm4.5$ & $-4.4\pm4.9$ \\
        F150W &  $106.5\pm1.9$ & $64.5\pm2.2$ & $42.0\pm2.7$ & $3.0\pm3.0$ \\
        F200W &  $143.4\pm1.8$ & $99.0\pm2.9$ & $44.4\pm3.4$ & $6.5\pm2.4$\\
        \hline
        F277W & $109.8\pm2.0$ & $66.3\pm3.9$ & $43.5\pm4.8$ & $19.6\pm2.8$ \\
        F335M & $112.8\pm4.0$ & $63.7\pm7.9$ & $49.1\pm9.5$ & $3.7\pm4.4$ \\
        F356W &  $108.6\pm1.8$ & $71.6\pm3.9$ & $37.0\pm4.7$ & $15.7\pm2.3$ \\
        F410M &  $113.7\pm2.9$ & $85.9\pm7.2$ & $27.9\pm8.7$ & $-2.9\pm3.2$ \\
        F444W &  $122.9\pm2.7$ & $75.8\pm7.0$ & $47.2\pm8.5$ & $12.5\pm3.3$ \\
        \hline
    \end{tabular}
    \end{center}\smallskip
    Notes: Photometry from ForcePho simultaneous PSF-convolved modeling of the scene (Fig.~\ref{fig:FP_residual}). The ``Galaxy'' consists of the combination of a point source (``PS'') and a extended S\'{e}rsic profile (``Extended'') component.  The mean of the Markov Chain of the latter yields a S\'{e}rsic index $n$ of $0.9\pm0.1$ with a half-light radius of $49\pm3$ mas and an axis ratio of $q=0.67\pm0.05$, slightly offset from the point source by 21 mas.  The ``Haze'' is a separate S\'{e}rsic component, which yields a mean result of $n=0.9\pm0.1$, a half-light radius of $0.11''\pm0.01''$ and an axis ratio of $q=0.57\pm0.15$, offset by $0.41''$ from the point source.
\end{table}

\subsection{Fitting methods}

To investigate the morphology of GN-z11, we need to assess the PSF and then fit models to our multi-band images.  We do this in several ways.

We perform fits using ForcePho (Johnson et al., in prep.), which fits multiple S\'{e}rsic components to our individual exposures across all filters, producing a Markov Chain of joint parameter fits. This method was used and is described further in \citet{tacchella22}, \citet{robertson23} and \citet{robertson23}. ForcePho uses Gaussian mixtures and graphics processing units to accelerate the convolution of S\'{e}rsic models with the point spread function, taken from WebbPSF \citep{webbpsf}. We note that the Gaussian mixture model of the PSF is designed to reproduce the inner core and smoothed light profile; it does not have enough flexibility to capture the Airy rings of the PSF.  In our analysis, we require that each component have the same light profile in all bands; only the amplitude changes per filter.  However, because all components are fit simultaneously in the Markov Chain, ForcePho captures the covariance between components, e.g,. how flux might be differently assigned to different components.  Multiple components of the same object can have their flux co-added in the Markov Chain outputs to yield total photometry that is more stable than each component separately.  Because ForcePho runs on the individual exposures, it avoids issues of how mosaicking smooths the PSF or introduces covariance between pixels.

We also do analysis on the mosaics.  For this, we need an estimate of the PSF after image combination.  We build the empirical PSF (ePSF) from bright stars in the NIRCam mosaic with the {\tt Photutils} package \citep{bradley22}. We first select bright stars from the public CANDELS catalog in the GOODS-N field \citep{barro19}, restricting to those with $\mathrm{CLASS\_STAR} > 0.75$ and $\mathrm{FLAGS} = 0$. Through visual inspection, we end up with 10 and 9 unsaturated stars that have the highest SNR in the individual SW (F090W, F115W, F150W, and F200W) and LW (F277W, F335M, F356W, F410M, and F444W) mosaics, respectively. The HST H-band magnitudes for these stars range from 19.2 to 22.4 AB mag. We use the {\tt EPSFBuilder} module in {\tt Photutils} to model ePSF from the stars, which follows the prescription of \cite{anderson00} and \cite{anderson16}. As expected, the resulting ePSF is mildly broader (see Figure \ref{fig:sbprofile}) than the WebbPSF, which is oversampled and un-mosaicked. We then fit both single S\'{e}rsic and point source models to the combined mosaicked data using ProFit \citep{robotham17} and Lenstronomy \citep{birrer18}. Both ProFit and Lenstronomy are configurable codes that produces consistent results as the commonly-used GALFIT fitting software \citep[e.g.,][]{kawinwanichakij21, robertson23_morph}.

\subsection{Results}

The first JWST images revealed a diversity of morphological and structural properties of high-$z$ galaxies \citep[e.g.,][]{ferreira22, jacobs22, kartaltepe23, nelson23, robertson23_morph, suess22}, including compact and clumpy structures \citep[e.g.,][]{chen23, tacchella22}. GN-z11 is not an exception and the images are morphologically complex because of the haze to the north-east.  Further, it became quickly clear that the core of GN-z11 is extremely compact.  Fitting a single S\'{e}rsic component yielded very large S\'{e}rsic indices using Forcepho, ProFit and Lenstronomy, raising concerns that the combination of diffuse light and a bright core might be at play.

In Figure \ref{fig:sbprofile}, we show the encircled flux as a function of radius, both for full annuli and for wedges on and off the haze. The haze causes a clear excess at angular separations of 0.2--0.6$''$.  We therefore introduce an off-center S\'{e}rsic component to fit this source; we call this the ``Haze''.  For the light near the core, the profile is mildly more extended than either PSF estimate. We opt to fit a central point source and a separate S\'{e}rsic component, the latter being named ``Extended''.  We also include a S\'{e}rsic component for the brighter galaxy just over $1''$ away to the north-east.  All four components are varied simultaneously by ForcePho, resulting in a Markov Chain that incorporates the joint covariances of the fits to these components.

The model photometry results are presented in Table \ref{tab:modelphotom}.  The Point Source and Extended component are both well detected.  As expected, the fluxes from these two very close components are anti-correlated, so that the sum, called ``Galaxy'', has substantially smaller errors than the quadrature sum.  Both components show a sharp Ly$\alpha$ dropout. The summed Galaxy photometry is a close match to the $0.2''$ diameter aperture photometry. The model images and residuals for the best-fit model are shown in Figure \ref{fig:FP_residual}, from which one can see that the three components do substantially explain the images in all filters.

The centroid of the Extended component is allowed to shift from the point source, and the best fit does give a small shift of 21 mas, less than one pixel. The best fit has a  S\'ersic index $n=0.9\pm0.1$ and half-light radius along the major axis of $49\pm3$ mas, corresponding to 200 pc at $z=10.6$.  The combined sizes (Point and Extended source) is $0.016\pm0.005$ arcsec ($64\pm20$ pc), which we obtain by flux-weighting the sizes of both models and taking into account the 21 mas spatial offset between the point source and extended components. We find for the Extended component an axis ratio of $q=0.67\pm0.05$ and a position angle PA of $34\pm5^{\circ}$.  The two components have similar SEDs, but it is interesting to note the variations at F410M and F444W, which straddle the Balmer break.  The Extended component shows a redder F410M--F444W color, while the point source is mildly blue.  This will be discussed more in the next section. 

The Haze component also fits to $n=0.9\pm0.1$ and a half-light radius of $0.11''\pm0.01''$, offset $0.41''$ from the point source. Its axis ratio is $q=0.57\pm0.15$ at a PA of $-55\pm16^{\circ}$. The Haze clearly differs in its spectrum from the other two components.  It is much redder in F200W--F277W color and is not detected in F150W.  We also find notable drops in flux in the two medium-band filters, F335M and F410M, which might be indicative of strong emission lines in the SED. Photometric redshift fits with EAZY  and Prospector substantially favor $4<z<5$ solutions compared to those at $z=10.6$.
From this, we conclude that it is more likely that the Haze is a chance projection with a lower redshift low-surface brightness galaxy.  The nature of the $4<z<5$ solution is probably that of a young, star-bursting galaxy with emission lines and a high dust attenuation, similar to the recently discussed high-$z$ interlopers \citep{naidu22, zavala23}.  That said, given the low signal-to-noise ratio, there remains some chance that the Haze could be associated with GN-z11, e.g., as the tidal spray from a merger, with a cessation of star formation dropping the far-UV emission.  

In order to investigate whether the second, ``Extended'' component is indeed present, we plot in Figure~\ref{fig:flux_ratio_size} the ForcePho posterior distribution of the fraction of $F200W$ flux in the Extended component as a function of its half-light size (i.e., major axis radius). There is a covariance between this fraction and the size: the larger the Extended component, the less flux it contributes. Importantly, there is no tail towards zero size or zero flux, emphasizing that the Extended component is clearly detected and indeed extended. 
Based upon a Bayesian Information Criterion model selection, the 2-component fit is preferred.

We confirm the compact size of GN-z11 by fitting both single S\'{e}rsic and point source models to the mosaic using Lenstronomy \citep{birrer18}. To ensure we account for the extended background from the bright neighbor $\sim1''$ to the NE, we fit GN-z11 and the bright neighbor simultaneously using single S\'{e}rsic profiles for both galaxies. In all bands, the best-fit model for GN-z11 has an intrinsic half-light radius $<$ 1 pixel and a S\'{e}rsic index of $>7.8$, indicating a source that is not significantly resolved with respect to the ePSF. Fitting GN-z11 with a point source model instead of a S\'{e}rsic profile yields slightly lower residuals, further indicating the compact nature of the source.

Using ProFit, we analyse the structure of GN-z11 in the F277W mosaic. using a multi-component fit including the central point source, the central extended source, and the Haze. Given the number of components, a nominal ProFit model would involve optimizing 24 free parameters. Unlike our ForcePho method, in using ProFit we are limited to the information provided by a single filter in the mosaic (see discussion above) and, hence, we restrict the number of free parameters to the centroids of each component, their brightnesses, the S\'{e}rsic indices, and the effective radii. The axis ratios and position angles are kept fixed at the value found in our ForcePho models, and we set the isophotal boxiness to be negligible in each case. Importantly, we find a better fit including a three component model (point source, extended component, and Haze) than for a point source + Haze two-component model. We see no evidence that the PSF adopted by ForcePho is artificially inflating the size of Extended component or the Haze, as we can find independently a good quality fit for the extended component and the Haze with ProFit with an effective radius of $0.04''\pm0.01''$ and $0.11''\pm0.06$, respectively. The S\'{ersic} index of both the extended component and the Haze are consistent with $n=1$. Similarly, the relative brightnesses of the Extended component and the Haze to the Point Source are very similar to the ForcePho fit (about two magnitudes for the latter). In summary, ProFit allows us to confirm the presence of the Extended component, albeit with less constraining power than what we can do with ForcePho. This is expected given the limited amount of data used in the fitting (single band and mosaic).

We note that our modeled size of GN-z11 is noticeably smaller than the $0.6\pm0.3$ kpc half-light radius reported in \citet{holwerda15} fitting HST WFC3 data using a S\'ersic index $n=1.5$. Indeed, even at JWST resolution, the unresolved component has over half of the total light (Figure~\ref{fig:flux_ratio_size}). We do not think the difference can be due to the Haze, as that component is very weak in F150W.  Clearly the angular resolution of JWST will be very important in probing the physical sizes of these very small high-redshift galaxies \citep[e.g.,][]{wu20, costantin23, ono22, suess22, tacchella22}.

\section{Interpretations of the Spectral Energy Distribution}
\label{sec:sed}

\begin{figure*}[t]
\noindent
\includegraphics[width=\textwidth]{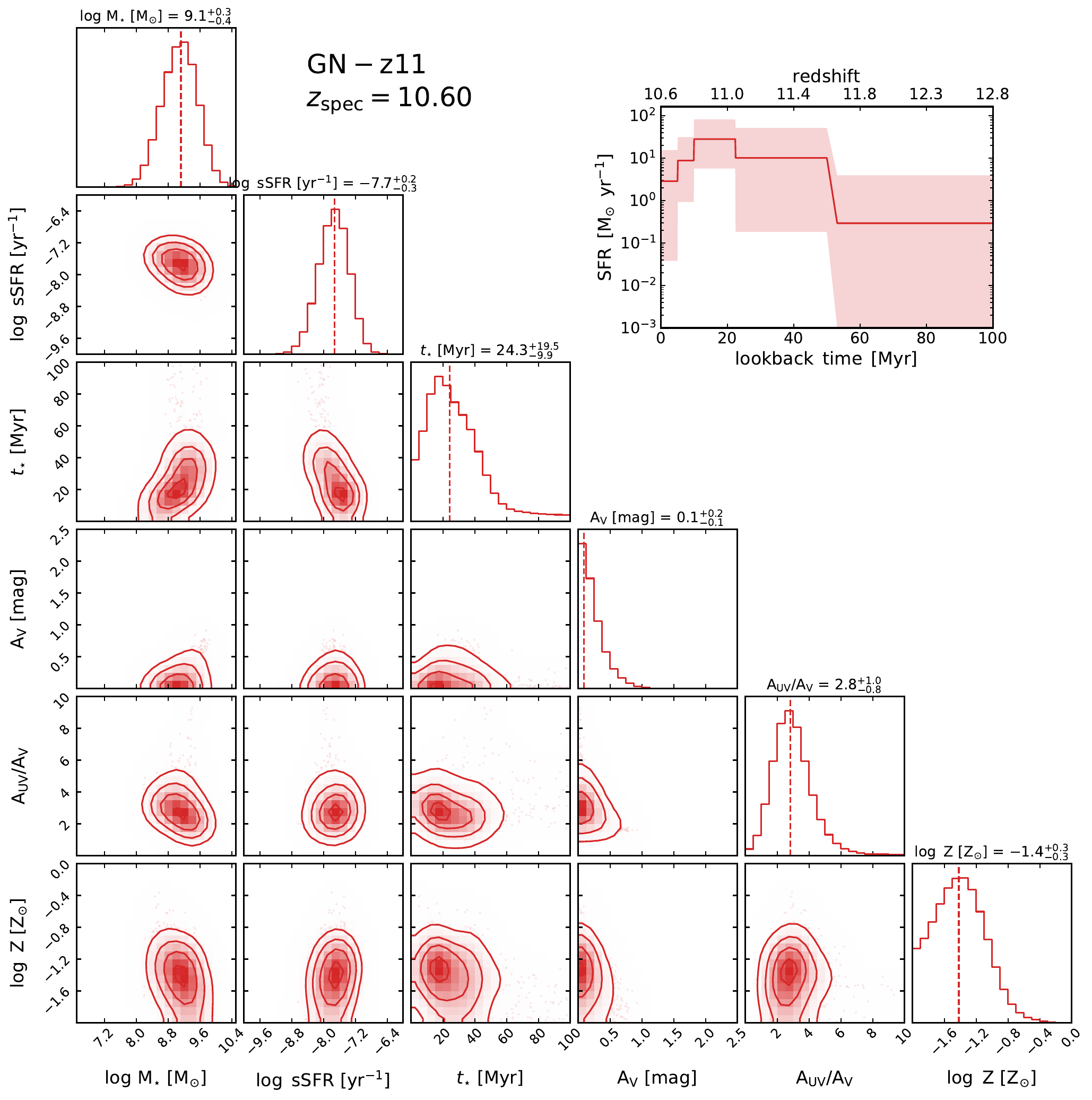} 
\caption{Corner plot of the 2-d projections of the posterior of the Markov Chain from the Prospector fits, along with an inset of the reconstructed star-formation history. These posterior distributions are obtained by fitting the combined SED of the Point Source and Extended Source (``Galaxy'' in Table~\ref{tab:modelphotom}). The posterior of the SED is shown in Fig.~\ref{fig:sed}. We find that GN-z11 has a stellar mass of $10^{9.1\pm0.3}~M_{\odot}$, is actively forming stars ($\mathrm{sSFR}=10^{-7.7\pm0.3}~\mathrm{yr}^{-1}$) with a young stellar age (half-mass time -- look back time at which 50\% of the stellar mass formed -- is $\sim24_{-10}^{+20}$ Myr). The SED of GN-z11 is consistent with being dust-free ($\mathrm{A}_{\rm V}=0.1_{-0.1}^{+0.2}$) and with a low stellar metallicity (about 10\% solar metallicity).
}
\label{fig:sed_corner}
\end{figure*}

\begin{figure*}[t]
\noindent
\includegraphics[width=1.0\textwidth]{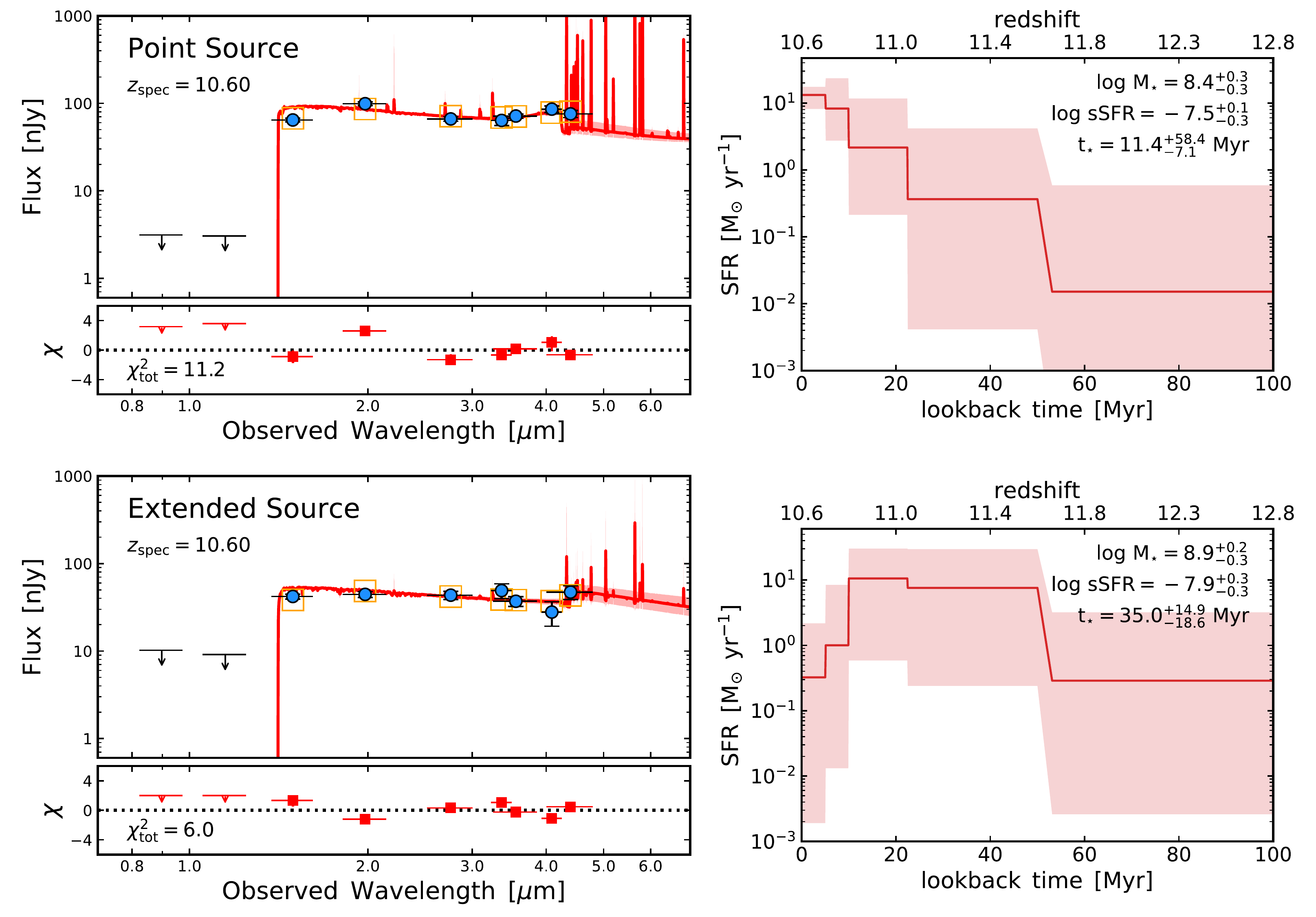}\\
\caption{Prospector fits of stellar population models to the photometry of the central point source (PS; upper two panels) and of the Extended component (bottom two panels), see also Table~\ref{tab:props}. The left panels show the SED of the best-fit model and the residuals, following the same layout as Figure~\ref{fig:sed}. The right panels plot the reconstructed star-formation history from the SED modeling: the solid line and shaded regions mark the median and 16-84th percentiles of the posterior distribution, respectively. One sees that the central point source, if dominated by starlight, is consistent with a young low-metallicity, and unreddened stellar population of about $10^{8.4\pm0.3}$~\Msun; note the nebular continuum just blueward of the Balmer break in the fitted model. Specifically, the stellar age, as defined by the lookback time when 50\% of the stellar mass was formed, is $t_{\star}=11_{-7}^{+58}~\mathrm{Myr}$). For the extended component the star-formation history is more extended ($t_{\star}=35_{-19}^{+15}~\mathrm{Myr}$), even dropping at the most recent time, due to the increased flux in F444W compared to F410M.  The fitted stellar mass is somewhat larger, $10^{8.9\pm0.3}$~\Msun, despite  this component contributing only 1/3 of the combined rest-UV flux.
}
\label{fig:prospector_by_component}
\end{figure*}

\begin{table*}
    \begin{center}
    \caption{\label{tab:props}Properties of GN-z11.}
    \noindent\begin{tabular}{l c c c}
    \hline 
    Property & Galaxy & PS & Extended \\
    \hline
    Stellar mass $\log(M_{\star}/M_{\odot})$ & $9.1_{-0.4}^{+0.3}$ & $8.4_{-0.3}^{+0.3 }$ & $8.9_{-0.3}^{+0.2}$ \\
    Observed M$_{\rm UV}$ [AB mag] & $-21.58_{-0.04}^{+0.03}$ & $-21.10_{-0.03}^{+0.04}$ & $-20.48_{-0.05}^{+0.05}$ \\
    Intrinsic M$_{\rm UV}$ [AB mag] & $-21.79_{-0.47}^{+0.24}$ & $-21.08_{-0.29}^{+0.23}$ & $-20.73_{-0.45}^{+0.21}$ \\
    UV continuum slope $\beta$ & $-2.41_{-0.07}^{+0.06}$ & $-2.48_{-0.11}^{+0.08}$ & $-2.43_{-0.09}^{+0.09}$ \\
    SFR $[\mathrm{M_{\odot}}/\mathrm{yr}]$ & $21_{-10}^{+22}$ & $6_{-2}^{+3}$ & $9_{-4}^{+7}$ \\
    $\log(\mathrm{sSFR})$ $[\mathrm{yr}^{-1}]$ & $-7.7_{-0.3}^{+0.2}$ & $-7.5_{-0.4}^{+0.1}$ & $-7.9_{-0.3}^{+0.3}$ \\
    Stellar age $t_{\star}$ $[\mathrm{Myr}]$ & $24_{-10}^{+20}$ & $11_{-7}^{+58}$ & $35_{-19}^{+15}$ \\
    Attenuation $\mathrm{A_{\rm V}}$ [mag] & $0.08_{-0.06}^{+0.23}$ & $0.05_{-0.05}^{+0.08}$ & $0.09_{-0.07}^{+0.26}$ \\
    Half-light size $R_{\rm e}$ [arcsec] & $0.016 \pm 0.005$ & point-like & $0.049 \pm 0.003$ \\
    Half-light size $R_{\rm e}$ [pc] & $64 \pm 20$ & point-like & $196 \pm 12$ \\
    \hline
    \end{tabular}\end{center}
    \smallskip
   Notes: The stellar population properties quoted here (median and $16^{\rm th}-84^{\rm th}$ percentiles) come from the SED modeling with Prospector of the combined photometry of the Point Source (PS) and Extended component (i.e. ``Galaxy'' in Table~\ref{tab:modelphotom}), the PS photometry and the Extended component photometry. The average half-light size (and its standard deviation) for the combined GN-z11 is derived by flux-weighting the sizes of both the PS and Extended models.
\end{table*}

We now move to interpret the measured SEDs of GN-z11, its sub-components and the Haze. Importantly, we model the SEDs assuming that the emission is powered by star light, either through direct stellar emission or reprocessed via gas (i.e. nebular emission) and dust. 
With regard of a possible AGN contribution, we can therefore only assess whether the SED is consistent with such stellar powered emission. 

\subsection{Methods}

We analyze the SEDs with three different spectral synthesis codes, fixing the redshift to the spectroscopic redshift of 10.60 throughout. The first is Prospector \citep{johnson21}, which we use for the figures in this paper.  Prospector computes stellar population synthesis combined with a model of nebular line and continuum emission as well as dust attenuation, comparing to the data with a Bayesian formalism and using Markov Chain Monte Carlo to quantify the posterior.  Here we utilize a similar setup as in \citet[][see also \citealt{tacchella22, whitler23_sfh}]{tacchella22_highz}. Specifically, we assume a non-parametric star-formation history with 6 time bins and a bursty continuity prior. We put them at $0-5$ Myr, $5-10$ Myr and the 4 bins are logarithmically-spaced up to $z=20$.  We adopt a single metallicity for both stars and gas, assuming a truncated log-normal centered on $\log(Z/Z_{\odot})=-1.5$ with width of 0.5, minimum of $-2.0$ and maximum of 0.0. We model dust attenuation using a two-component dust attenuation model with a  flexible attenuation curve. The first component is a birth-cloud component in our model that attenuates nebular emission and stellar emission only from stars formed in the last 10 Myr (attenuation law is a power law with a slope of $-1$). The second component is a diffuse component that has a variable attenuation curve and attenuates all stellar and nebular emission from the galaxy. The variation in the attenuation law is modeled as a multiplicative factor of the \citet{calzetti07} law in order to account for uncertainties in the geometry of dust extinction. For the stellar population synthesis we adopt the MIST isochrones \citep{choi16} that include effects of stellar rotation but not binaries, and assume a \citet{chabrier03} initial mass function (IMF) between 0.08 and 120 M$_\odot$. No Ly-$\alpha$ emission line is added to the model, to account for resonant absorption effects. This assumption might be too simplistic given the Ly-$\alpha$ detection in the spectrum \citep{bunker23}. However, we estimate that this will not affect our SED fits significantly since this line contributes only at the 0.05 mag level, while we put an error floor of 5\% on the photometry. The rest of the nebular emission (emission lines and continuum) is self-consistently modeled \citep{byler17} with two parameters, the gas-phase metallicity (tied to the stellar metallicity) and the ionization parameter (uniform prior in $-4<\log(U)<-1$).

As these data for GN-z11 fall mostly in the rest-UV, the broadband photometry is less sensitive to the presence of very strong emission lines.  In particular, the strong [OIII], H$\beta$, and H$\alpha$ lines fall redward of our wavelength range.  However,  some contribution from bluer lines in the F444W band is expected; this will be quantified in \citet{bunker23}.

A key component not included in this modeling is the possibility of luminosity from an AGN.  Clearly the very compact morphology of GN-z11, with about 2/3 of the emission coming from an unresolved nucleus, permits this.  As the SED from an AGN component is highly flexible, we opt here to present results based on the null hypothesis that the light is dominated by stars.  Clearly a luminous AGN would decrease the inferred star-formation rate and stellar mass. 
Whether or not GN-z11 contains a luminous AGN is a question that will be investigated with the spectroscopy \citet{bunker23} and \citet{maiolino23}.

In addition to the effects of metallicity, dust attenuation, and star-formation history on the stellar mass-to-light ratio, several other model assumptions may affect the stellar mass inferred from the SED \citep[e.g.,][]{conroy13_rev}. We highlight some of these, while noting that the effects on the inferred masses can be complex since other inferred parameters may compensate to predict the same SED with a similar stellar mass under different assumptions. If the IMF had fewer very low mass stars than we assume then the inferred stellar masses would be overestimated \citep[e.g.,][]{steinhardt22}.  The slope of the upper IMF affects the UV-optical color and mass-to-light ratio for a given star-formation history.  The evolution of high-mass stars at low metallicity is important to the UV and ionizing continuum of galaxies, but is not well constrained from observations in the local universe \citep{eldridge22}.  Binary stellar evolution and stellar rotation may increase the lifetimes of massive, UV bright stars, leading to lower mass-to-light ratios for a given star-formation history \citep[e.g.][]{choi17}. Furthermore, non-solar abundance ratios may be common in the early universe \citep[e.g.][]{steidel16} leading to changes in stellar evolutionary tracks and stellar SEDs at a variety of ages. Our nebular emission models may not capture the effects of complex geometries on the emergent nebular continuum and nebular line ratios \citep[e.g.][]{jin22}, which is important for the shape of the Balmer break and hence mass-to-light ratios \citep[e.g.,][]{papovich22}.  We do not include potential contributions from Population III stars.

We compared the Prospector results to those from BEAGLE \citep{chevallard16} and BAGPIPES \citep{carnall18}, which provide similar functionality but differ in numerous modeling aspects. We fit the observed ForcePho flux with both BEAGLE and BAGPIPES using a delayed exponential star-formation history, fixing the redshift to the spectroscopic value. For both BEAGLE and BAGPIPES we use the \citet[][with the 2016 update]{bruzual03} stellar templates and \citet{kroupa01} IMF. We assume the \citet{charlot00} and the \citet{calzetti00} dust attenuation law for BEAGLE and BAGPIPES, respectively.

\subsection{Results}

We first focus on the combined photometry of the Point Source and the Extended Component, which we refer to as ``Galaxy'' GN-z11 (Table~\ref{tab:modelphotom}). We show the resulting posterior distributions of several key stellar population parameters in Figure~\ref{fig:sed_corner}, including the stellar mass, specific SFR, stellar age, dust attenuation, and stellar metallicity (see also Table~\ref{tab:props}). We find a formed stellar mass of $10^{9.1\pm0.3}~M_{\odot}$. We find that GN-z11 is actively forming stars with $\mathrm{SFR}_{\rm 30Myr}=21_{-10}^{+22}~M_{\odot}~\mathrm{yr}^{-1}$ ($\mathrm{SFR}_{\rm 10Myr}=12_{-3}^{+10}~M_{\odot}~\mathrm{yr}^{-1}$) and a specific SFR of $\mathrm{sSFR}_{\rm 30Myr}=10^{-7.7\pm0.3}~\mathrm{yr}^{-1}$, indicating this galaxy is doubling it stellar mass roughly every $\sim50$ Myr. (s)$\mathrm{SFR}_{\rm 30Myr}$ refers to the (specific) SFR averaged over the past 30 Myr. Consistent with this, we find that the galaxy has a stellar age (half-mass time: look back time at which 50\% of the stellar mass formed) of $\sim24_{-10}^{+20}$ Myr. The inset on the top-right of Figure \ref{fig:sed_corner} shows the posterior of the star-formation history. We find that the SFR has increased $\sim60$ Myr ago ($z\approx12$), peaked at a lookback time of $10-20$ Myr, and has slightly decreased in the recent 10 Myr. 

These Prospector-inferred parameters are in overall good agreement with the parameters inferred with BEAGLE and BAGPIPES. Specifically, using BEAGLE with a parametric star formation history, we find log(M$_{\star,\mathrm{BEAGLE}}/M_{\odot}) = 8.9 \pm 0.1$,  SFR$_{10\mathrm{Myr}}$ = $22\pm 5$ $M_{\odot}/\mathrm{yr}$, and attenuation $A_{\rm V} = 0.09 \pm 0.08$ mag. With BAGPIPES, we obtain log(M$_{\star,\mathrm{BAGPIPES}}/M_{\odot}) = 9.3_{-0.8}^{+0.1}$,  SFR$_{10\mathrm{Myr}}$ = $16\pm 6$ $M_{\odot}/\mathrm{yr}$, and attenuation $A_{\rm V} = 0.14_{-0.07}^{+0.03}$ mag. 
All inferred parameters are consistent with previous inferences in \citet{jiang21}, who used a single stellar population to model the broad-band HST+Spitzer photometry.  
Our results are also within $\sim$1$\sigma$ of those obtained in \citet{bunker23} by fitting the NIRSpec spectroscopy with BEAGLE.  We consider this good agreement given the differences in the parameterization of the star formation histories and associated priors \citep[e.g.,][]{leja19_nonparm, whitler23_sfh}.

Figure~\ref{fig:prospector_by_component} shows the Prospector results for the two components of GN-z11 (Point Source and Extended) individually. The top panels are the results for the Point Source, while the bottom panels are for the Extended Source. The individual component (Point Source and Extended) stellar population parameters agree between Prospector, BEAGLE and BAGPIPES.

The SEDs of those two components show an important difference around the observed wavelength of 4~$\mu$m, which is around the rest-frame Balmer jump. The Point Source shows a blue F410M--F444W color and a red F356W--F410M color, indicating strong nebular emission. Contrarily, the Extended component shows a red F410M--F444W and blue F356W-F410M color, consistent with less strong nebular emission and weak Balmer/4000 \AA\ break. From the ForcePho Markov Chains, we find that the F410M--F444W color of the Extended component is redder than that of the Point Source in 95\% of cases, with a median color difference of 0.7 mag.
Not surprisingly, Prospector prefers a younger age and an increasing star-formation history for the central Point Source, while the Extended Source is consistent with a more extended star-formation history (even decreasing in the recent 10 Myr). The Extended Source is slightly more massive than the Point Source ($10^{8.9\pm0.3}~M_{\odot}$ versus $10^{8.4\pm0.3}~M_{\odot}$), despite that this component is only 1/3 of the combined rest-UV flux. 

In summary, the SED modeling presented here does not reveal a strong support for the AGN scenario. We find that that the central emission can be powered by intense and compact star formation, thereby outshining the underlying extended emission of the galaxy. However, we cannot rule out the presence of a luminous AGN, and we expect that detailed analysis of the NIRSpec spectroscopy (\citealt{bunker23} and \citealt{maiolino23}) will be needed to fully explore this scenario.

\section{The Environment of GN-z11}
\label{sec:environment}

\begin{figure*}[t]
\includegraphics[width=\textwidth]{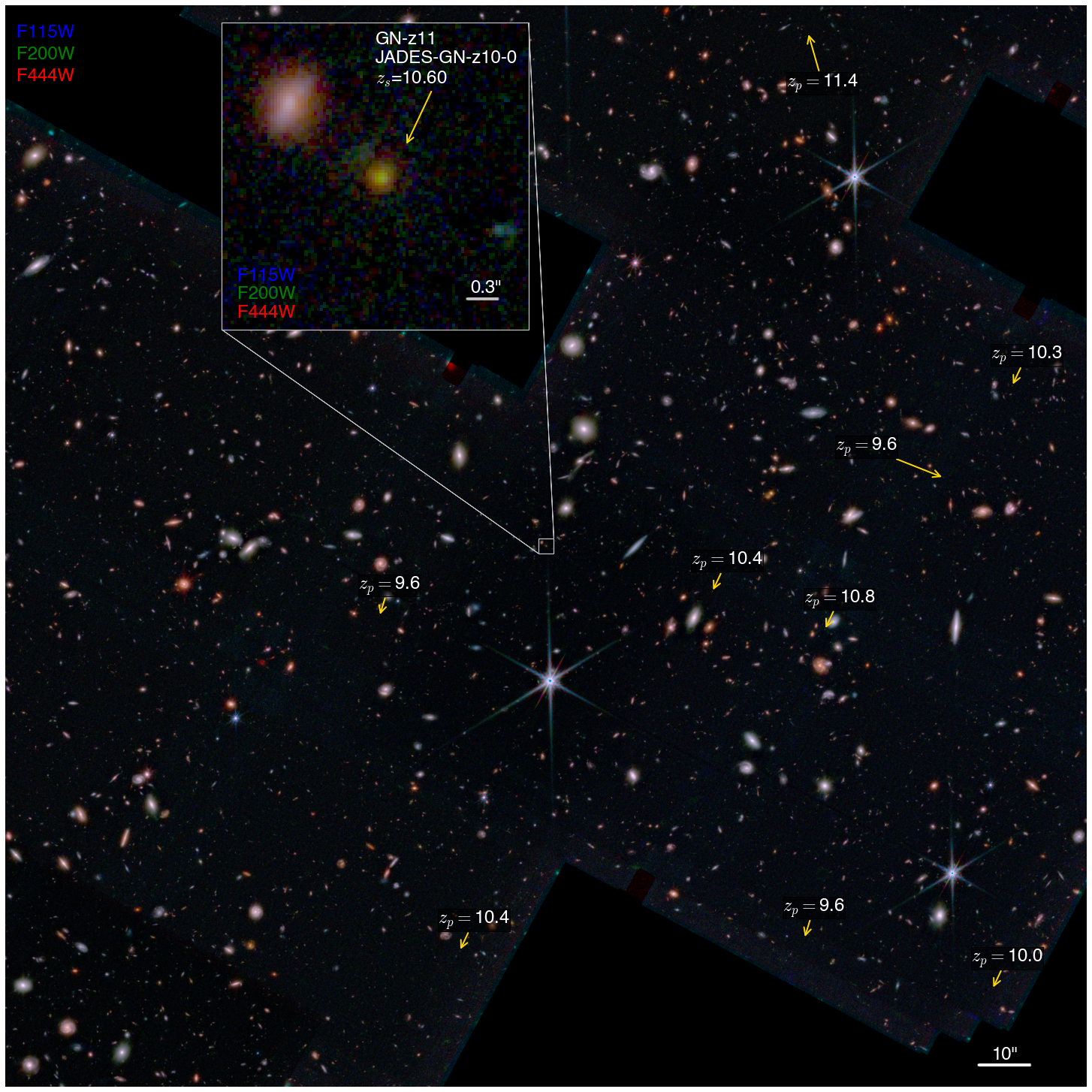}
\caption{The region around GN-z11.  This image is $210''$ on a side, which is 10 comoving Mpc at $z=10.6$.  It uses F115W, F200W, and F444W as the blue, green, and red colors.  The location and redshift of the 9 nearest objects with photometric redshifts around $z\sim10$ are marked.}
\label{fig:field}
\end{figure*}

\begin{table*}[t]
\begin{center}
\caption{\label{tab:neighbors}Galaxy candidates with photometric redshifts consistent with $z=10.6$}
\noindent\begin{tabular}{ccccccr}
\hline
ID & Name       & RA  & Dec  & Distance ($''$) & SNR$_{\rm F200W}$ & $z_{\rm phot}$ \\
\hline
465 & JADES-GN-189.11621+62.22008  & 189.116208 & 62.220076  & 81 & 4.7 & 10.45 \\
544 & JADES-GN-189.07604+62.22072  & 189.076035 & 62.220716  & 92 &  5.0 & 9.57 \\
4155 &JADES-GN-189.07357+62.23749  & 189.073566 & 62.237487  & 57 & 15.6 & 10.80 \\
4418 &JADES-GN-189.12549+62.23826  & 189.125491 & 62.238263  & 35 &  5.7 & 9.58 \\
4811 &JADES-GN-189.08668+62.23957  & 189.086676 & 62.239566  & 34 & 8.2 & 10.45 \\
6862 &JADES-GN-189.05971+62.24572  & 189.059714 & 62.245717  & 79 & 8.2 & 9.57 \\
8597 &JADES-GN-189.05166+62.25070  & 189.051657 & 62.250698  & 96 & 10.2 & 10.34 \\
13453&JADES-GN-189.07537+62.26988  & 189.075370 & 62.269878  & 113 & 5.5 & 11.40 \\
62240&JADES-GN-189.05413+62.21795  & 189.054128 & 62.217950  & 123 & 3.0 & 10.00 \\
\hline
\end{tabular}
\end{center}\smallskip
Notes: The angular distance from GN-z11 is given, along with the signal-to-noise ratio in the F200W band and the photometric redshift.  The ID number provides a short name for cross-referencing the tables and figures. The photometric redshift we report is EAZY $z_a$, the value at the minimum $\chi^2$ across all redshifts.  We remind that 1$''$ is 47 comoving kpc at $z=10.6$.
\end{table*}

\begin{table*}[t]
\begin{center}
\caption{\label{tab:neighborsPhot}Aperture photometry for the candidates listed in Table \ref{tab:neighbors}}
\noindent\begin{tabular}{rrrrrrrrrr}
\hline
ID       & F090W & F115W & F150W & F200W & F277W & F335M & F356W & F410M & F444W \\
\hline
   465& $ 1.7\pm3.0$& $ 0.8\pm2.2$& $ 7.5\pm2.4$& $ 9.9\pm2.1$& $10.5\pm1.5$& $ 8.4\pm2.1$& $ 6.9\pm1.3$& $13.1\pm2.3$& $ 9.3\pm1.9$\\
   544& $-2.0\pm3.0$& $-0.2\pm2.4$& $14.9\pm2.4$& $10.1\pm2.0$& $11.9\pm1.5$&$-$& $12.4\pm1.4$& $10.1\pm2.0$& $13.1\pm1.8$\\
  4155& $ 6.0\pm3.3$& $ 0.7\pm2.2$& $26.7\pm2.5$& $36.3\pm2.3$& $28.5\pm1.2$&$-$& $20.7\pm1.1$& $19.4\pm1.7$& $22.2\pm1.6$\\
  4418& $-1.6\pm2.2$& $-1.4\pm1.6$& $ 7.1\pm1.8$& $ 8.8\pm1.5$& $ 5.7\pm1.2$& $ 2.5\pm1.9$& $ 4.8\pm1.1$& $ 5.4\pm2.1$& $ 4.1\pm1.5$\\
  4811& $ 1.6\pm2.0$& $ 1.6\pm1.5$& $ 8.3\pm1.5$& $11.5\pm1.4$& $13.4\pm1.1$&$-$& $13.2\pm1.2$& $21.7\pm1.8$& $21.7\pm1.5$\\
  6862& $ 3.2\pm2.6$& $-2.2\pm2.0$& $12.5\pm2.1$& $14.7\pm1.8$& $10.3\pm1.2$&$-$& $11.5\pm1.2$& $11.8\pm1.9$& $12.1\pm1.6$\\
  8597& $ 0.8\pm2.4$& $-1.5\pm1.7$& $15.9\pm2.0$& $18.4\pm1.8$& $12.8\pm1.2$&$-$& $15.4\pm1.1$& $17.6\pm1.7$& $22.7\pm1.6$\\
 13453& $ 0.4\pm1.9$& $-2.0\pm1.3$& $ 3.9\pm1.5$& $ 7.2\pm1.3$& $ 6.0\pm1.3$& $ 7.2\pm1.9$& $ 3.0\pm1.2$& $ 6.8\pm1.9$& $ 4.6\pm1.5$\\
 62240& $-3.4\pm3.7$& $ 3.4\pm2.7$& $10.2\pm2.9$& $ 7.3\pm2.4$& $ 8.8\pm1.4$&$-$& $ 5.0\pm1.4$& $10.6\pm2.3$& $ 8.3\pm2.2$\\
 \hline
\end{tabular}
\end{center}\smallskip
Notes:
The flux and 1--$\sigma$ errors are given in nJy.  The aperture is $0.2''$ diameter, with point-source aperture corrections.  Six of these galaxies fall outside of the F335M imaging footprint. 
\end{table*}

\begin{figure*}[t!]
\noindent
\includegraphics[width=0.33\textwidth]{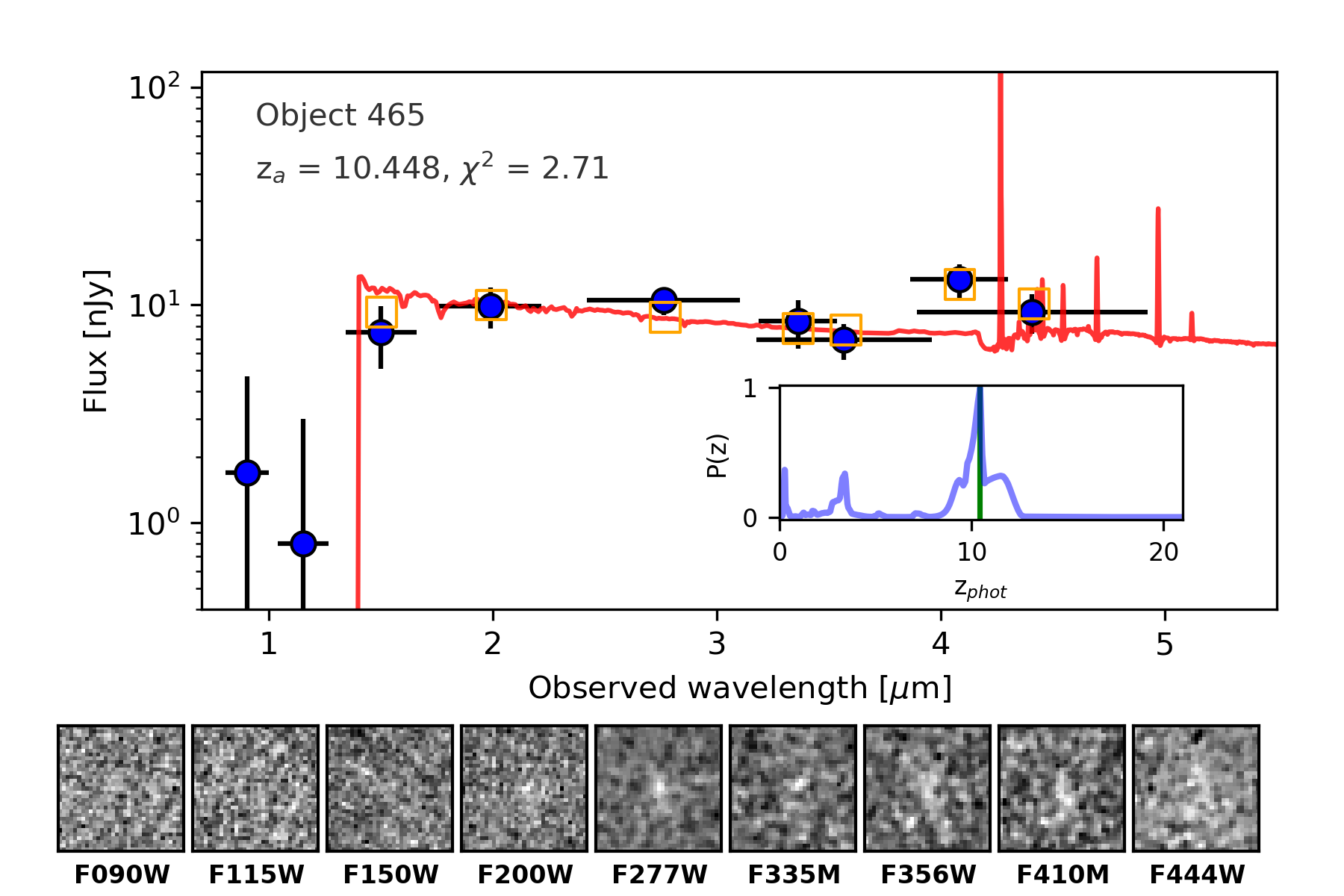}\hfill
\includegraphics[width=0.33\textwidth]{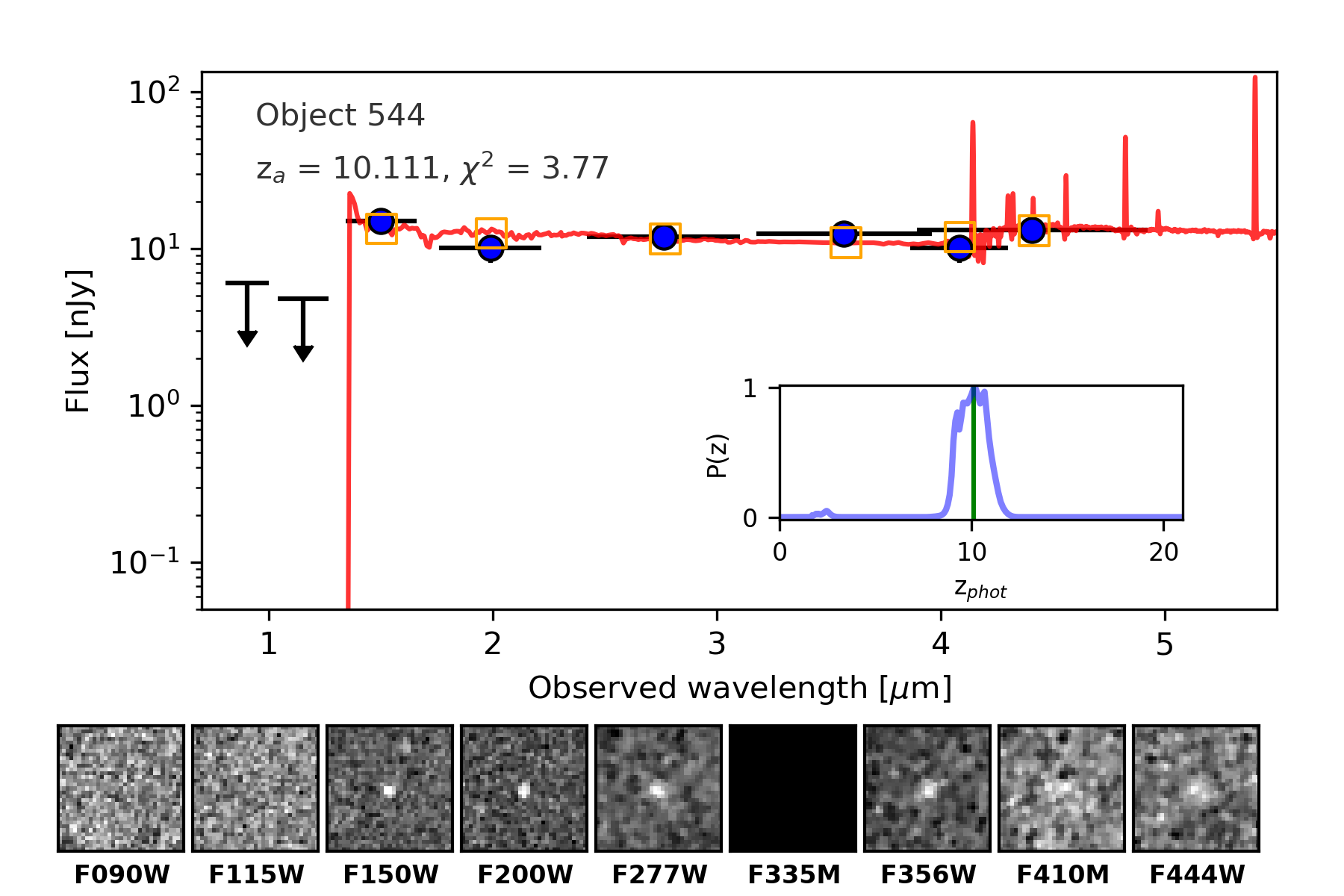}\hfill
\includegraphics[width=0.33\textwidth]{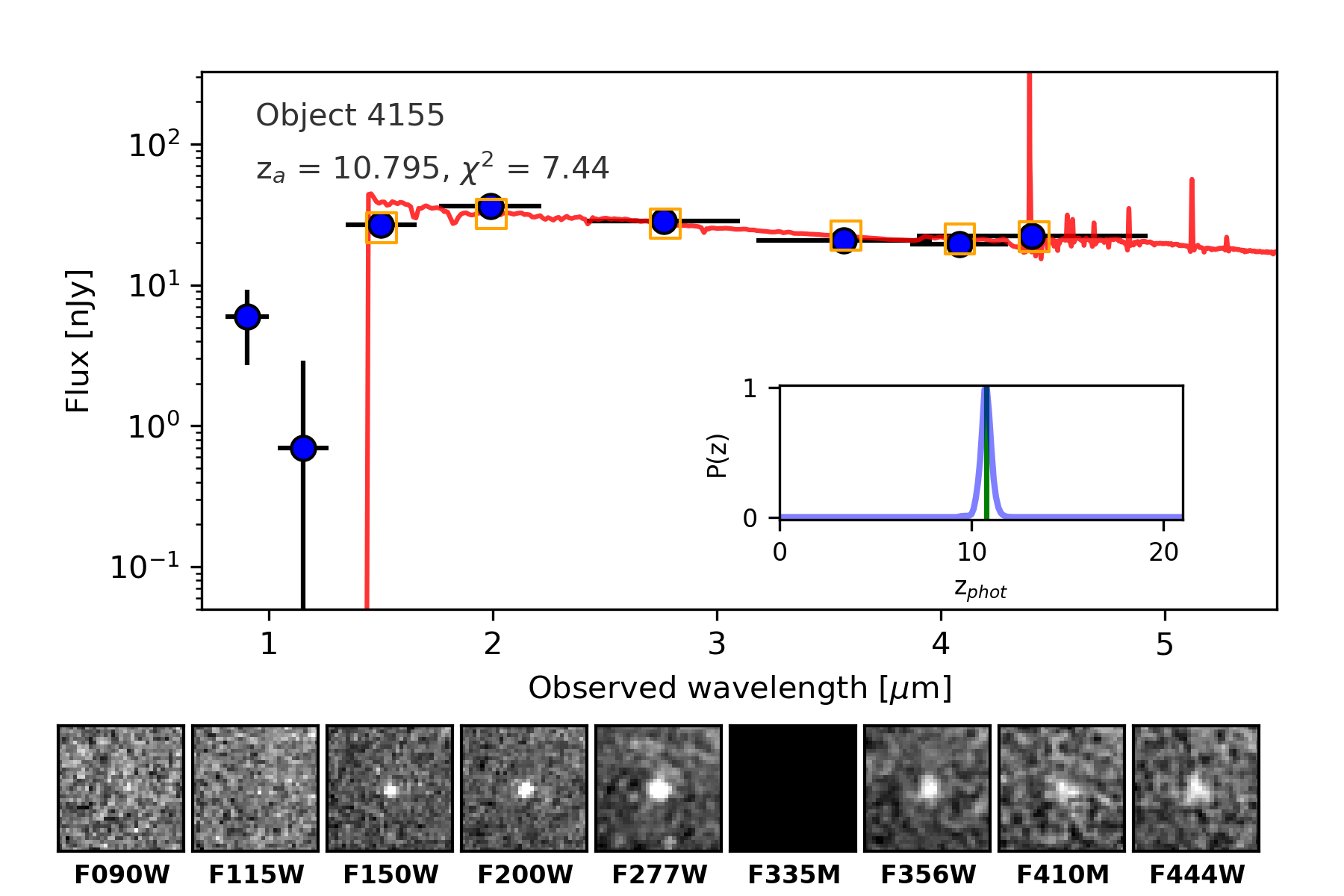}\\[12pt]
\noindent
\includegraphics[width=0.33\textwidth]{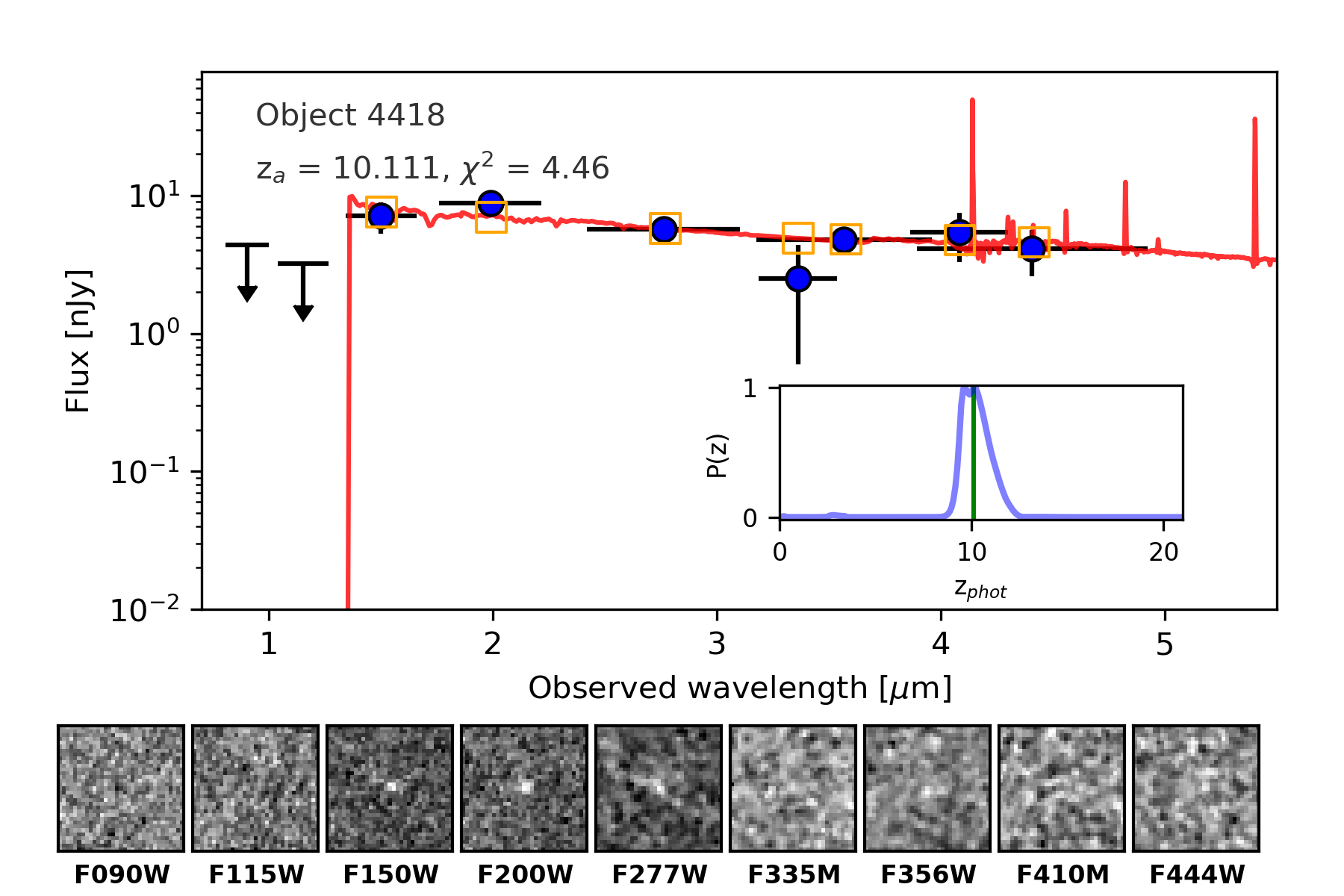}\hfill
\includegraphics[width=0.33\textwidth]{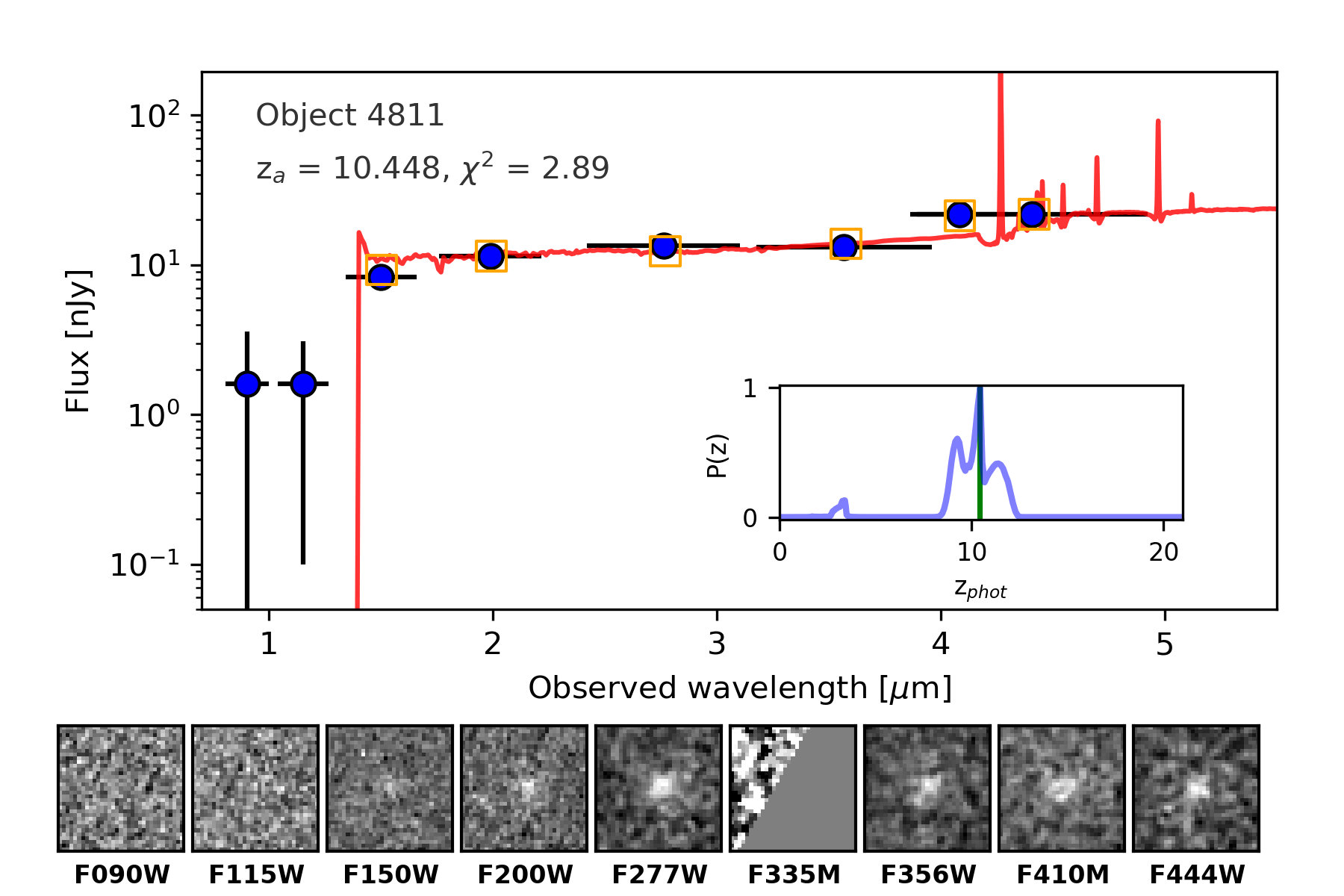}\hfill
\includegraphics[width=0.33\textwidth]{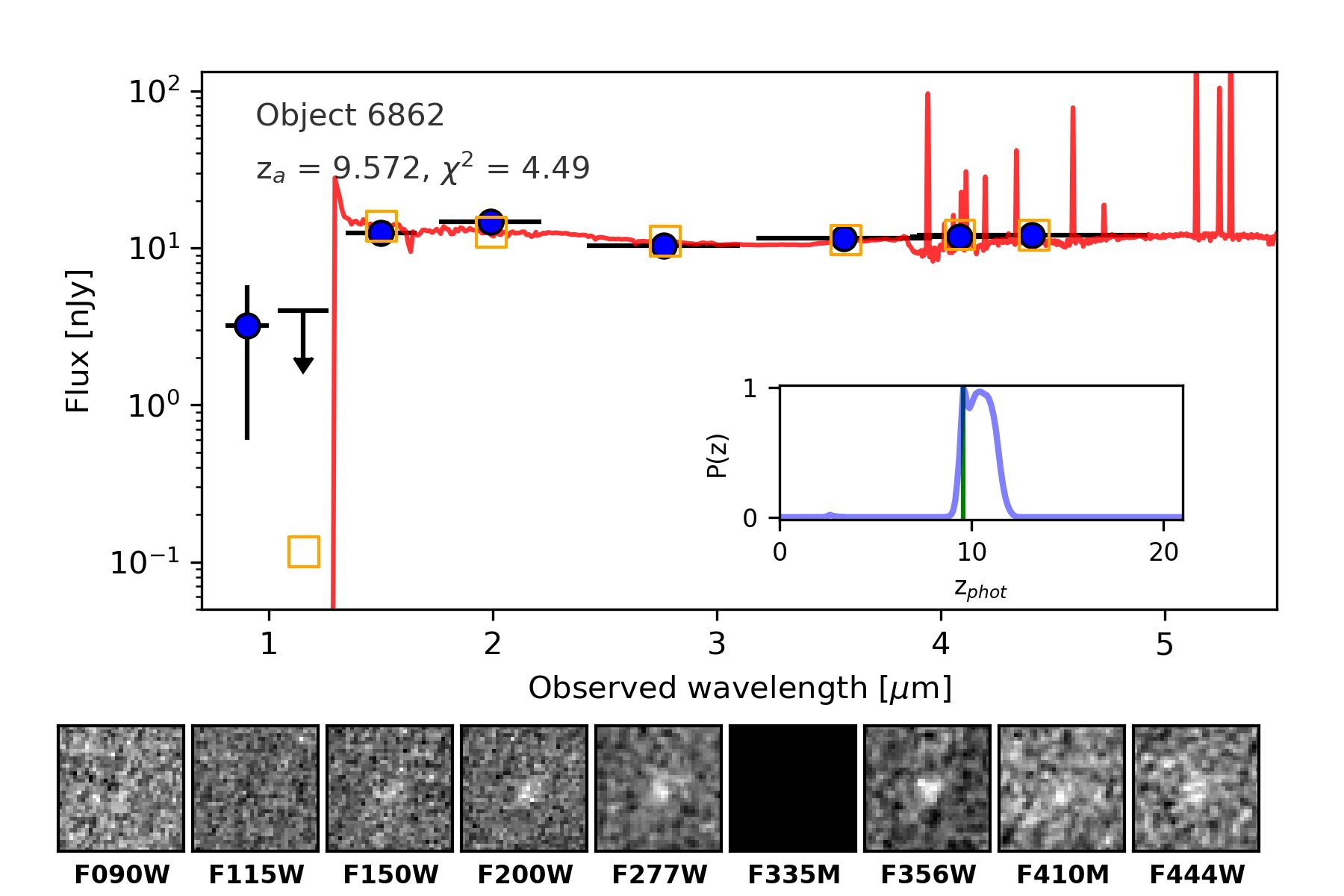}\\[12pt]
\noindent
\includegraphics[width=0.33\textwidth]{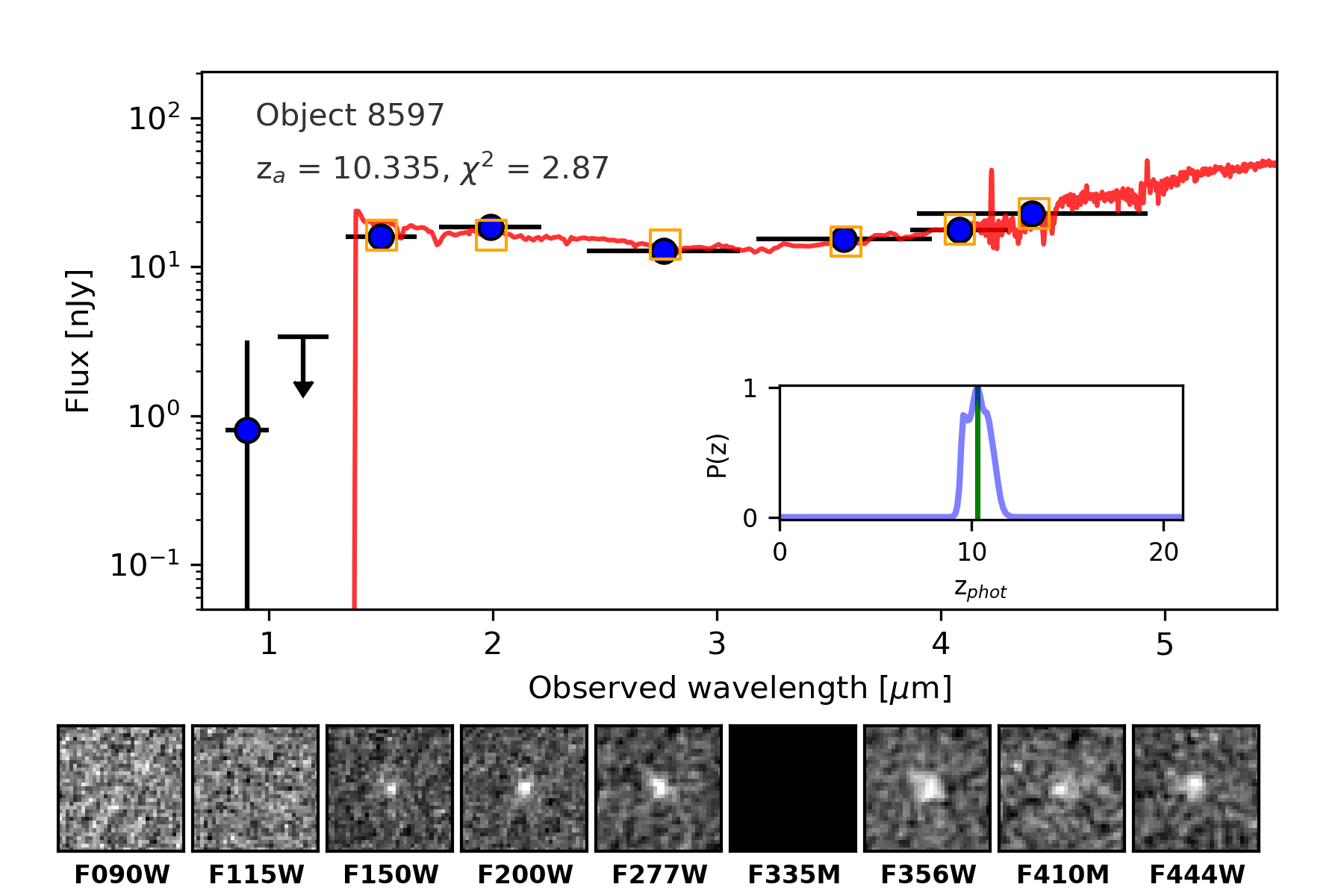}\hfill
\includegraphics[width=0.33\textwidth]{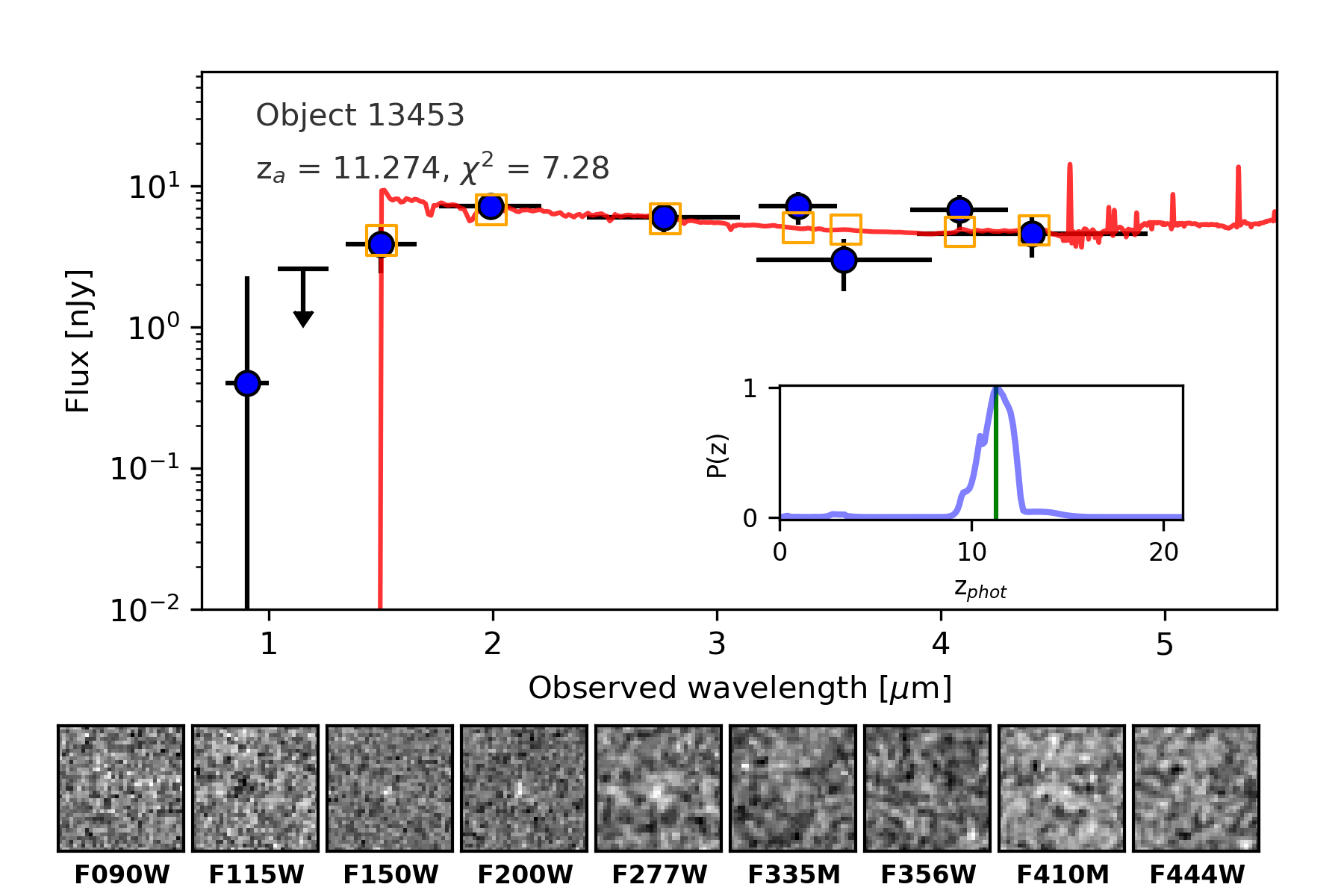}\hfill
\includegraphics[width=0.33\textwidth]{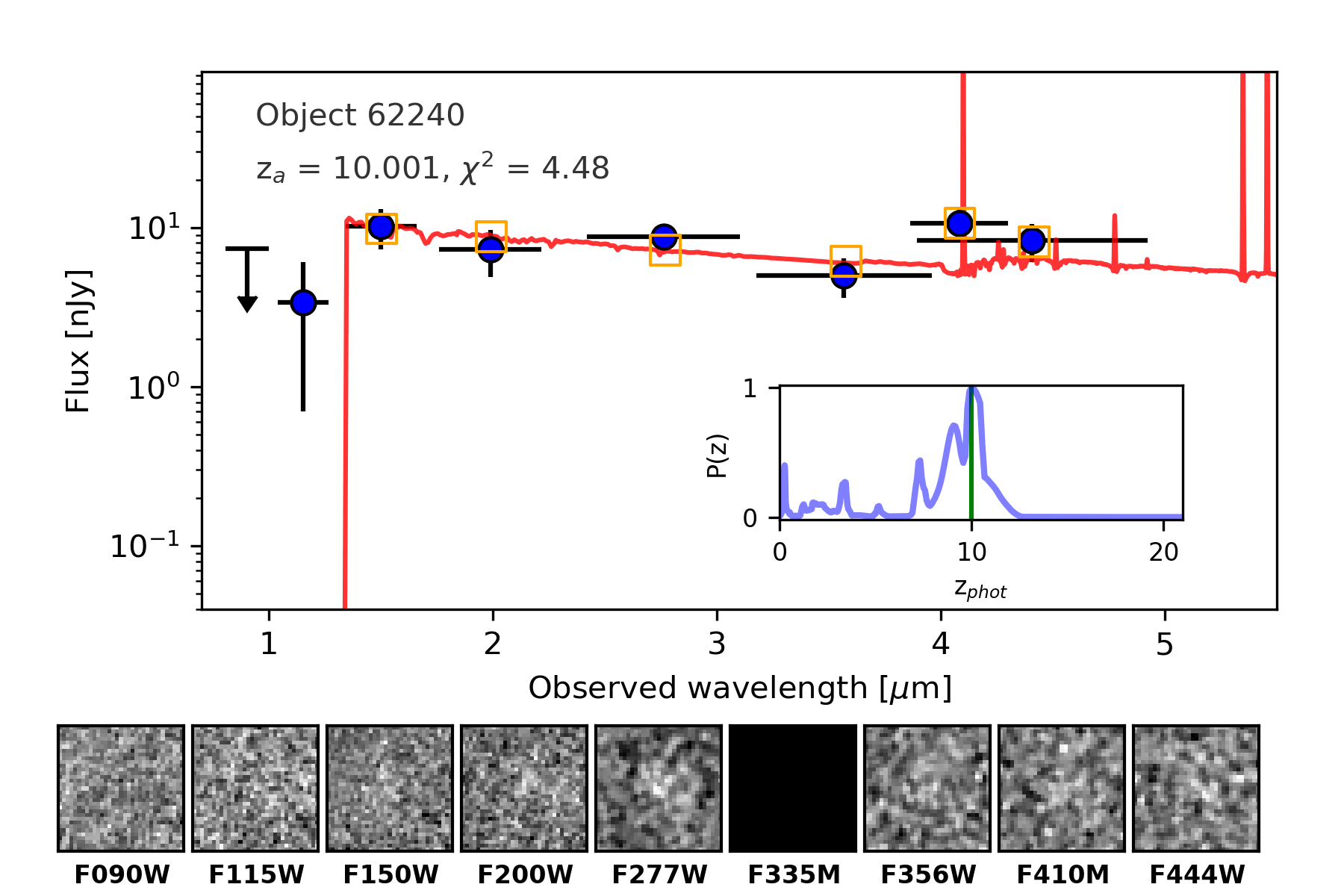}\\[12pt]
\caption{Spectral energy distributions, photometric redshift probabilities, and thumbnail images of candidates near to GN-z11 and consistent with its redshift.  The photometric redshift probability distributions are based on $\exp[-\chi^2(z)/2]$ assuming a uniform redshift prior.  Arrows correspond to 2$\sigma$ upper limits on negative flux measurements. The thumbnails are $1.0''$ on a side.}
\label{fig:neighborSED}
\end{figure*}

Massive galaxies at high redshift are expected to be substantially clustered \citep[e.g.,][]{larson22, leonova22, tang23}, and therefore we look for neighbors of GN-z11 in our faint multi-band imaging.  
The discovery of Ly$\alpha$ emission from GN-z11 \citep{bunker23} further motivates the search for neighbors that might impact the photoionization of the surrounding intergalactic medium.
We use the code EAZY \citep{brammer08} which estimates photometric redshifts using a template-fitting approach. We fit the 0.2$''$ aperture JADES NIRCam photometry for the full catalog of detected sources across the JADES GOODS-N footprint. For the fits, we let redshift vary uniformly between $z = 0.01 - 22$ assuming a uniform redshift prior and adopted as the final EAZY photometric redshift the value corresponding to the minimum $\chi^2$ fit at all redshifts, $z_a$. To select potential neighbors of GN-z11, we chose galaxies with $z_a = 9.5 - 11.5$, required that the object had a flux SNR $> 5$ in either the F200W or F277W filters, and that the summed probability of the galaxy being at a redshift above 7, $P(z > 7)$, is greater than $0.7$. We focus here on candidates that lie within the Figure \ref{fig:field} footprint, a 10 comoving Mpc (212$''$) square centered on GN-z11.

In this footprint, we find a population of nine objects with photometric redshifts consistent with 10.6 in the vicinity of GN-z11.
These are shown in Figure \ref{fig:field}, with astrometry and photometry reported in Tables \ref{tab:neighbors} and \ref{tab:neighborsPhot}.  
Figure \ref{fig:neighborSED} shows the spectral energy distributions, 
thumbnail images in each filter, and 
photometric redshift probability versus redshifts (computed from the $\chi^2$ of the EAZY fit, with a uniform redshift prior).
These candidates are much fainter than GN-z11, typically around 10 nJy (AB mag 29). 
JADES-GN-189.07357+62.23749 (ID 4155) is notably brighter, around 30 nJy.
JADES-GN-189.05413+62.21795 (ID 62240) is probably the most tentative physical association.
We have conducted a preliminary search using JADES imaging of the larger GOODS-N JADES footprint, about 50 square arcminutes at this writing, finding many other candidates, which we will report on in a future paper.  However, we do note that this region has more candidates than other portions of the footprint, despite being mildly shallower imaging. Hence, there is an indication of angular clustering.  

The dark matter halos of massive galaxies at these redshifts are likely only a few arcseconds in radius.  We therefore look very close to GN-z11, seeking yet fainter candidates.  We find only one close object, at $\alpha=189.105482$ and $\delta=62.241200$, only 3.2$''$ away, that visually could be a F115W-dropout.  
This object is faint and noisy, only 4 nJy (AB mag 30) but marginally detected in F150W, F200W, and the stack of longer filters.
The photometric redshift fitting to its current photometry mildly favors a mid-redshift solution.   However, its proximity to GN-z11 could reasonably boost one's prior for a high-redshift solution.  We consider this object to be worthy of further attention, but do not consider it to be a highly robust candidate.

\section{Discussion and Conclusions}
\label{sec:conclusions}

We have presented deep JADES 9-band NIRCam imaging of GN-z11, a particularly bright galaxy now known to be at a spectroscopic redshift of $z=10.60$. Our photometry is deeper than and consistent with past GOODS-N imaging of GN-z11.  We  find a strong UV continuum abruptly cut off as a Ly$\alpha$ dropout. The morphology of GN-z11 is very compact, but mildly resolved.  We fit the images with the combination of a point source and a nearly exponential disk with a half-light radius of 49 mas or 200 pc.  About 2/3 of the emission is from the point source, leading to a half-light size of 16 mas or 64 pc.  A faint haze about $0.4''$ away is likely to be a lower redshift galaxy, but might yet be another component of GN-z11.

We fit the point source, the extended emission, and the combination with galaxy spectral synthesis models.  Omitting the possibility of a luminous AGN, our fits argue for an unattenuated low-metallicity star forming galaxy, with about 20~\Msun/yr of star formation and a total stellar mass of $10^{9.1}$~\Msun.  Fitting the two components separately indicates that the extended component has an older star formation history and a lower current specific star formation rate.  The point source component is best fit as a young burst.  We note that such modeling carries substantial uncertainties, particularly when deblending marginally resolved components.  Nevertheless, it is intriguing to see how the subtle color gradient in the F410M--F444W color can inform us about the formation history of the galaxy. This nuclear starburst seems to outshine the galaxy, making up for $\sim2/3$ of the total rest-UV flux, but only contributing $\sim1/5$ of the stellar mass. Theoretically, in cosmological simulations, such events are expected when the gas is rapidly fueled into the central region of early, gas-rich galaxies \citep{dekel14_nugget, zolotov15, tacchella16_profile, dekel23}.

We then search for other galaxies that might be associated to GN-z11.  One candidate $3''$ away is very faint and tentative.  Searching more broadly, nine other galaxies with separations from $0.5'$ to $2'$ appear to be F115W dropouts with photometric redshifts consistent with $z=10.6$.  Our initial impression is that this is a mild angular overdensity, but we leave this study to future work.

GN-z11 is the brightest $z\sim11$ candidate known in the GOODS-S or GOODS-N fields of JADES.  Taking the 320 square arcminute GOODS fields as a lower bound on the search area and considering the redshift range of $10<z<11$, we can view this galaxy as being indicative of the brightest galaxy in a volume of at least $5\times10^5$~Mpc$^3$.  Using an N-body simulation of a $300h^{-1}$~Mpc box with particle mass of $10^7h^{-1}$~M$_\odot$ based on the AbacusSummit methodologies \citep{maksimova21}, we estimate that the most massive halo at this redshift in such a volume would typically be about $8\times10^{10}$~M$_\odot$ and would have a 90\% mass radius of about 8 proper kpc, which is $2''$ on the sky.  

The contextual interpretation of GN-z11 depends on the reason for its unusual brightness.  Is it brighter because it is in a particularly massive halo, or because of some other cause, such as a luminous AGN or some extreme starburst?  If due to its host halo, then the halo mass cannot be much different than $8\times10^{10}$ $M_\odot$, because the halo mass function is very steep.  Even a factor of 3 in mass would be about 30-fold in number density, yielding many other halos that could outshine GN-z11.

The stellar masses inferred in Section~\ref{sec:sed}, about $10^{9.1} \Msun$,  would be compatible with such a halo mass only if around 10\% of the baryons in the halo have been formed into stars.  This is rather efficient \citep{behroozi18, tacchella18, boylan-kolchin22, lovell23}, comparable to that of $L^*$ galaxies today, and perhaps surprising given the low metallicity.  However, the early Universe is much denser than today, and perhaps cooling is indeed very rapid \citep[e.g.,][]{krumholz12, ostriker11, somerville15}.  We note that globular clusters provide a similar behavior: they must consume much of their gas (else they would become unbound when the remaining gas is expelled) and they show evidence of self-enrichment, yet they remain at low metallicity. Perhaps high-redshift galaxies like GN-z11 follow a similar path.  Of course, an alternative explanation would be that the stellar mass is highly overestimated due to incorrect modeling assumptions, such as the stellar initial mass function or the role of binaries in stellar populations \citep[e.g.,][]{eldridge22}. 

If GN-z11 is unusually bright for a reason not primarily driven by its presence in a high-mass halo, then the halo mass would usually be lower. Such a situation would predict that the large-scale clustering of such galaxies would be lower, due to the increase in clustering bias with halo mass. Of course, it also reduces the baryon supply, exacerbating the concerns about the inferred stellar mass.

The compact morphology of GN-z11 clearly could permit a luminous AGN, which would be a plausible route to reduce the fitted stellar mass and ease the tensions with the halo mass. Specifically, the point source component could be interpreted as an AGN, while the extended component would be the host galaxy. Such a scenario is investigated using the JADES NIRSpec data: a variety of emission line ratios (including CIII]$\lambda$1908, CIV$\lambda$1550 and HeII$\lambda$1640) do not obviously favour photoionization due to AGN or star formation \citep{bunker23}. A potential AGN contribution will be considered in a forthcoming paper \citep{maiolino23}. We note that a luminous AGN could also be indicated by photometric variability or multi-wavelength imaging.

JWST continues to transform our view of the high-redshift Universe. Not only does it reveal multiple $z>10$ candidates in every moderately deep pointing, but we are starting to see the diversity of these candidates: in morphology, SEDs, and spectral line diagnostics. GN-z11 is an exemplar of the bright end of the HST redshift frontier, but the great detail we have been able to uncover in it through JADES imaging and spectroscopy shows the marvelous opportunity that JWST will convey. It is most remarkable that we now look to $z\sim10$ galaxies in a manner past limited to $z<2$.

\section{Acknowledgements}

We are grateful to the many people who worked for decades to turn JWST into a reality and in particular to our other colleagues on the NIRCam and NIRSpec instrument teams. We thank the referee for a constructive report that has helped to improve multiple aspects of the paper.

All of the data presented in this paper were obtained from the Mikulski Archive for Space Telescopes (MAST) at the Space Telescope Science Institute. The specific observations analyzed can be accessed via \dataset[https://doi.org/10.17909/247y-bk69]{https://doi.org/10.17909/247y-bk69}.

This research made use of the {\it lux} supercomputer at UC Santa Cruz, funded by NSF MRI grant AST 1828315, NASA's Astrophysics Data System (ADS), the arXiv.org preprint server, the Python plotting library \texttt{matplotlib} \citep{hunter07}, \texttt{astropy}, a community-developed core Python package for Astronomy \citep{astropy-collaboration22}, the python binding of \texttt{FSPS} \citep{foreman_mackey14}, the SED fitting code \texttt{Prospector} \citep{johnson19}, the dynamic nested sampling algorithm \texttt{dynesty} \citep{speagle20}, the {\tt Photutils} package \citep{bradley22}, and the SED handling tool \texttt{sedpy} \citep{johnson19_sedpy}.

D.J.E. is supported as a Simons Investigator. D.J.E., B.D.J., B.E.R., F.S., E.E., M.R., G.R., C.N.A.W. and T.J.L. acknowledge support from the NIRCam Science Team contract to the University of Arizona, NAS5-02015. W.B., R.M., J.W., L.S. and J.S. acknowledges support by the Science and Technology Facilities Council (STFC) and the ERC Advanced Grant 695671 ``QUENCH''. D.P. acknowledges support by the Huo Family Foundation through a P.C. Ho PhD Studentship. R.H. acknowledges funding by the Johns Hopkins University, Institute for Data Intensive Engineering and Science (IDIES). The research of C.C.W. is supported by NOIRLab, which is managed by the Association of Universities for Research in Astronomy (AURA) under a cooperative agreement with the National Science Foundation. E.C.L. acknowledges support of an STFC Webb Fellowship (ST/W001438/1). S.C. acknowledges support by European Union’s HE ERC Starting Grant No. 101040227 - WINGS. R.M. acknowledges funding from a research professorship from the Royal Society. A.J.B., A.J.C., J.C., A.S. and G.C.J. acknowledge funding from the ``FirstGalaxies'' Advanced Grant from the European Research Council (ERC) under the European Union's Horizon 2020 research and innovation programme (Grant agreement No. 789056). A.L.D. thanks the University of Cambridge Harding Distinguished Postgraduate Scholars Programme and Technology Facilities Council (STFC) Center for Doctoral Training (CDT) in Data intensive science at the University of Cambridge (STFC grant number 2742605) for a PhD studentship. J.W. acknowledges support from Fondation MERAC. R.S. acknowledges support from a STFC Ernest Rutherford Fellowship (ST/S004831/1). H{\"U} gratefully acknowledges support by the Isaac Newton Trust and by the Kavli Foundation through a Newton-Kavli Junior Fellowship. K.B. is supported in part by the Australian Research Council Centre of Excellence for All Sky Astrophysics in 3 Dimensions (ASTRO 3D), through project number CE170100013. L.W. acknowledges support from the National Science Foundation Graduate Research Fellowship under Grant No. DGE-2137419.


\begin{thebibliography}{}
\expandafter\ifx\csname natexlab\endcsname\relax\def\natexlab#1{#1}\fi
\providecommand{\url}[1]{\href{#1}{#1}}
\providecommand{\dodoi}[1]{doi:~\href{http://doi.org/#1}{\nolinkurl{#1}}}
\providecommand{\doeprint}[1]{\href{http://ascl.net/#1}{\nolinkurl{http://ascl.net/#1}}}
\providecommand{\doarXiv}[1]{\href{https://arxiv.org/abs/#1}{\nolinkurl{https://arxiv.org/abs/#1}}}

\bibitem[{{Adams} {et~al.}(2023){Adams}, {Conselice}, {Ferreira}, {Austin},
  {Trussler}, {Juod{\v{z}}balis}, {Wilkins}, {Caruana}, {Dayal}, {Verma}, \&
  {Vijayan}}]{adams23}
{Adams}, N.~J., {Conselice}, C.~J., {Ferreira}, L., {et~al.} 2023, \mnras, 518,
  4755, \dodoi{10.1093/mnras/stac3347}

\bibitem[{{Anderson}(2016)}]{anderson16}
{Anderson}, J. 2016, {Empirical Models for the WFC3/IR PSF}, Instrument Science
  Report WFC3 2016-12, 42 pages

\bibitem[{{Anderson} \& {King}(2000)}]{anderson00}
{Anderson}, J., \& {King}, I.~R. 2000, \pasp, 112, 1360, \dodoi{10.1086/316632}

\bibitem[{{Ashby} {et~al.}(2013){Ashby}, {Willner}, {Fazio}, {Huang}, {Arendt},
  {Barmby}, {Barro}, {Bell}, {Bouwens}, {Cattaneo}, {Croton}, {Dav{\'e}},
  {Dunlop}, {Egami}, {Faber}, {Finlator}, {Grogin}, {Guhathakurta},
  {Hernquist}, {Hora}, {Illingworth}, {Kashlinsky}, {Koekemoer}, {Koo},
  {Labb{\'e}}, {Li}, {Lin}, {Moseley}, {Nand ra}, {Newman}, {Noeske}, {Ouchi},
  {Peth}, {Rigopoulou}, {Robertson}, {Sarajedini}, {Simard}, {Smith}, {Wang},
  {Wechsler}, {Weiner}, {Wilson}, {Wuyts}, {Yamada}, \& {Yan}}]{ashby13}
{Ashby}, M.~L.~N., {Willner}, S.~P., {Fazio}, G.~G., {et~al.} 2013, \apj, 769,
  80, \dodoi{10.1088/0004-637X/769/1/80}

\bibitem[{{Astropy Collaboration} {et~al.}(2022){Astropy Collaboration},
  {Price-Whelan}, {Lim}, {Earl}, {Starkman}, {Bradley}, {Shupe}, {Patil},
  {Corrales}, {Brasseur}, {N{\"o}the}, {Donath}, {Tollerud}, {Morris},
  {Ginsburg}, {Vaher}, {Weaver}, {Tocknell}, {Jamieson}, {van Kerkwijk},
  {Robitaille}, {Merry}, {Bachetti}, {G{\"u}nther}, {Aldcroft},
  {Alvarado-Montes}, {Archibald}, {B{\'o}di}, {Bapat}, {Barentsen},
  {Baz{\'a}n}, {Biswas}, {Boquien}, {Burke}, {Cara}, {Cara}, {Conroy},
  {Conseil}, {Craig}, {Cross}, {Cruz}, {D'Eugenio}, {Dencheva}, {Devillepoix},
  {Dietrich}, {Eigenbrot}, {Erben}, {Ferreira}, {Foreman-Mackey}, {Fox},
  {Freij}, {Garg}, {Geda}, {Glattly}, {Gondhalekar}, {Gordon}, {Grant},
  {Greenfield}, {Groener}, {Guest}, {Gurovich}, {Handberg}, {Hart},
  {Hatfield-Dodds}, {Homeier}, {Hosseinzadeh}, {Jenness}, {Jones}, {Joseph},
  {Kalmbach}, {Karamehmetoglu}, {Ka{\l}uszy{\'n}ski}, {Kelley}, {Kern},
  {Kerzendorf}, {Koch}, {Kulumani}, {Lee}, {Ly}, {Ma}, {MacBride}, {Maljaars},
  {Muna}, {Murphy}, {Norman}, {O'Steen}, {Oman}, {Pacifici}, {Pascual},
  {Pascual-Granado}, {Patil}, {Perren}, {Pickering}, {Rastogi}, {Roulston},
  {Ryan}, {Rykoff}, {Sabater}, {Sakurikar}, {Salgado}, {Sanghi}, {Saunders},
  {Savchenko}, {Schwardt}, {Seifert-Eckert}, {Shih}, {Jain}, {Shukla}, {Sick},
  {Simpson}, {Singanamalla}, {Singer}, {Singhal}, {Sinha}, {Sip{\H{o}}cz},
  {Spitler}, {Stansby}, {Streicher}, {{\v{S}}umak}, {Swinbank}, {Taranu},
  {Tewary}, {Tremblay}, {de Val-Borro}, {Van Kooten}, {Vasovi{\'c}}, {Verma},
  {de Miranda Cardoso}, {Williams}, {Wilson}, {Winkel}, {Wood-Vasey}, {Xue},
  {Yoachim}, {Zhang}, {Zonca}, \& {Astropy Project
  Contributors}}]{astropy-collaboration22}
{Astropy Collaboration}, {Price-Whelan}, A.~M., {Lim}, P.~L., {et~al.} 2022,
  \apj, 935, 167, \dodoi{10.3847/1538-4357/ac7c74}

\bibitem[{{Atek} {et~al.}(2023){Atek}, {Shuntov}, {Furtak}, {Richard}, {Kneib},
  {Mahler}, {Zitrin}, {McCracken}, {Charlot}, {Chevallard}, \&
  {Chemerynska}}]{atek23}
{Atek}, H., {Shuntov}, M., {Furtak}, L.~J., {et~al.} 2023, \mnras, 519, 1201,
  \dodoi{10.1093/mnras/stac3144}

\bibitem[{{Barro} {et~al.}(2019){Barro}, {P{\'e}rez-Gonz{\'a}lez}, {Cava},
  {Brammer}, {Pandya}, {Eliche Moral}, {Esquej}, {Dom{\'\i}nguez-S{\'a}nchez},
  {Alcalde Pampliega}, {Guo}, {Koekemoer}, {Trump}, {Ashby}, {Cardiel},
  {Castellano}, {Conselice}, {Dickinson}, {Dolch}, {Donley}, {Espino Briones},
  {Faber}, {Fazio}, {Ferguson}, {Finkelstein}, {Fontana}, {Galametz},
  {Gardner}, {Gawiser}, {Giavalisco}, {Grazian}, {Grogin}, {Hathi}, {Hemmati},
  {Hern{\'a}n-Caballero}, {Kocevski}, {Koo}, {Kodra}, {Lee}, {Lin}, {Lucas},
  {Mobasher}, {McGrath}, {Nandra}, {Nayyeri}, {Newman}, {Pforr}, {Peth},
  {Rafelski}, {Rodr{\'\i}guez-Munoz}, {Salvato}, {Stefanon}, {van der Wel},
  {Willner}, {Wiklind}, \& {Wuyts}}]{barro19}
{Barro}, G., {P{\'e}rez-Gonz{\'a}lez}, P.~G., {Cava}, A., {et~al.} 2019, \apjs,
  243, 22, \dodoi{10.3847/1538-4365/ab23f2}

\bibitem[{{Behroozi} \& {Silk}(2018)}]{behroozi18}
{Behroozi}, P., \& {Silk}, J. 2018, \mnras, 477, 5382,
  \dodoi{10.1093/mnras/sty945}

\bibitem[{{Birrer} \& {Amara}(2018)}]{birrer18}
{Birrer}, S., \& {Amara}, A. 2018, Physics of the Dark Universe, 22, 189,
  \dodoi{10.1016/j.dark.2018.11.002}

\bibitem[{{Bouwens} {et~al.}(2019){Bouwens}, {Stefanon}, {Oesch},
  {Illingworth}, {Nanayakkara}, {Roberts-Borsani}, {Labb{\'e}}, \&
  {Smit}}]{bouwens19}
{Bouwens}, R.~J., {Stefanon}, M., {Oesch}, P.~A., {et~al.} 2019, \apj, 880, 25,
  \dodoi{10.3847/1538-4357/ab24c5}

\bibitem[{{Bouwens} {et~al.}(2010){Bouwens}, {Illingworth}, {Gonz{\'a}lez},
  {Labb{\'e}}, {Franx}, {Conselice}, {Blakeslee}, {van Dokkum}, {Holden},
  {Magee}, {Marchesini}, \& {Zheng}}]{bouwens10_nicmos}
{Bouwens}, R.~J., {Illingworth}, G.~D., {Gonz{\'a}lez}, V., {et~al.} 2010,
  \apj, 725, 1587, \dodoi{10.1088/0004-637X/725/2/1587}

\bibitem[{{Bouwens} {et~al.}(2021){Bouwens}, {Oesch}, {Stefanon},
  {Illingworth}, {Labb{\'e}}, {Reddy}, {Atek}, {Montes}, {Naidu},
  {Nanayakkara}, {Nelson}, \& {Wilkins}}]{bouwens21}
{Bouwens}, R.~J., {Oesch}, P.~A., {Stefanon}, M., {et~al.} 2021, \aj, 162, 47,
  \dodoi{10.3847/1538-3881/abf83e}

\bibitem[{{Boyer} {et~al.}(2022){Boyer}, {Anderson}, {Gennaro}, {Geha},
  {Wingfield McQuinn}, {Tollerud}, {Correnti}, {Brenner Newman}, {Cohen},
  {Kallivayalil}, {Beaton}, {Cole}, {Dolphin}, {Kalirai}, {Sandstrom},
  {Savino}, {Skillman}, {Weisz}, \& {Williams}}]{boyer22}
{Boyer}, M.~L., {Anderson}, J., {Gennaro}, M., {et~al.} 2022, Research Notes of
  the American Astronomical Society, 6, 191, \dodoi{10.3847/2515-5172/ac923a}

\bibitem[{{Boylan-Kolchin}(2022)}]{boylan-kolchin22}
{Boylan-Kolchin}, M. 2022, arXiv e-prints, arXiv:2208.01611,
  \dodoi{10.48550/arXiv.2208.01611}

\bibitem[{{Bradley} {et~al.}(2022){Bradley}, {Sip{\H{o}}cz}, {Robitaille},
  {Tollerud}, {Vin{\'\i}cius}, {Deil}, {Barbary}, {Wilson}, {Busko}, {Donath},
  {G{\"u}nther}, {Cara}, {Lim}, {Me{\ss}linger}, {Conseil}, {Bostroem},
  {Droettboom}, {Bray}, {Andersen Bratholm}, {Barentsen}, {Craig}, {Rathi},
  {Pascual}, {Perren}, {Georgiev}, {De Val-Borro}, {Kerzendorf}, {Bach},
  {Quint}, \& {Souchereau}}]{bradley22}
{Bradley}, L., {Sip{\H{o}}cz}, B., {Robitaille}, T., {et~al.} 2022,
  {astropy/photutils: 1.5.0}, 1.5.0, Zenodo,  Zenodo,
  \dodoi{10.5281/zenodo.6825092}

\bibitem[{{Brammer} {et~al.}(2008){Brammer}, {van Dokkum}, \&
  {Coppi}}]{brammer08}
{Brammer}, G.~B., {van Dokkum}, P.~G., \& {Coppi}, P. 2008, \apj, 686, 1503,
  \dodoi{10.1086/591786}

\bibitem[{{Bruzual} \& {Charlot}(2003)}]{bruzual03}
{Bruzual}, G., \& {Charlot}, S. 2003, \mnras, 344, 1000,
  \dodoi{10.1046/j.1365-8711.2003.06897.x}

\bibitem[{{Bunker} {et~al.}(2023){Bunker}, {Saxena}, {Cameron}, {Willott},
  {Curtis-Lake}, {Jakobsen}, {Carniani}, {Smit}, {Maiolino}, {Witstok},
  {Curti}, {D'Eugenio}, {Jones}, {Ferruit}, {Arribas}, {Charlot}, {Chevallard},
  {Giardino}, {de Graaff}, {Looser}, {Luetzgendorf}, {Maseda}, {Rawle}, {Rix},
  {Rodriguez Del Pino}, {Alberts}, {Egami}, {Eisenstein}, {Endsley},
  {Hainline}, {Hausen}, {Johnson}, {Rieke}, {Rieke}, {Robertson}, {Shivaei},
  {Stark}, {Sun}, {Tacchella}, {Tang}, {Williams}, {Willmer}, {Baker}, {Baum},
  {Bhatawdekar}, {Bowler}, {Boyett}, {Chen}, {Circosta}, {Helton}, {Ji}, {Lyu},
  {Nelson}, {Parlanti}, {Perna}, {Sandles}, {Scholtz}, {Suess}, {Topping},
  {Uebler}, {Wallace}, \& {Whitler}}]{bunker23}
{Bunker}, A.~J., {Saxena}, A., {Cameron}, A.~J., {et~al.} 2023, arXiv e-prints,
  arXiv:2302.07256, \dodoi{10.48550/arXiv.2302.07256}

\bibitem[{{Byler} {et~al.}(2017){Byler}, {Dalcanton}, {Conroy}, \&
  {Johnson}}]{byler17}
{Byler}, N., {Dalcanton}, J.~J., {Conroy}, C., \& {Johnson}, B.~D. 2017, \apj,
  840, 44, \dodoi{10.3847/1538-4357/aa6c66}

\bibitem[{{Calzetti} {et~al.}(2000){Calzetti}, {Armus}, {Bohlin}, {Kinney},
  {Koornneef}, \& {Storchi-Bergmann}}]{calzetti00}
{Calzetti}, D., {Armus}, L., {Bohlin}, R.~C., {et~al.} 2000, \apj, 533, 682,
  \dodoi{10.1086/308692}

\bibitem[{{Calzetti} {et~al.}(2007){Calzetti}, {Kennicutt}, {Engelbracht},
  {Leitherer}, {Draine}, {Kewley}, {Moustakas}, {Sosey}, {Dale}, {Gordon},
  {Helou}, {Hollenbach}, {Armus}, {Bendo}, {Bot}, {Buckalew}, {Jarrett}, {Li},
  {Meyer}, {Murphy}, {Prescott}, {Regan}, {Rieke}, {Roussel}, {Sheth}, {Smith},
  {Thornley}, \& {Walter}}]{calzetti07}
{Calzetti}, D., {Kennicutt}, R.~C., {Engelbracht}, C.~W., {et~al.} 2007, \apj,
  666, 870, \dodoi{10.1086/520082}

\bibitem[{{Carnall} {et~al.}(2018){Carnall}, {McLure}, {Dunlop}, \&
  {Dav{\'e}}}]{carnall18}
{Carnall}, A.~C., {McLure}, R.~J., {Dunlop}, J.~S., \& {Dav{\'e}}, R. 2018,
  \mnras, 480, 4379, \dodoi{10.1093/mnras/sty2169}

\bibitem[{{Chabrier}(2003)}]{chabrier03}
{Chabrier}, G. 2003, \pasp, 115, 763, \dodoi{10.1086/376392}

\bibitem[{{Charlot} \& {Fall}(2000)}]{charlot00}
{Charlot}, S., \& {Fall}, S.~M. 2000, \apj, 539, 718, \dodoi{10.1086/309250}

\bibitem[{{Chen} {et~al.}(2023){Chen}, {Stark}, {Endsley}, {Topping},
  {Whitler}, \& {Charlot}}]{chen23}
{Chen}, Z., {Stark}, D.~P., {Endsley}, R., {et~al.} 2023, \mnras, 518, 5607,
  \dodoi{10.1093/mnras/stac3476}

\bibitem[{{Chevallard} \& {Charlot}(2016)}]{chevallard16}
{Chevallard}, J., \& {Charlot}, S. 2016, \mnras, 462, 1415,
  \dodoi{10.1093/mnras/stw1756}

\bibitem[{{Choi} {et~al.}(2017){Choi}, {Conroy}, \& {Byler}}]{choi17}
{Choi}, J., {Conroy}, C., \& {Byler}, N. 2017, \apj, 838, 159,
  \dodoi{10.3847/1538-4357/aa679f}

\bibitem[{{Choi} {et~al.}(2016){Choi}, {Dotter}, {Conroy}, {Cantiello},
  {Paxton}, \& {Johnson}}]{choi16}
{Choi}, J., {Dotter}, A., {Conroy}, C., {et~al.} 2016, \apj, 823, 102,
  \dodoi{10.3847/0004-637X/823/2/102}

\bibitem[{{Conroy}(2013)}]{conroy13_rev}
{Conroy}, C. 2013, \araa, 51, 393, \dodoi{10.1146/annurev-astro-082812-141017}

\bibitem[{{Conselice} {et~al.}(2011){Conselice}, {Bluck}, {Buitrago}, {Bauer},
  {Gr{\"u}tzbauch}, {Bouwens}, {Bevan}, {Mortlock}, {Dickinson}, {Daddi},
  {Yan}, {Scott}, {Chapman}, {Chary}, {Ferguson}, {Giavalisco}, {Grogin},
  {Illingworth}, {Jogee}, {Koekemoer}, {Lucas}, {Mobasher}, {Moustakas},
  {Papovich}, {Ravindranath}, {Siana}, {Teplitz}, {Trujillo}, {Urry}, \&
  {Weinzirl}}]{conselice11}
{Conselice}, C.~J., {Bluck}, A.~F.~L., {Buitrago}, F., {et~al.} 2011, \mnras,
  413, 80, \dodoi{10.1111/j.1365-2966.2010.18113.x}

\bibitem[{{Costantin} {et~al.}(2023){Costantin}, {P{\'e}rez-Gonz{\'a}lez},
  {Vega-Ferrero}, {Huertas-Company}, {Bisigello}, {Buitrago}, {Bagley},
  {Cleri}, {Cooper}, {Finkelstein}, {Holwerda}, {Kartaltepe}, {Koekemoer},
  {Nelson}, {Papovich}, {Pillepich}, {Pirzkal}, {Tacchella}, \&
  {Yung}}]{costantin23}
{Costantin}, L., {P{\'e}rez-Gonz{\'a}lez}, P.~G., {Vega-Ferrero}, J., {et~al.}
  2023, \apj, 946, 71, \dodoi{10.3847/1538-4357/acb926}

\bibitem[{{Curtis-Lake} {et~al.}(2023){Curtis-Lake}, {Carniani}, {Cameron},
  {Charlot}, {Jakobsen}, {Maiolino}, {Bunker}, {Witstok}, {Smit}, {Chevallard},
  {Willott}, {Ferruit}, {Arribas}, {Bonaventura}, {Curti}, {D'Eugenio},
  {Franx}, {Giardino}, {Looser}, {L{\"u}tzgendorf}, {Maseda}, {Rawle}, {Rix},
  {Rodr{\'\i}guez del Pino}, {{\"U}bler}, {Sirianni}, {Dressler}, {Egami},
  {Eisenstein}, {Endsley}, {Hainline}, {Hausen}, {Johnson}, {Rieke},
  {Robertson}, {Shivaei}, {Stark}, {Tacchella}, {Williams}, {Willmer},
  {Bhatawdekar}, {Bowler}, {Boyett}, {Chen}, {de Graaff}, {Helton}, {Hviding},
  {Jones}, {Kumari}, {Lyu}, {Nelson}, {Perna}, {Sandles}, {Saxena}, {Suess},
  {Sun}, {Topping}, {Wallace}, \& {Whitler}}]{curtis-lake23}
{Curtis-Lake}, E., {Carniani}, S., {Cameron}, A., {et~al.} 2023, Nature
  Astronomy, \dodoi{10.1038/s41550-023-01918-w}

\bibitem[{{Dayal} \& {Ferrara}(2018)}]{dayal18}
{Dayal}, P., \& {Ferrara}, A. 2018, \physrep, 780, 1,
  \dodoi{10.1016/j.physrep.2018.10.002}

\bibitem[{{Dayal} {et~al.}(2015){Dayal}, {Mesinger}, \& {Pacucci}}]{dayal15}
{Dayal}, P., {Mesinger}, A., \& {Pacucci}, F. 2015, \apj, 806, 67,
  \dodoi{10.1088/0004-637X/806/1/67}

\bibitem[{{Dekel} \& {Burkert}(2014)}]{dekel14_nugget}
{Dekel}, A., \& {Burkert}, A. 2014, \mnras, 438, 1870,
  \dodoi{10.1093/mnras/stt2331}

\bibitem[{{Dekel} {et~al.}(2023){Dekel}, {Sarkar}, {Birnboim}, {Mandelker}, \&
  {Li}}]{dekel23}
{Dekel}, A., {Sarkar}, K.~S., {Birnboim}, Y., {Mandelker}, N., \& {Li}, Z.
  2023, arXiv e-prints, arXiv:2303.04827, \dodoi{10.48550/arXiv.2303.04827}

\bibitem[{{Donnan} {et~al.}(2023){Donnan}, {McLeod}, {Dunlop}, {McLure},
  {Carnall}, {Begley}, {Cullen}, {Hamadouche}, {Bowler}, {Magee}, {McCracken},
  {Milvang-Jensen}, {Moneti}, \& {Targett}}]{donnan23}
{Donnan}, C.~T., {McLeod}, D.~J., {Dunlop}, J.~S., {et~al.} 2023, \mnras, 518,
  6011, \dodoi{10.1093/mnras/stac3472}

\bibitem[{{Eldridge} \& {Stanway}(2022)}]{eldridge22}
{Eldridge}, J.~J., \& {Stanway}, E.~R. 2022, \araa, 60, 455,
  \dodoi{10.1146/annurev-astro-052920-100646}

\bibitem[{{Ellis} {et~al.}(2013){Ellis}, {McLure}, {Dunlop}, {Robertson},
  {Ono}, {Schenker}, {Koekemoer}, {Bowler}, {Ouchi}, {Rogers}, {Curtis-Lake},
  {Schneider}, {Charlot}, {Stark}, {Furlanetto}, \& {Cirasuolo}}]{ellis13}
{Ellis}, R.~S., {McLure}, R.~J., {Dunlop}, J.~S., {et~al.} 2013, \apjl, 763,
  L7, \dodoi{10.1088/2041-8205/763/1/L7}

\bibitem[{{Ferreira} {et~al.}(2022){Ferreira}, {Conselice}, {Sazonova},
  {Ferrari}, {Caruana}, {Tohill}, {Lucatelli}, {Adams}, {Irodotou}, {Marshall},
  {Roper}, {Lovell}, {Verma}, {Austin}, {Trussler}, \& {Wilkins}}]{ferreira22}
{Ferreira}, L., {Conselice}, C.~J., {Sazonova}, E., {et~al.} 2022, arXiv
  e-prints, arXiv:2210.01110, \dodoi{10.48550/arXiv.2210.01110}

\bibitem[{{Finkelstein} {et~al.}(2015){Finkelstein}, {Ryan}, {Papovich},
  {Dickinson}, {Song}, {Somerville}, {Ferguson}, {Salmon}, {Giavalisco},
  {Koekemoer}, {Ashby}, {Behroozi}, {Castellano}, {Dunlop}, {Faber}, {Fazio},
  {Fontana}, {Grogin}, {Hathi}, {Jaacks}, {Kocevski}, {Livermore}, {McLure},
  {Merlin}, {Mobasher}, {Newman}, {Rafelski}, {Tilvi}, \&
  {Willner}}]{finkelstein15}
{Finkelstein}, S.~L., {Ryan}, Jr., R.~E., {Papovich}, C., {et~al.} 2015, \apj,
  810, 71, \dodoi{10.1088/0004-637X/810/1/71}

\bibitem[{{Finkelstein} {et~al.}(2022{\natexlab{a}}){Finkelstein}, {Bagley},
  {Song}, {Larson}, {Papovich}, {Dickinson}, {Finkelstein}, {Koekemoer},
  {Pirzkal}, {Somerville}, {Yung}, {Behroozi}, {Ferguson}, {Giavalisco},
  {Grogin}, {Hathi}, {Hutchison}, {Jung}, {Kocevski}, {Kawinwanichakij},
  {Rojas-Ruiz}, {Ryan}, {Snyder}, \& {Tacchella}}]{finkelstein22_hst}
{Finkelstein}, S.~L., {Bagley}, M., {Song}, M., {et~al.} 2022{\natexlab{a}},
  \apj, 928, 52, \dodoi{10.3847/1538-4357/ac3aed}

\bibitem[{{Finkelstein} {et~al.}(2022{\natexlab{b}}){Finkelstein}, {Bagley},
  {Haro}, {Dickinson}, {Ferguson}, {Kartaltepe}, {Papovich}, {Burgarella},
  {Kocevski}, {Huertas-Company}, {Iyer}, {Koekemoer}, {Larson},
  {P{\'e}rez-Gonz{\'a}lez}, {Rose}, {Tacchella}, {Wilkins}, {Chworowsky},
  {Medrano}, {Morales}, {Somerville}, {Yung}, {Fontana}, {Giavalisco},
  {Grazian}, {Grogin}, {Kewley}, {Kirkpatrick}, {Kurczynski}, {Lotz},
  {Pentericci}, {Pirzkal}, {Ravindranath}, {Ryan}, {Trump}, {Yang}, {Almaini},
  {Amor{\'\i}n}, {Annunziatella}, {Backhaus}, {Barro}, {Behroozi}, {Bell},
  {Bhatawdekar}, {Bisigello}, {Bromm}, {Buat}, {Buitrago}, {Calabr{\`o}},
  {Casey}, {Castellano}, {Ch{\'a}vez Ortiz}, {Ciesla}, {Cleri}, {Cohen},
  {Cole}, {Cooke}, {Cooper}, {Cooray}, {Costantin}, {Cox}, {Croton}, {Daddi},
  {Dav{\'e}}, {de La Vega}, {Dekel}, {Elbaz}, {Estrada-Carpenter}, {Faber},
  {Fern{\'a}ndez}, {Finkelstein}, {Freundlich}, {Fujimoto},
  {Garc{\'\i}a-Argum{\'a}nez}, {Gardner}, {Gawiser}, {G{\'o}mez-Guijarro},
  {Guo}, {Hamblin}, {Hamilton}, {Hathi}, {Holwerda}, {Hirschmann}, {Hutchison},
  {Jaskot}, {Jha}, {Jogee}, {Juneau}, {Jung}, {Kassin}, {Le Bail}, {Leung},
  {Lucas}, {Magnelli}, {Mantha}, {Matharu}, {McGrath}, {McIntosh}, {Merlin},
  {Mobasher}, {Newman}, {Nicholls}, {Pandya}, {Rafelski}, {Ronayne}, {Santini},
  {Seill{\'e}}, {Shah}, {Shen}, {Simons}, {Snyder}, {Stanway}, {Straughn},
  {Teplitz}, {Vanderhoof}, {Vega-Ferrero}, {Wang}, {Weiner}, {Willmer},
  {Wuyts}, {Zavala}, \& {CEERS Team}}]{finkelstein22}
{Finkelstein}, S.~L., {Bagley}, M.~B., {Haro}, P.~A., {et~al.}
  2022{\natexlab{b}}, \apjl, 940, L55, \dodoi{10.3847/2041-8213/ac966e}

\bibitem[{{Foreman-Mackey} {et~al.}(2014){Foreman-Mackey}, {Sick}, \&
  {Johnson}}]{foreman_mackey14}
{Foreman-Mackey}, D., {Sick}, J., \& {Johnson}, B. 2014, Zenodo

\bibitem[{{Gaia Collaboration} {et~al.}(2021){Gaia Collaboration}, {Brown},
  {Vallenari}, {Prusti}, {de Bruijne}, {Babusiaux}, {Biermann}, {Creevey},
  {Evans}, {Eyer}, {Hutton}, {Jansen}, {Jordi}, {Klioner}, {Lammers},
  {Lindegren}, {Luri}, {Mignard}, {Panem}, {Pourbaix}, {Randich}, {Sartoretti},
  {Soubiran}, {Walton}, {Arenou}, {Bailer-Jones}, {Bastian}, {Cropper},
  {Drimmel}, {Katz}, {Lattanzi}, {van Leeuwen}, {Bakker}, {Cacciari},
  {Casta{\~n}eda}, {De Angeli}, {Ducourant}, {Fabricius}, {Fouesneau},
  {Fr{\'e}mat}, {Guerra}, {Guerrier}, {Guiraud}, {Jean-Antoine Piccolo},
  {Masana}, {Messineo}, {Mowlavi}, {Nicolas}, {Nienartowicz}, {Pailler},
  {Panuzzo}, {Riclet}, {Roux}, {Seabroke}, {Sordo}, {Tanga}, {Th{\'e}venin},
  {Gracia-Abril}, {Portell}, {Teyssier}, {Altmann}, {Andrae}, {Bellas-Velidis},
  {Benson}, {Berthier}, {Blomme}, {Brugaletta}, {Burgess}, {Busso}, {Carry},
  {Cellino}, {Cheek}, {Clementini}, {Damerdji}, {Davidson}, {Delchambre},
  {Dell'Oro}, {Fern{\'a}ndez-Hern{\'a}ndez}, {Galluccio}, {Garc{\'\i}a-Lario},
  {Garcia-Reinaldos}, {Gonz{\'a}lez-N{\'u}{\~n}ez}, {Gosset}, {Haigron},
  {Halbwachs}, {Hambly}, {Harrison}, {Hatzidimitriou}, {Heiter},
  {Hern{\'a}ndez}, {Hestroffer}, {Hodgkin}, {Holl}, {Jan{\ss}en}, {Jevardat de
  Fombelle}, {Jordan}, {Krone-Martins}, {Lanzafame}, {L{\"o}ffler}, {Lorca},
  {Manteiga}, {Marchal}, {Marrese}, {Moitinho}, {Mora}, {Muinonen}, {Osborne},
  {Pancino}, {Pauwels}, {Petit}, {Recio-Blanco}, {Richards}, {Riello},
  {Rimoldini}, {Robin}, {Roegiers}, {Rybizki}, {Sarro}, {Siopis}, {Smith},
  {Sozzetti}, {Ulla}, {Utrilla}, {van Leeuwen}, {van Reeven}, {Abbas}, {Abreu
  Aramburu}, {Accart}, {Aerts}, {Aguado}, {Ajaj}, {Altavilla}, {{\'A}lvarez},
  {{\'A}lvarez Cid-Fuentes}, {Alves}, {Anderson}, {Anglada Varela}, {Antoja},
  {Audard}, {Baines}, {Baker}, {Balaguer-N{\'u}{\~n}ez}, {Balbinot}, {Balog},
  {Barache}, {Barbato}, {Barros}, {Barstow}, {Bartolom{\'e}}, {Bassilana},
  {Bauchet}, {Baudesson-Stella}, {Becciani}, {Bellazzini}, {Bernet}, {Bertone},
  {Bianchi}, {Blanco-Cuaresma}, {Boch}, {Bombrun}, {Bossini}, {Bouquillon},
  {Bragaglia}, {Bramante}, {Breedt}, {Bressan}, {Brouillet}, {Bucciarelli},
  {Burlacu}, {Busonero}, {Butkevich}, {Buzzi}, {Caffau}, {Cancelliere},
  {C{\'a}novas}, {Cantat-Gaudin}, {Carballo}, {Carlucci}, {Carnerero},
  {Carrasco}, {Casamiquela}, {Castellani}, {Castro-Ginard}, {Castro Sampol},
  {Chaoul}, {Charlot}, {Chemin}, {Chiavassa}, {Cioni}, {Comoretto}, {Cooper},
  {Cornez}, {Cowell}, {Crifo}, {Crosta}, {Crowley}, {Dafonte}, {Dapergolas},
  {David}, {David}, {de Laverny}, {De Luise}, {De March}, {De Ridder}, {de
  Souza}, {de Teodoro}, {de Torres}, {del Peloso}, {del Pozo}, {Delbo},
  {Delgado}, {Delgado}, {Delisle}, {Di Matteo}, {Diakite}, {Diener},
  {Distefano}, {Dolding}, {Eappachen}, {Edvardsson}, {Enke}, {Esquej}, {Fabre},
  {Fabrizio}, {Faigler}, {Fedorets}, {Fernique}, {Fienga}, {Figueras},
  {Fouron}, {Fragkoudi}, {Fraile}, {Franke}, {Gai}, {Garabato},
  {Garcia-Gutierrez}, {Garc{\'\i}a-Torres}, {Garofalo}, {Gavras}, {Gerlach},
  {Geyer}, {Giacobbe}, {Gilmore}, {Girona}, {Giuffrida}, {Gomel}, {Gomez},
  {Gonzalez-Santamaria}, {Gonz{\'a}lez-Vidal}, {Granvik},
  {Guti{\'e}rrez-S{\'a}nchez}, {Guy}, {Hauser}, {Haywood}, {Helmi}, {Hidalgo},
  {Hilger}, {H{\l}adczuk}, {Hobbs}, {Holland}, {Huckle}, {Jasniewicz},
  {Jonker}, {Juaristi Campillo}, {Julbe}, {Karbevska}, {Kervella}, {Khanna},
  {Kochoska}, {Kontizas}, {Kordopatis}, {Korn}, {Kostrzewa-Rutkowska},
  {Kruszy{\'n}ska}, {Lambert}, {Lanza}, {Lasne}, {Le Campion}, {Le Fustec},
  {Lebreton}, {Lebzelter}, {Leccia}, {Leclerc}, {Lecoeur-Taibi}, {Liao},
  {Licata}, {Lindstr{\o}m}, {Lister}, {Livanou}, {Lobel}, {Madrero Pardo},
  {Managau}, {Mann}, {Marchant}, {Marconi}, {Marcos Santos}, {Marinoni},
  {Marocco}, {Marshall}, {Martin Polo}, {Mart{\'\i}n-Fleitas}, {Masip},
  {Massari}, {Mastrobuono-Battisti}, {Mazeh}, {McMillan}, {Messina},
  {Michalik}, {Millar}, {Mints}, {Molina}, {Molinaro}, {Moln{\'a}r},
  {Montegriffo}, {Mor}, {Morbidelli}, {Morel}, {Morris}, {Mulone}, {Munoz},
  {Muraveva}, {Murphy}, {Musella}, {Noval}, {Ord{\'e}novic}, {Orr{\`u}},
  {Osinde}, {Pagani}, {Pagano}, {Palaversa}, {Palicio}, {Panahi}, {Pawlak},
  {Pe{\~n}alosa Esteller}, {Penttil{\"a}}, {Piersimoni}, {Pineau}, {Plachy},
  {Plum}, {Poggio}, {Poretti}, {Poujoulet}, {Pr{\v{s}}a}, {Pulone}, {Racero},
  {Ragaini}, {Rainer}, {Raiteri}, {Rambaux}, {Ramos}, {Ramos-Lerate}, {Re
  Fiorentin}, {Regibo}, {Reyl{\'e}}, {Ripepi}, {Riva}, {Rixon}, {Robichon},
  {Robin}, {Roelens}, {Rohrbasser}, {Romero-G{\'o}mez}, {Rowell}, {Royer},
  {Rybicki}, {Sadowski}, {Sagrist{\`a} Sell{\'e}s}, {Sahlmann}, {Salgado},
  {Salguero}, {Samaras}, {Sanchez Gimenez}, {Sanna}, {Santove{\~n}a},
  {Sarasso}, {Schultheis}, {Sciacca}, {Segol}, {Segovia}, {S{\'e}gransan},
  {Semeux}, {Shahaf}, {Siddiqui}, {Siebert}, {Siltala}, {Slezak}, {Smart},
  {Solano}, {Solitro}, {Souami}, {Souchay}, {Spagna}, {Spoto}, {Steele},
  {Steidelm{\"u}ller}, {Stephenson}, {S{\"u}veges}, {Szabados}, {Szegedi-Elek},
  {Taris}, {Tauran}, {Taylor}, {Teixeira}, {Thuillot}, {Tonello}, {Torra},
  {Torra}, {Turon}, {Unger}, {Vaillant}, {van Dillen}, {Vanel}, {Vecchiato},
  {Viala}, {Vicente}, {Voutsinas}, {Weiler}, {Wevers}, {Wyrzykowski}, {Yoldas},
  {Yvard}, {Zhao}, {Zorec}, {Zucker}, {Zurbach}, \& {Zwitter}}]{gaia_2021}
{Gaia Collaboration}, {Brown}, A.~G.~A., {Vallenari}, A., {et~al.} 2021, \aap,
  649, A1, \dodoi{10.1051/0004-6361/202039657}

\bibitem[{{Gandolfi} {et~al.}(2022){Gandolfi}, {Lapi}, {Ronconi}, \&
  {Danese}}]{gandolfi22}
{Gandolfi}, G., {Lapi}, A., {Ronconi}, T., \& {Danese}, L. 2022, Universe, 8,
  589, \dodoi{10.3390/universe8110589}

\bibitem[{{Harikane} {et~al.}(2023){Harikane}, {Ouchi}, {Oguri}, {Ono},
  {Nakajima}, {Isobe}, {Umeda}, {Mawatari}, \& {Zhang}}]{harikane23_uvlf}
{Harikane}, Y., {Ouchi}, M., {Oguri}, M., {et~al.} 2023, \apjs, 265, 5,
  \dodoi{10.3847/1538-4365/acaaa9}

\bibitem[{{Holwerda} {et~al.}(2015){Holwerda}, {Bouwens}, {Oesch}, {Smit},
  {Illingworth}, \& {Labbe}}]{holwerda15}
{Holwerda}, B.~W., {Bouwens}, R., {Oesch}, P., {et~al.} 2015, \apj, 808, 6,
  \dodoi{10.1088/0004-637X/808/1/6}

\bibitem[{Hunter(2007)}]{hunter07}
Hunter, J.~D. 2007, Computing In Science \& Engineering, 9, 90

\bibitem[{{Illingworth} {et~al.}(2013){Illingworth}, {Magee}, {Oesch},
  {Bouwens}, {Labb{\'e}}, {Stiavelli}, {van Dokkum}, {Franx}, {Trenti},
  {Carollo}, \& {Gonzalez}}]{illingworth13}
{Illingworth}, G.~D., {Magee}, D., {Oesch}, P.~A., {et~al.} 2013, \apjs, 209,
  6, \dodoi{10.1088/0067-0049/209/1/6}

\bibitem[{{Jacobs} {et~al.}(2022){Jacobs}, {Glazebrook}, {Calabr{\`o}}, {Treu},
  {Nanayakkara}, {Jones}, {Merlin}, {Abraham}, {Stevens}, {Vulcani}, {Yang},
  {Bonchi}, {Bradac}, {Castellano}, {Fontana}, {Mason}, {Morishita}, {Paris},
  {Trenti}, {Marchesini}, {Wang}, \& {Santini}}]{jacobs22}
{Jacobs}, C., {Glazebrook}, K., {Calabr{\`o}}, A., {et~al.} 2022, arXiv
  e-prints, arXiv:2208.06516, \dodoi{10.48550/arXiv.2208.06516}

\bibitem[{{Jiang} {et~al.}(2021){Jiang}, {Kashikawa}, {Wang}, {Walth}, {Ho},
  {Cai}, {Egami}, {Fan}, {Ito}, {Liang}, {Schaerer}, \& {Stark}}]{jiang21}
{Jiang}, L., {Kashikawa}, N., {Wang}, S., {et~al.} 2021, Nature Astronomy, 5,
  256, \dodoi{10.1038/s41550-020-01275-y}

\bibitem[{{Jin} {et~al.}(2022){Jin}, {Kewley}, \& {Sutherland}}]{jin22}
{Jin}, Y., {Kewley}, L.~J., \& {Sutherland}, R.~S. 2022, \apjl, 934, L8,
  \dodoi{10.3847/2041-8213/ac80f3}

\bibitem[{{Johnson}(2019)}]{johnson19_sedpy}
{Johnson}, B.~D. 2019, {SEDPY: Modules for storing and operating on
  astronomical source spectral energy distribution}, Astrophysics Source Code
  Library, record ascl:1905.026.
\newblock \doeprint{1905.026}

\bibitem[{{Johnson} {et~al.}(2021){Johnson}, {Leja}, {Conroy}, \&
  {Speagle}}]{johnson21}
{Johnson}, B.~D., {Leja}, J., {Conroy}, C., \& {Speagle}, J.~S. 2021, \apjs,
  254, 22, \dodoi{10.3847/1538-4365/abef67}

\bibitem[{{Johnson} {et~al.}(2019){Johnson}, {Leja}, {Conroy}, \&
  {Speagle}}]{johnson19}
{Johnson}, B.~D., {Leja}, J.~L., {Conroy}, C., \& {Speagle}, J.~S. 2019,
  {Prospector: Stellar population inference from spectra and SEDs}.
\newblock \doeprint{1905.025}

\bibitem[{{Kartaltepe} {et~al.}(2023){Kartaltepe}, {Rose}, {Vanderhoof},
  {McGrath}, {Costantin}, {Cox}, {Yung}, {Kocevski}, {Wuyts}, {Ferguson},
  {Bagley}, {Finkelstein}, {Amor{\'\i}n}, {Andrews}, {Aarabal Haro},
  {Backhaus}, {Behroozi}, {Bisigello}, {Calabr{\`o}}, {Casey}, {Coogan},
  {Cooper}, {Croton}, {de la Vega}, {Dickinson}, {Fontana}, {Franco},
  {Grazian}, {Grogin}, {Hathi}, {Holwerda}, {Huertas-Company}, {Iyer}, {Jogee},
  {Jung}, {Kewley}, {Kirkpatrick}, {Koekemoer}, {Liu}, {Lotz}, {Lucas},
  {Newman}, {Pacifici}, {Pandya}, {Papovich}, {Pentericci},
  {P{\'e}rez-Gonz{\'a}lez}, {Petersen}, {Pirzkal}, {Rafelski}, {Ravindranath},
  {Simons}, {Snyder}, {Somerville}, {Stanway}, {Straughn}, {Tacchella},
  {Trump}, {Vega-Ferrero}, {Wilkins}, {Yang}, \& {Zavala}}]{kartaltepe23}
{Kartaltepe}, J.~S., {Rose}, C., {Vanderhoof}, B.~N., {et~al.} 2023, \apjl,
  946, L15, \dodoi{10.3847/2041-8213/acad01}

\bibitem[{{Kawinwanichakij} {et~al.}(2021){Kawinwanichakij}, {Silverman},
  {Ding}, {George}, {Damjanov}, {Sawicki}, {Tanaka}, {Taranu}, {Birrer},
  {Huang}, {Li}, {Onodera}, {Shibuya}, \& {Yasuda}}]{kawinwanichakij21}
{Kawinwanichakij}, L., {Silverman}, J.~D., {Ding}, X., {et~al.} 2021, \apj,
  921, 38, \dodoi{10.3847/1538-4357/ac1f21}

\bibitem[{{Khimey} {et~al.}(2021){Khimey}, {Bose}, \& {Tacchella}}]{khimey21}
{Khimey}, D., {Bose}, S., \& {Tacchella}, S. 2021, \mnras, 506, 4139,
  \dodoi{10.1093/mnras/stab2019}

\bibitem[{{Kroupa}(2001)}]{kroupa01}
{Kroupa}, P. 2001, \mnras, 322, 231, \dodoi{10.1046/j.1365-8711.2001.04022.x}

\bibitem[{{Krumholz} \& {Thompson}(2012)}]{krumholz12}
{Krumholz}, M.~R., \& {Thompson}, T.~A. 2012, \apj, 760, 155,
  \dodoi{10.1088/0004-637X/760/2/155}

\bibitem[{{Larson} {et~al.}(2022){Larson}, {Finkelstein}, {Hutchison},
  {Papovich}, {Bagley}, {Dickinson}, {Rojas-Ruiz}, {Ferguson}, {Jung},
  {Giavalisco}, {Grazian}, {Pentericci}, \& {Tacchella}}]{larson22}
{Larson}, R.~L., {Finkelstein}, S.~L., {Hutchison}, T.~A., {et~al.} 2022, \apj,
  930, 104, \dodoi{10.3847/1538-4357/ac5dbd}

\bibitem[{{Leja} {et~al.}(2019){Leja}, {Carnall}, {Johnson}, {Conroy}, \&
  {Speagle}}]{leja19_nonparm}
{Leja}, J., {Carnall}, A.~C., {Johnson}, B.~D., {Conroy}, C., \& {Speagle},
  J.~S. 2019, \apj, 876, 3, \dodoi{10.3847/1538-4357/ab133c}

\bibitem[{{Leonova} {et~al.}(2022){Leonova}, {Oesch}, {Qin}, {Naidu}, {Wyithe},
  {de Barros}, {Bouwens}, {Ellis}, {Endsley}, {Hutter}, {Illingworth},
  {Kerutt}, {Labb{\'e}}, {Laporte}, {Magee}, {Mutch}, {Roberts-Borsani},
  {Smit}, {Stark}, {Stefanon}, {Tacchella}, \& {Zitrin}}]{leonova22}
{Leonova}, E., {Oesch}, P.~A., {Qin}, Y., {et~al.} 2022, \mnras, 515, 5790,
  \dodoi{10.1093/mnras/stac1908}

\bibitem[{{Lovell} {et~al.}(2023){Lovell}, {Harrison}, {Harikane}, {Tacchella},
  \& {Wilkins}}]{lovell23}
{Lovell}, C.~C., {Harrison}, I., {Harikane}, Y., {Tacchella}, S., \& {Wilkins},
  S.~M. 2023, \mnras, 518, 2511, \dodoi{10.1093/mnras/stac3224}

\bibitem[{{Maiolino} {et~al.}(2023){Maiolino}, {Scholtz}, {Witstok},
  {Carniani}, {D'Eugenio}, {de Graaff}, {Uebler}, {Tacchella}, {Curtis-Lake},
  {Arribas}, {Bunker}, {Charlot}, {Chevallard}, {Curti}, {Looser}, {Maseda},
  {Rawle}, {Rodriguez Del Pino}, {Willott}, {Egami}, {Eisenstein}, {Hainline},
  {Robertson}, {Williams}, {Willmer}, {Baker}, {Boyett}, {DeCoursey}, {Fabian},
  {Helton}, {Ji}, {Jones}, {Kumari}, {Laporte}, {Nelson}, {Perna}, {Sandles},
  {Shivaei}, \& {Sun}}]{maiolino23}
{Maiolino}, R., {Scholtz}, J., {Witstok}, J., {et~al.} 2023, arXiv e-prints,
  arXiv:2305.12492, \dodoi{10.48550/arXiv.2305.12492}

\bibitem[{{Maksimova} {et~al.}(2021){Maksimova}, {Garrison}, {Eisenstein},
  {Hadzhiyska}, {Bose}, \& {Satterthwaite}}]{maksimova21}
{Maksimova}, N.~A., {Garrison}, L.~H., {Eisenstein}, D.~J., {et~al.} 2021,
  \mnras, 508, 4017, \dodoi{10.1093/mnras/stab2484}

\bibitem[{{McLeod} {et~al.}(2016){McLeod}, {McLure}, \& {Dunlop}}]{mcleod16}
{McLeod}, D.~J., {McLure}, R.~J., \& {Dunlop}, J.~S. 2016, \mnras, 459, 3812,
  \dodoi{10.1093/mnras/stw904}

\bibitem[{{Naidu} {et~al.}(2022{\natexlab{a}}){Naidu}, {Oesch}, {van Dokkum},
  {Nelson}, {Suess}, {Brammer}, {Whitaker}, {Illingworth}, {Bouwens},
  {Tacchella}, {Matthee}, {Allen}, {Bezanson}, {Conroy}, {Labbe}, {Leja},
  {Leonova}, {Magee}, {Price}, {Setton}, {Strait}, {Stefanon}, {Toft},
  {Weaver}, \& {Weibel}}]{naidu22_highz}
{Naidu}, R.~P., {Oesch}, P.~A., {van Dokkum}, P., {et~al.} 2022{\natexlab{a}},
  \apjl, 940, L14, \dodoi{10.3847/2041-8213/ac9b22}

\bibitem[{{Naidu} {et~al.}(2022{\natexlab{b}}){Naidu}, {Oesch}, {Setton},
  {Matthee}, {Conroy}, {Johnson}, {Weaver}, {Bouwens}, {Brammer}, {Dayal},
  {Illingworth}, {Barrufet}, {Belli}, {Bezanson}, {Bose}, {Heintz}, {Leja},
  {Leonova}, {Marques-Chaves}, {Stefanon}, {Toft}, {van der Wel}, {van Dokkum},
  {Weibel}, \& {Whitaker}}]{naidu22}
{Naidu}, R.~P., {Oesch}, P.~A., {Setton}, D.~J., {et~al.} 2022{\natexlab{b}},
  arXiv e-prints, arXiv:2208.02794, \dodoi{10.48550/arXiv.2208.02794}

\bibitem[{{Nelson} {et~al.}(2023){Nelson}, {Suess}, {Bezanson}, {Price}, {van
  Dokkum}, {Leja}, {Wang}, {Whitaker}, {Labb{\'e}}, {Barrufet}, {Brammer},
  {Eisenstein}, {Gibson}, {Hartley}, {Johnson}, {Heintz}, {Mathews}, {Miller},
  {Oesch}, {Sandles}, {Setton}, {Speagle}, {Tacchella}, {Tadaki}, {{\"U}bler},
  \& {Weaver}}]{nelson23}
{Nelson}, E.~J., {Suess}, K.~A., {Bezanson}, R., {et~al.} 2023, \apjl, 948,
  L18, \dodoi{10.3847/2041-8213/acc1e1}

\bibitem[{{Oesch} {et~al.}(2013){Oesch}, {Bouwens}, {Illingworth}, {Labb{\'e}},
  {Franx}, {van Dokkum}, {Trenti}, {Stiavelli}, {Gonzalez}, \&
  {Magee}}]{oesch13b}
{Oesch}, P.~A., {Bouwens}, R.~J., {Illingworth}, G.~D., {et~al.} 2013, \apj,
  773, 75, \dodoi{10.1088/0004-637X/773/1/75}

\bibitem[{{Oesch} {et~al.}(2014){Oesch}, {Bouwens}, {Illingworth}, {Labb{\'e}},
  {Smit}, {Franx}, {van Dokkum}, {Momcheva}, {Ashby}, {Fazio}, {Huang},
  {Willner}, {Gonzalez}, {Magee}, {Trenti}, {Brammer}, {Skelton}, \&
  {Spitler}}]{oesch14}
---. 2014, \apj, 786, 108, \dodoi{10.1088/0004-637X/786/2/108}

\bibitem[{{Oesch} {et~al.}(2016){Oesch}, {Brammer}, {van Dokkum},
  {Illingworth}, {Bouwens}, {Labb{\'e}}, {Franx}, {Momcheva}, {Ashby}, {Fazio},
  {Gonzalez}, {Holden}, {Magee}, {Skelton}, {Smit}, {Spitler}, {Trenti}, \&
  {Willner}}]{oesch16}
{Oesch}, P.~A., {Brammer}, G., {van Dokkum}, P.~G., {et~al.} 2016, \apj, 819,
  129, \dodoi{10.3847/0004-637X/819/2/129}

\bibitem[{{Ono} {et~al.}(2022){Ono}, {Harikane}, {Ouchi}, {Yajima}, {Abe},
  {Isobe}, {Shibuya}, {Zhang}, {Nakajima}, \& {Umeda}}]{ono22}
{Ono}, Y., {Harikane}, Y., {Ouchi}, M., {et~al.} 2022, arXiv e-prints,
  arXiv:2208.13582, \dodoi{10.48550/arXiv.2208.13582}

\bibitem[{{Ostriker} \& {Shetty}(2011)}]{ostriker11}
{Ostriker}, E.~C., \& {Shetty}, R. 2011, \apj, 731, 41,
  \dodoi{10.1088/0004-637X/731/1/41}

\bibitem[{{Papovich} {et~al.}(2022){Papovich}, {Cole}, {Yang}, {Finkelstein},
  {Barro}, {Buat}, {Burgarella}, {P{\'e}rez-Gonz{\'a}lez}, {Santini},
  {Seill{\'e}}, {Shen}, {Arrabal Haro}, {Bagley}, {Bell}, {Bisigello},
  {Calabr{\`o}}, {Casey}, {Castellano}, {Chworowsky}, {Cleri}, {Cooper},
  {Costantin}, {Dickinson}, {Ferguson}, {Fontana}, {Giavalisco}, {Grazian},
  {Grogin}, {Hathi}, {Holwerda}, {Hutchison}, {Kartaltepe}, {Kewley},
  {Kirkpatrick}, {Kocevski}, {Koekemoer}, {Larson}, {Long}, {Lucas},
  {Pentericci}, {Pirzkal}, {Ravindranath}, {Somerville}, {Trump}, {Urbano
  Stawinski}, {Weiner}, {Wilkins}, {Yung}, \& {Zavala}}]{papovich22}
{Papovich}, C., {Cole}, J., {Yang}, G., {et~al.} 2022, arXiv e-prints,
  arXiv:2301.00027, \dodoi{10.48550/arXiv.2301.00027}

\bibitem[{{Perrin} {et~al.}(2014){Perrin}, {Sivaramakrishnan}, {Lajoie},
  {Elliott}, {Pueyo}, {Ravindranath}, \& {Albert}}]{webbpsf}
{Perrin}, M.~D., {Sivaramakrishnan}, A., {Lajoie}, C.-P., {et~al.} 2014, in
  Society of Photo-Optical Instrumentation Engineers (SPIE) Conference Series,
  Vol. 9143, Space Telescopes and Instrumentation 2014: Optical, Infrared, and
  Millimeter Wave, ed. J.~{Oschmann}, Jacobus~M., M.~{Clampin}, G.~G. {Fazio},
  \& H.~A. {MacEwen}, 91433X, \dodoi{10.1117/12.2056689}

\bibitem[{{Planck Collaboration} {et~al.}(2020){Planck Collaboration},
  {Aghanim}, {Akrami}, {Ashdown}, {Aumont}, {Baccigalupi}, {Ballardini},
  {Banday}, {Barreiro}, {Bartolo}, {Basak}, {Battye}, {Benabed}, {Bernard},
  {Bersanelli}, {Bielewicz}, {Bock}, {Bond}, {Borrill}, {Bouchet}, {Boulanger},
  {Bucher}, {Burigana}, {Butler}, {Calabrese}, {Cardoso}, {Carron},
  {Challinor}, {Chiang}, {Chluba}, {Colombo}, {Combet}, {Contreras}, {Crill},
  {Cuttaia}, {de Bernardis}, {de Zotti}, {Delabrouille}, {Delouis}, {Di
  Valentino}, {Diego}, {Dor{\'e}}, {Douspis}, {Ducout}, {Dupac}, {Dusini},
  {Efstathiou}, {Elsner}, {En{\ss}lin}, {Eriksen}, {Fantaye}, {Farhang},
  {Fergusson}, {Fernandez-Cobos}, {Finelli}, {Forastieri}, {Frailis},
  {Fraisse}, {Franceschi}, {Frolov}, {Galeotta}, {Galli}, {Ganga},
  {G{\'e}nova-Santos}, {Gerbino}, {Ghosh}, {Gonz{\'a}lez-Nuevo}, {G{\'o}rski},
  {Gratton}, {Gruppuso}, {Gudmundsson}, {Hamann}, {Handley}, {Hansen},
  {Herranz}, {Hildebrandt}, {Hivon}, {Huang}, {Jaffe}, {Jones}, {Karakci},
  {Keih{\"a}nen}, {Keskitalo}, {Kiiveri}, {Kim}, {Kisner}, {Knox},
  {Krachmalnicoff}, {Kunz}, {Kurki-Suonio}, {Lagache}, {Lamarre}, {Lasenby},
  {Lattanzi}, {Lawrence}, {Le Jeune}, {Lemos}, {Lesgourgues}, {Levrier},
  {Lewis}, {Liguori}, {Lilje}, {Lilley}, {Lindholm}, {L{\'o}pez-Caniego},
  {Lubin}, {Ma}, {Mac{\'\i}as-P{\'e}rez}, {Maggio}, {Maino}, {Mandolesi},
  {Mangilli}, {Marcos-Caballero}, {Maris}, {Martin}, {Martinelli},
  {Mart{\'\i}nez-Gonz{\'a}lez}, {Matarrese}, {Mauri}, {McEwen}, {Meinhold},
  {Melchiorri}, {Mennella}, {Migliaccio}, {Millea}, {Mitra},
  {Miville-Desch{\^e}nes}, {Molinari}, {Montier}, {Morgante}, {Moss}, {Natoli},
  {N{\o}rgaard-Nielsen}, {Pagano}, {Paoletti}, {Partridge}, {Patanchon},
  {Peiris}, {Perrotta}, {Pettorino}, {Piacentini}, {Polastri}, {Polenta},
  {Puget}, {Rachen}, {Reinecke}, {Remazeilles}, {Renzi}, {Rocha}, {Rosset},
  {Roudier}, {Rubi{\~n}o-Mart{\'\i}n}, {Ruiz-Granados}, {Salvati}, {Sandri},
  {Savelainen}, {Scott}, {Shellard}, {Sirignano}, {Sirri}, {Spencer},
  {Sunyaev}, {Suur-Uski}, {Tauber}, {Tavagnacco}, {Tenti}, {Toffolatti},
  {Tomasi}, {Trombetti}, {Valenziano}, {Valiviita}, {Van Tent}, {Vibert},
  {Vielva}, {Villa}, {Vittorio}, {Wandelt}, {Wehus}, {White}, {White},
  {Zacchei}, \& {Zonca}}]{planck-collaboration20}
{Planck Collaboration}, {Aghanim}, N., {Akrami}, Y., {et~al.} 2020, \aap, 641,
  A6, \dodoi{10.1051/0004-6361/201833910}

\bibitem[{{Rigby} {et~al.}(2022){Rigby}, {Perrin}, {McElwain}, {Kimble},
  {Friedman}, {Lallo}, {Doyon}, {Feinberg}, {Ferruit}, {Glasse}, \&
  et~al.}]{rigby22}
{Rigby}, J., {Perrin}, M., {McElwain}, M., {et~al.} 2022, arXiv e-prints,
  arXiv:2207.05632.
\newblock \doarXiv{2207.05632}

\bibitem[{{Robertson}(2022)}]{robertson22}
{Robertson}, B.~E. 2022, \araa, 60, 121,
  \dodoi{10.1146/annurev-astro-120221-044656}

\bibitem[{{Robertson} {et~al.}(2023{\natexlab{a}}){Robertson}, {Tacchella},
  {Johnson}, {Hainline}, {Whitler}, {Eisenstein}, {Endsley}, {Rieke}, {Stark},
  {Alberts}, {Dressler}, {Egami}, {Hausen}, {Rieke}, {Shivaei}, {Williams},
  {Willmer}, {Arribas}, {Bonaventura}, {Bunker}, {Cameron}, {Carniani},
  {Charlot}, {Chevallard}, {Curti}, {Curtis-Lake}, {D'Eugenio}, {Jakobsen},
  {Looser}, {L{\"u}tzgendorf}, {Maiolino}, {Maseda}, {Rawle}, {Rix}, {Smit},
  {{\"U}bler}, {Willott}, {Witstok}, {Baum}, {Bhatawdekar}, {Boyett}, {Chen},
  {de Graaff}, {Florian}, {Helton}, {Hviding}, {Ji}, {Kumari}, {Lyu}, {Nelson},
  {Sandles}, {Saxena}, {Suess}, {Sun}, {Topping}, \& {Wallace}}]{robertson23}
{Robertson}, B.~E., {Tacchella}, S., {Johnson}, B.~D., {et~al.}
  2023{\natexlab{a}}, Nature Astronomy, \dodoi{10.1038/s41550-023-01921-1}

\bibitem[{{Robertson} {et~al.}(2023{\natexlab{b}}){Robertson}, {Tacchella},
  {Johnson}, {Hausen}, {Alabi}, {Boyett}, {Bunker}, {Carniani}, {Egami},
  {Eisenstein}, {Hainline}, {Helton}, {Ji}, {Kumari}, {Lyu}, {Maiolino},
  {Nelson}, {Rieke}, {Shivaei}, {Sun}, {{\"U}bler}, {Williams}, {Willmer}, \&
  {Witstok}}]{robertson23_morph}
---. 2023{\natexlab{b}}, \apjl, 942, L42, \dodoi{10.3847/2041-8213/aca086}

\bibitem[{{Robotham} {et~al.}(2017){Robotham}, {Taranu}, {Tobar}, {Moffett}, \&
  {Driver}}]{robotham17}
{Robotham}, A.~S.~G., {Taranu}, D.~S., {Tobar}, R., {Moffett}, A., \& {Driver},
  S.~P. 2017, \mnras, 466, 1513, \dodoi{10.1093/mnras/stw3039}

\bibitem[{{Schlawin} {et~al.}(2020){Schlawin}, {Leisenring}, {Misselt},
  {Greene}, {McElwain}, {Beatty}, \& {Rieke}}]{schlawin20}
{Schlawin}, E., {Leisenring}, J., {Misselt}, K., {et~al.} 2020, \aj, 160, 231,
  \dodoi{10.3847/1538-3881/abb811}

\bibitem[{{Somerville} {et~al.}(2015){Somerville}, {Popping}, \&
  {Trager}}]{somerville15}
{Somerville}, R.~S., {Popping}, G., \& {Trager}, S.~C. 2015, \mnras, 453, 4337,
  \dodoi{10.1093/mnras/stv1877}

\bibitem[{{Speagle}(2020)}]{speagle20}
{Speagle}, J.~S. 2020, \mnras, 493, 3132, \dodoi{10.1093/mnras/staa278}

\bibitem[{{Stark}(2016)}]{stark16}
{Stark}, D.~P. 2016, \araa, 54, 761,
  \dodoi{10.1146/annurev-astro-081915-023417}

\bibitem[{{Steidel} {et~al.}(2016){Steidel}, {Strom}, {Pettini}, {Rudie},
  {Reddy}, \& {Trainor}}]{steidel16}
{Steidel}, C.~C., {Strom}, A.~L., {Pettini}, M., {et~al.} 2016, \apj, 826, 159,
  \dodoi{10.3847/0004-637X/826/2/159}

\bibitem[{{Steinhardt} {et~al.}(2022){Steinhardt}, {Kokorev}, {Rusakov},
  {Garcia}, \& {Sneppen}}]{steinhardt22}
{Steinhardt}, C.~L., {Kokorev}, V., {Rusakov}, V., {Garcia}, E., \& {Sneppen},
  A. 2022, arXiv e-prints, arXiv:2208.07879, \dodoi{10.48550/arXiv.2208.07879}

\bibitem[{{Suess} {et~al.}(2022){Suess}, {Bezanson}, {Nelson}, {Setton},
  {Price}, {van Dokkum}, {Brammer}, {Labb{\'e}}, {Leja}, {Miller}, {Robertson},
  {Wel}, {Weaver}, \& {Whitaker}}]{suess22}
{Suess}, K.~A., {Bezanson}, R., {Nelson}, E.~J., {et~al.} 2022, \apjl, 937,
  L33, \dodoi{10.3847/2041-8213/ac8e06}

\bibitem[{{Tacchella} {et~al.}(2018){Tacchella}, {Bose}, {Conroy},
  {Eisenstein}, \& {Johnson}}]{tacchella18}
{Tacchella}, S., {Bose}, S., {Conroy}, C., {Eisenstein}, D.~J., \& {Johnson},
  B.~D. 2018, \apj, 868, 92, \dodoi{10.3847/1538-4357/aae8e0}

\bibitem[{{Tacchella} {et~al.}(2016){Tacchella}, {Dekel}, {Carollo},
  {Ceverino}, {DeGraf}, {Lapiner}, {Mandelker}, \&
  {Primack}}]{tacchella16_profile}
{Tacchella}, S., {Dekel}, A., {Carollo}, C.~M., {et~al.} 2016, \mnras, 458,
  242, \dodoi{10.1093/mnras/stw303}

\bibitem[{{Tacchella} {et~al.}(2022{\natexlab{a}}){Tacchella}, {Johnson},
  {Robertson}, {Carniani}, {D'Eugenio}, {Kumar}, {Maiolino}, {Nelson}, {Suess},
  {{\"U}bler}, {Williams}, {Adebusola}, {Alberts}, {Arribas}, {Bhatawdekar},
  {Bonaventura}, {Bowler}, {Bunker}, {Cameron}, {Curti}, {Egami}, {Eisenstein},
  {Frye}, {Hainline}, {Helton}, {Ji}, {Looser}, {Lyu}, {Perna}, {Rawle},
  {Rieke}, {Rieke}, {Saxena}, {Sandles}, {Shivaei}, {Simmonds}, {Sun},
  {Willmer}, {Willott}, \& {Witstok}}]{tacchella22}
{Tacchella}, S., {Johnson}, B.~D., {Robertson}, B.~E., {et~al.}
  2022{\natexlab{a}}, arXiv e-prints, arXiv:2208.03281.
\newblock \doarXiv{2208.03281}

\bibitem[{{Tacchella} {et~al.}(2022{\natexlab{b}}){Tacchella}, {Finkelstein},
  {Bagley}, {Dickinson}, {Ferguson}, {Giavalisco}, {Graziani}, {Grogin},
  {Hathi}, {Hutchison}, {Jung}, {Koekemoer}, {Larson}, {Papovich}, {Pirzkal},
  {Rojas-Ruiz}, {Song}, {Schneider}, {Somerville}, {Wilkins}, \&
  {Yung}}]{tacchella22_highz}
{Tacchella}, S., {Finkelstein}, S.~L., {Bagley}, M., {et~al.}
  2022{\natexlab{b}}, \apj, 927, 170, \dodoi{10.3847/1538-4357/ac4cad}

\bibitem[{{Tang} {et~al.}(2023){Tang}, {Stark}, {Chen}, {Mason}, {Topping},
  {Endsley}, {Senchyna}, {Plat}, {Lu}, {Whitler}, {Robertson}, \&
  {Charlot}}]{tang23}
{Tang}, M., {Stark}, D.~P., {Chen}, Z., {et~al.} 2023, arXiv e-prints,
  arXiv:2301.07072, \dodoi{10.48550/arXiv.2301.07072}

\bibitem[{{Whitaker} {et~al.}(2019){Whitaker}, {Ashas}, {Illingworth}, {Magee},
  {Leja}, {Oesch}, {van Dokkum}, {Mowla}, {Bouwens}, {Franx}, {Holden},
  {Labb{\'e}}, {Rafelski}, {Teplitz}, \& {Gonzalez}}]{whitaker19}
{Whitaker}, K.~E., {Ashas}, M., {Illingworth}, G., {et~al.} 2019, \apjs, 244,
  16, \dodoi{10.3847/1538-4365/ab3853}

\bibitem[{{Whitler} {et~al.}(2023){Whitler}, {Stark}, {Endsley}, {Leja},
  {Charlot}, \& {Chevallard}}]{whitler23_sfh}
{Whitler}, L., {Stark}, D.~P., {Endsley}, R., {et~al.} 2023, \mnras, 519, 5859,
  \dodoi{10.1093/mnras/stad004}

\bibitem[{{Williams} {et~al.}(2023){Williams}, {Tacchella}, {Maseda},
  {Robertson}, {Johnson}, {Willott}, {Eisenstein}, {Willmer}, {Ji}, {Hainline},
  {Helton}, {Alberts}, {Baum}, {Bhatawdekar}, {Boyett}, {Bunker}, {Carniani},
  {Charlot}, {Chevallard}, {Curtis-Lake}, {de Graaf}, {Egami}, {Franx},
  {Kumari}, {Maiolino}, {Nelson}, {Rieke}, {Sandles}, {Shivaei}, {Simmonds},
  {Smit}, {Suess}, {Sun}, {Ubler}, \& {Witstok}}]{williams23}
{Williams}, C.~C., {Tacchella}, S., {Maseda}, M.~V., {et~al.} 2023, arXiv
  e-prints, arXiv:2301.09780, \dodoi{10.48550/arXiv.2301.09780}

\bibitem[{{Wu} {et~al.}(2020){Wu}, {Dav{\'e}}, {Tacchella}, \& {Lotz}}]{wu20}
{Wu}, X., {Dav{\'e}}, R., {Tacchella}, S., \& {Lotz}, J. 2020, \mnras, 494,
  5636, \dodoi{10.1093/mnras/staa1044}

\bibitem[{{Zavala} {et~al.}(2023){Zavala}, {Buat}, {Casey}, {Finkelstein},
  {Burgarella}, {Bagley}, {Ciesla}, {Daddi}, {Dickinson}, {Ferguson}, {Franco},
  {Jim{\'e}nez-Andrade}, {Kartaltepe}, {Koekemoer}, {Bail}, {Murphy},
  {Papovich}, {Tacchella}, {Wilkins}, {Aretxaga}, {Behroozi}, {Champagne},
  {Fontana}, {Giavalisco}, {Grazian}, {Grogin}, {Kewley}, {Kocevski},
  {Kirkpatrick}, {Lotz}, {Pentericci}, {P{\'e}rez-Gonz{\'a}lez}, {Pirzkal},
  {Ravindranath}, {Somerville}, {Trump}, {Yang}, {Yung}, {Almaini},
  {Amor{\'\i}n}, {Annunziatella}, {Haro}, {Backhaus}, {Barro}, {Bell},
  {Bhatawdekar}, {Bisigello}, {Buitrago}, {Calabr{\`o}}, {Castellano},
  {Ch{\'a}vez Ortiz}, {Chworowsky}, {Cleri}, {Cohen}, {Cole}, {Cooke},
  {Cooper}, {Cooray}, {Costantin}, {Cox}, {Croton}, {Dav{\'e}}, {de La Vega},
  {Dekel}, {Elbaz}, {Estrada-Carpenter}, {Fern{\'a}ndez}, {Finkelstein},
  {Freundlich}, {Fujimoto}, {Garc{\'\i}a-Argum{\'a}nez}, {Gardner}, {Gawiser},
  {G{\'o}mez-Guijarro}, {Guo}, {Hamilton}, {Hathi}, {Holwerda}, {Hirschmann},
  {Huertas-Company}, {Hutchison}, {Iyer}, {Jaskot}, {Jha}, {Jogee}, {Juneau},
  {Jung}, {Kassin}, {Kurczynski}, {Larson}, {Leung}, {Long}, {Lucas},
  {Magnelli}, {Mantha}, {Matharu}, {McGrath}, {McIntosh}, {Medrano}, {Merlin},
  {Mobasher}, {Morales}, {Newman}, {Nicholls}, {Pandya}, {Rafelski}, {Ronayne},
  {Rose}, {Ryan}, {Santini}, {Seill{\'e}}, {Shah}, {Shen}, {Simons}, {Snyder},
  {Stanway}, {Straughn}, {Teplitz}, {Vanderhoof}, {Vega-Ferrero}, {Wang},
  {Weiner}, {Willmer}, {Wuyts}, \& {(The Ceers Team)}}]{zavala23}
{Zavala}, J.~A., {Buat}, V., {Casey}, C.~M., {et~al.} 2023, \apjl, 943, L9,
  \dodoi{10.3847/2041-8213/acacfe}

\bibitem[{{Zolotov} {et~al.}(2015){Zolotov}, {Dekel}, {Mandelker}, {Tweed},
  {Inoue}, {DeGraf}, {Ceverino}, {Primack}, {Barro}, \& {Faber}}]{zolotov15}
{Zolotov}, A., {Dekel}, A., {Mandelker}, N., {et~al.} 2015, \mnras, 450, 2327,
  \dodoi{10.1093/mnras/stv740}

\end{thebibliography}

\end{document}